\documentclass[prd,aps,twocolumn,a4paper,floatfix,showpacs,nofootinbib]{revtex4-1}

\usepackage{graphicx,psfrag}
\usepackage{mathrsfs}
\usepackage{amsmath,amsfonts,amssymb}
\usepackage{multirow}
\usepackage{comment,hyperref}
\usepackage{color}

\newcommand{\be}{\begin{equation}}
\newcommand{\ee}{\end{equation}}
\newcommand{\bea}{\begin{eqnarray}}
\newcommand{\eea}{\end{eqnarray}}
\newcommand{\bel}{\begin{align}}
\newcommand{\eel}{\end{align}}

\newcommand{\mean}[1]{\langle#1\rangle}

\def\GMc2{{\rm G M_{\odot} c^{-2}}}
\def\Mpc{{\rm Mpc}}

\usepackage{color}

\definecolor{cyan}{rgb}{0,0.9,0.9}
\definecolor{orange}{rgb}{0.9,0.5,0}
\definecolor{magenta}{rgb}{1,0,1}
\definecolor{purple}{rgb}{0.8,0.4,0.8}
\definecolor{gray}{rgb}{0.5,0.5,0.5}

\newcommand{\PE}[2]{${{#1} \times 10^{-{#2}}}$}

\begin{document}

\title{Gravitational waves and mass ejecta from binary neutron star
  mergers:\\ Effect of the mass-ratio}

 \author{Tim \surname{Dietrich}${}^{1}$}
 \author{Maximiliano \surname{Ujevic}$^{2}$}
 \author{Wolfgang \surname{Tichy}$^{3}$}
 \author{Sebastiano \surname{Bernuzzi}$^{4}$}
 \author{Bernd \surname{Br\"ugmann}$^{5}$}
 \affiliation{${}^1$ Max-Planck-Institut for Gravitational Physics, Albert-Einstein-Institut, D-14476 Golm, Germany}
 \affiliation{${}^2$ Centro de Ci\^encias Naturais e Humanas, Universidade Federal do ABC, 09210-170, Santo Andr\'e, S\~ao Paulo, Brazil}
 \affiliation{${}^3$ Department of Physics, Florida Atlantic University, Boca Raton, FL 33431 USA}
 \affiliation{${}^4$ DiFeST, University of Parma, and INFN Parma  I-43124 Parma, Italy}  
 \affiliation{${}^5$ Theoretical Physics Institute, University of Jena, 07743 Jena, Germany}
\date{\today}

\begin{abstract}
We present new (3+1)D numerical relativity simulations 
of the binary neutron star (BNS) merger and postmerger phase.
We focus on a previously inaccessible region of the binary parameter space 
spanning the binary's mass-ratio $q\sim1.00-1.75$ for different total
masses and equations of state, and up to $q\sim2$ for a stiff BNS system. 
We study the mass-ratio effect on the gravitational waves (GWs) and on
the possible electromagnetic emission associated to dynamical mass 
ejecta. 
We compute waveforms, spectra, and spectrograms of the GW strain
including all the multipoles up to $l=4$. The mass-ratio has
a specific imprint on the GW multipoles in the late-inspiral-merger
signal, and it affects qualitatively the spectra of the merger remnant.
The multipole effect is also studied by considering the dependency of
the GW spectrograms on the source's sky location.
Unequal mass BNSs produce more ejecta than equal mass systems with
ejecta masses and kinetic energies depending almost linearly on $q$. 
We estimate luminosity peaks and light curves of macronovae events
associated to the mergers using a simple approach. For $q\sim2$ the
luminosity peak is delayed for several days and can be  
up to four times larger than for the $q=1$ cases.
The macronova emission associated with the $q\sim2$ BNS is more persistent in time and 
could be observed for weeks instead of few days ($q=1$) in the near infrared.
Finally, we estimate the flux of possible radio flares produced by the
interaction of relativistic outflows with the surrounding medium. Also
in this case a large $q$ can significantly enhance the emission and 
delay the peak luminosity.
Overall, our results indicate that BNS merger with large mass ratio
have EM signatures distinct from the equal mass case and
more similar to black hole - neutron star binaries.  
\end{abstract}


\maketitle

\section{Introduction}
\label{sec:intro}

Binary neutron star (BNS) mergers are primary sources of gravitational
waves (GWs) and are associated with a variety of electromagnetic (EM) 
emissions. 
GW observations of BNS are eagerly expected in the upcoming
LIGO-Virgo runs, after the first binary black hole (BBH) GW detections
GW150914~\cite{Abbott:2016blz} and GW151226~\cite{Abbott:2016nmj}. 
Such GWs will allow us to place constraints on
the nature of matter at densities above nuclear density,
e.g.~\cite{DelPozzo:2013ala}, and to identify the origin of EM
emissions like kilo/macronovae and short gamma-ray burst (SGRB),
e.g.~\cite{Rosswog:2015nja,Fernandez:2015use}.  
Kilo/macronovae events are transient emissions in the optical or
near-infrared band observed in e.g.~\cite{Tanvir:2013pia}. They are believed to be
triggered by the radioactive decay of r-process nuclei 
in the neutron-rich material ejected during a BNS merger.
SGRBs models are instead based on highly relativistic outflows, powered
e.g.~by the merger remnant accretion disk~\cite{Paczynski:1986px,Eichler:1989ve}. 
Their joint observation with GWs might be challenging due to the short duration of
the burst and to the fact that they are highly collimated 
emissions~\cite{Soderberg:2006bn}. 
Additionally, the interaction of mildly or sub- relativistic outflows with
the surrounding material generates synchrotron radiation known as radio
flares~\cite{Nakar:2011cw}. This emission can persist from months to
years after merger, depending on the composition, which makes radio
flares a particularly attractive EM counterpart to detect.  

Understanding the dependency of the GW and EM emissions on the
source's parameters is of key importance for GW astronomy and
multimessenger astrophysics. The BNS parameter space is composed by the
component masses (and spins), and by a choice of equation of state
(EOS) describing the NS matter. The ranges of the mass and spins
parameters are rather uncertain.    

The expected NS mass range is $\sim 0.9 - 3 M_\odot$. The lower bound is
inferred from the formation scenario (gravitational-collapse) and from
current observations, although those measurements have typically large uncertainties, 
see e.g.~\cite{Rawls:2011jw,Ozel:2012ax}. The upper bound is inferred
from stability argument (maximum theoretical mass), from precise
measurements of $\sim 2M_\odot$ NSs in double NS 
systems~\cite{Demorest:2010bx,Antoniadis:2013pzd}, and from models of
SGRBs, which suggest a maximum mass of
$2.2M_\odot$, e.g.~\cite{Lawrence:2015oka}. There exist observations
of larger NS masses but with large uncertainties, so they cannot give
strong constraints on the NS maximum mass \cite{Lattimer:2012nd}. 
Also, the maximum mass of a NS is determined by the particular
EOS. Most tabulated EOS compatible with astrophysical constraints
support NSs with maximum masses in a range of $\sim 2-3M_\odot$. 

The above considerations suggest that the BNS mass-ratio 
\be
q := M^A/M^B\geq1 \ ,
\ee
where $M^{A,B}$ are the individual gravitational masses of the NSs (in
isolation), is most likely constrained to $q\lesssim 2-3$.
Observations suggest that BNS systems consist of equal mass NSs with
masses of around $\sim 1.35 M_\odot$ and that the mass-ratio is close
to one e.g.~\cite{Swiggum:2015yras,Lattimer:2012nd}. However, only
approximately a dozen BNS systems are known so far and only $6$ of these
systems have well determined masses and will merge within a Hubble time.
The lack of very unequal mass configurations might only be a selection
effect. For example Ref.~\cite{Martinez:2015mya} discovered a compact
binary system with a mass ratio of $q \approx 1.3$, suggesting that
BNS with larger mass ratios exist. Population synthesis models for binaries 
formed ``\emph{in situ}'' predict a wider range of masses and 
mass ratios up to $q\approx 1.9$~\cite{Dominik:2012kk,Dietrich:2015pxa}.

Also NS spins are constrained by theoretical arguments and observations,
e.g. \cite{Burgay:2003jj,Lynch:2011aa,Miller:2014aaa,Dietrich:2015pxa}. 
However, we do not consider here the NS rotation
and we remind to the above references and to future work \cite{Dietrich:2016prep2}.

Parameter space investigation of BNS are challenging due to the
unknown EOS, and the need of simulating each masses (and spins)
configuration with different EOS.
Most numerical relativity studies of BNS systems have focused on equal
masses and irrotational configurations. The first simulations of unequal
mass systems have been presented in 
\cite{Shibata:2003ga,Rezzolla:2010fd,Hotokezaka:2012ze} using
polytropic and piecewise polytropic EOS. 
Mass ejecta in $q\neq1$ BNS simulations have been studied in 
e.g.~\cite{Bauswein:2013yna,Hotokezaka:2012ze,Dietrich:2015iva}.
Unequal mass simulations with microphysical EOS and neutrino cooling
have been presented in
e.g.~\cite{Bernuzzi:2015opx,Lehner:2016wjg,Lehner:2016lxy} 
and with radiation-hydrodynamics in~\cite{Sekiguchi:2016bjd}. 
Previous works were restricted to mass ratios $q\leq 1.35$. 
Overall, the main results are that i) asymmetric mergers produce more
massive ejecta with smaller electron fractions than the corresponding equal mass setups
ii) unequal mass systems are likely to produce kilonovae and 
iii) the remnant disk mass increases for an increasing mass ratio. 

In \cite{Dietrich:2015pxa} we reported an upgrade of the SGRID code
able to generate generic initial data for BNS simulations together
with few preliminary evolutions. Among other results, we showed the
possibility of generating ``large mass-ratio'' configurations with
$q\sim2$. The test evolution of a $q=2$ BNS showed interesting features,
including large mass ejection and mass transfer from one star to the
other during the last revolutions. 

In this work we study the effect of the binary's mass-ratio $q$ on the
GWs and on the characteristics of possible EM emission
associated to dynamical mass ejecta, in particular macronovae and radio flares. 
We present a new set of (3+1)D numerical relativity simulations 
of the merger and postmerger phase, and focus on a previously
inaccessible region of the binary parameter space spanning
$q\in[1,1.75]$ for different masses and equations of state, and a
setup with $q=2$. 

The article is structured as follows: 
In Sec.~\ref{sec:simeth}, we give a short description of the numerical methods 
and describe important quantities used to analyze our simulations. 
Section~\ref{sec:config} summarizes our configurations and the investigated part 
of the BNS parameter space. Section~\ref{sec:dynamics} deals with the dynamics of the simulation, 
where in particular we focus on the mass-transfer, the ejecta, and the final remnant. 
The GW signal is investigated in Sec.~\ref{sec:GW} with respect to spectrograms, 
the sky location and the emitted GW energy per mode. 
Sec.~\ref{sec:EM} focuses on EM counterparts and 
we conclude in Sec.~\ref{sec:conclusion}. 
In Appendix~\ref{sec:accuracy} we test the accuracy of our simulations with respect to conserved quantities, 
convergence, the constraints. 

Throughout this work we use geometric units, setting $c=G=M_\odot=1$,
though we will sometimes include $M_\odot$ explicitly or quote values
in CGS units for better understanding.  Spatial indices are denoted by
Latin letters running from 1 to 3 and Greek letters are used for
spacetime indices running from 0 to 3.

\section{Simulation methods}
\label{sec:simeth}

\subsection{Initial configurations}

Our initial configurations are constructed with the SGRID 
code~\cite{Tichy:2006qn,Tichy:2009yr,Tichy:2009zr}.
SGRID uses pseudospectral methods to accurately compute spatial derivatives.
To obtain stationary configurations that are appropriate
as initial data, we use the conformal thin sandwich equations~\cite{Wilson:1995uh,Wilson:1996ty,York:1998hy}
together with assumptions of the constant rotational velocity approach~\cite{Tichy:2011gw,Tichy:2012rp}, 
which allows to construct generic NS binaries in hydrodynamical equilibrium. 
The computational domain is divided into six patches (Fig.~1 of \cite{Dietrich:2015pxa}). 
The domain reaches spatial infinity and thus allows to impose exact boundary conditions at spatial infinity. 
Recent changes presented in~\cite{Dietrich:2015pxa} allow us to construct 
configurations in almost all corners of the BNS parameter space 
including high mass ratios, spinning configurations, low and high eccentricity inspirals, 
as well as more realistic EOSs and highly compact stars. 
In this work we do not study the influence of the eccentricity and spin, but 
focus on the high-mass ratio configuration and the influence of the EOS. 

We employ $n_A=28,n_B=28,n_\varphi=8,n_{\rm Cart} = 24$ points for the spectral grid, 
cf.~\cite{Dietrich:2015pxa} for more details. In Tab.~\ref{tab:ID_and_grid_irr} we report 
the initial parameters, in particular, the initial ADM-mass and angular momentum of the system as well as 
the initial GW frequency.

\subsection{Evolutions}
  
Dynamical simulations are performed with the BAM 
code~\cite{Brugmann:2008zz,Thierfelder:2011yi,Dietrich:2015iva}. 
We use the Z4c scheme~\cite{Bernuzzi:2009ex,Hilditch:2012fp} and 
employ the 1+log and gamma-driver conditions for the evolution of the lapse 
and shift~\cite{Bona:1994a,Alcubierre:2002kk,vanMeter:2006vi}. 

The equations of general-relativistic hydrodynamics (GRHD) are solved in
conservative form by defining Eulerian conservative variables from the
rest-mass density $\rho$, pressure $p$, internal energy $\epsilon$, and
3-velocity, $v^i$. The system is closed by an EOS.
We model the EOS with piecewise polytropic fits of~\cite{Read:2008iy},
see Tab.~\ref{tab:EOS} below. 
Thermal effects are included during the simulation by an additive pressure
contribution given by
$p_{\rm th} = (\Gamma_{\rm th}-1)\rho\epsilon$~\cite{Shibata:2005ss,Bauswein:2010dn}.
As in our previous work we set $\Gamma_{\rm th}=1.75$.

The evolution algorithm is based on the method-of-lines with explicit
4th order Runge-Kutta time integrators. Finite difference stencils of 4th order 
are employed for the spatial derivatives of the metric. 
GRHD is solved by means of a high-resolution-shock-capturing 
method~\cite{Thierfelder:2011yi} based on primitive reconstruction
and the Local-Lax-Friedrichs (LLF) central scheme for the numerical
fluxes. Primitive reconstruction is performed with the 5th order WENOZ
scheme of~\cite{Borges20083191}. We do not employ the higher-order algorithm 
presented recently in~\cite{Bernuzzi:2016pie}, because the new method was 
not tested at the time these simulations were performed.

The numerical domain is made of a hierarchy of cell-centered nested
Cartesian grids. The hierarchy consists of $L$ levels of refinement labeled
by~$l = 0,...,L-1$. A refinement level $l$ has one or more
Cartesian grids with constant grid spacing $h_l$ and $n$ points per
direction. The refinement factor is two such that $h_l = h_0/2^l$. The grids
are properly nested, i.e., the coordinate extent of any grid at
level~$l$, $l > 0$, is completely covered by the grids at level~$l-1$.
Some of the mesh refinement levels $l>l^{\rm mv}$ can be dynamically moved and adapted
during the time evolution according to the technique of~``moving boxes''.
The BAM grid setups considered in this work consist of 7 refinement levels. 
We distinguish between two different grid setups. 
For the first setup, we substitute 
the outermost level ($l=0$) by a multipatch cubed-sphere
grid~\cite{Ronchi:1996,Thornburg:2004dv,Pollney:2009yz,Hilditch:2012fp} 
on which we do not solve the GRHD equations. 
The spheres allow us to apply constraint preserving boundary conditions~\cite{Ruiz:2010qj}. 
This `shell' setup is denoted in Tab.~\ref{tab:ID_and_grid_irr} with the ending `s'. 
The second grid setup does not make use of the multipatch cubed-spheres, so
all the refinement levels are Cartesian boxes. While this has the disadvantage that only 
Sommerfeld boundary conditions are used, it increases the computational speed 
and the matter fields are evolved on a larger region. We denote this `box'
setup by the ending `b' in Tab.~\ref{tab:ID_and_grid_irr}. 

The Berger-Oliger algorithm is employed for the time stepping~\cite{Berger:1984zza}
on the inner levels. As in~\cite{Dietrich:2014wja,Dietrich:2015iva,Dietrich:2015pxa}
we make use of an additional refluxing algorithm to enforce mass conservation 
across mesh refinement boundaries based on~\cite{East:2011aa,Berger:1989}. Restriction and prolongation between the 
refinement levels is performed with an average scheme and 2nd order essentially non-oscillatory
scheme, respectively.

\subsection{Simulation analysis} 
\label{sec:diagno}
 
We briefly want to summarize important quantities for our simulation analysis. 
For this purpose we follow the description of~\cite{Dietrich:2015iva} including 
also quantities to monitor the mass transfer and the GW spectrogram. \\

\paragraph*{Mass transfer:}
Because of the high mass ratios considered in this work, 
one might expect mass transport between the two NSs during the inspiral prior to merger. 
In our last work~\cite{Dietrich:2015iva} we estimated the amount of material 
flowing from one star to the other by measuring how
much mass is inside the Cartesian refinement boxes. 
Now, we estimate the mass transfer by computing the change of baryonic mass 
inside a coordinate sphere around the center of the NS. 
The center of the NS is defined as the minimum of the lapse. 
We perform the integration for different coordinate radii $r_{c,{\rm min}}=8.0$ 
up to $r_{c,{\rm max}}=13.0$ in steps of $\Delta r_c=0.5$. 
While a radius chosen too small does not cover all of the tidally deformed
star, if the radius is chosen too large the mass measurement will be effected by the other star a 
significant time before the actual merger.
We will focus on the radii $r_{c1}= r_{c2}=10.5$.\\

\paragraph*{Merger remnant:}
In agreement with the literature, e.g.~\cite{Baumgarte:1999cq}, 
we define the merger remnant as a hypermassive neutron star 
(HMNS) if its rest-mass is larger than the maximum rest-mass of 
a stable uniformly rotating star with the same EOS; 
or as a supramassive neutron star (SMNS)
if its rest-mass is smaller than the maximum rest-mass 
of a stable uniformly rotating star, but above the 
mass of a stable TOV-star.
In case its mass is also below the maximum supported mass 
of a TOV-star, we call it simply a massive neutron star (MNS).
Notice that these definitions \textit{cannot} be used strictly and should be seen as a 
qualitative description since they refer to equilibrium configurations
assuming barotropic EOS and axisymmetry.

For a HMNS the merger remnant collapses on a dynamical timescale to a BH. 
Typically the lifetime $\tau$ is the time from the moment of merger 
to the time an apparent horizon forms. 
The final BH is characterized by its horizon 
mass $M_{\rm BH}$ and dimensionless spin $j_{\rm BH}$.

The accretion disk around the BH has a mass of 
\begin{equation}
M_\text{disk} = \int_{r>r_{\rm AH}} \text{d}^3 x \ q^{(D)} \ , 
\end{equation}
with $ q^{(D)} = \sqrt{\gamma} D$ where $D$ is the fluid's rest frame baryonic mass, $\gamma$ the
determinant of the 3-metric, and the domain of
integration excludes the spherical region inside the apparent horizon.\\

\paragraph*{Mass ejecta:}
As in~\cite{Dietrich:2015iva} we label material as ejecta when 
\begin{equation}
\label{eq:unbound}
u_t<-1 \ \ \text{and} \ \ \bar{v}_r = v^i x_i >0 \ ,
\end{equation}
where $u_t = -W (\alpha - \beta_i v^i) $ is
the first lower component of the fluid 4-velocity
with the lapse $\alpha$, the shift $\beta^i$, and the Lorentz factor $W$.
and $x^i = (x,y,z)$. 
Equation~\eqref{eq:unbound} assumes that the fluid elements follow geodesics
and requires that the orbit is unbound and has an outward pointing velocity, 
cf.~also~\cite{East:2012ww}. Other ways of estimating the ejecta mass 
can be found in e.g.~\cite{Kastaun:2014fna,Radice:2016dwd}.
The total ejecta mass is given by 
\begin{equation}
M_\text{ej} = 
\int_{\mathcal{U}} \text{d}^3 x \ q^{(D)} \ ,
\end{equation}
 where the integral is computed on the region, 
\begin{equation}
\mathcal{U}=\{ x^i=(x,y,z)\, : \, u_t<-1 \ \ \text{and} \ \ \bar{v}_r >0 \} \ ,
\end{equation}
on which material is unbound according to Eq.~\eqref{eq:unbound}.
The kinetic energy of the ejecta can be approximated as the difference
between the total energy $E_\text{ejecta}$ (excluding gravitational
potential energy), the rest-mass, and the total internal energy
$U_\text{ejecta}$~\cite{Hotokezaka:2012ze,Dietrich:2015iva},
\begin{equation}
T_\text{ej} = E_\text{ej} - ( M_\text{ej} + U_\text{ej} )  = 
\int_{\mathcal{U}} \text{d}^3 x  q^{(D)} (e-1-\epsilon) , \label{eq:Tejecta}
\end{equation}
with $e=\alpha u^t h - p/(\rho \alpha u^t)$.
Additionally, we compute the $D$-weighted integral of $v^2=v_i v^i$ inside 
the orbital plane and in the $x$-$z$-plane,
\begin{eqnarray}
 \mean{\bar{v}}_{\rho}&=&\sqrt{\frac{\int_{\mathcal{U}_{z=0}} \text{d}^3 x q^{(D)} v^2}
                              {\int_{\mathcal{U}_{z=0}} \text{d}^3 x q^{(D)} }} , \label{eq:meanbarvrho} \\
 \mean{\bar{v}}_{z}&=&\sqrt{\frac{\int_{\mathcal{U}_{y=0}} \text{d}^3 x q^{(D)} v^2}
                              {\int_{\mathcal{U}_{y=0}} \text{d}^3 x q^{(D)} }} , \label{eq:meanbarvz}                            
\end{eqnarray}
and the $D$-weighted integrals
\begin{equation}
 \mean{v}_{\rho}  =  \sqrt{ \left( \frac{\int_{\mathcal{U}_{z=0}} \text{d}^3 x q^{(D)} v^{x}}
                              {\int_{\mathcal{U}_{z=0}} \text{d}^3 x q^{(D)} } \right)^2 + 
                        \left( \frac{\int_{\mathcal{U}_{z=0}} \text{d}^3 x q^{(D)} v^{y}}
                              {\int_{\mathcal{U}_{z=0}} \text{d}^3 x q^{(D)} }\right)^2 } \label{eq:barvrho}
\end{equation}
and 
\begin{eqnarray}
 \mean{|v|}_{\rho} & = & \sqrt{  \frac{\int_{\mathcal{U}_{z=0}} \text{d}^3 x q^{(D)} (v^{x})^2}
                              {\int_{\mathcal{U}_{z=0}} \text{d}^3 x q^{(D)} } + 
                         \frac{\int_{\mathcal{U}_{z=0}} \text{d}^3 x q^{(D)} (v^{y})^2}
                              {\int_{\mathcal{U}_{z=0}} \text{d}^3 x q^{(D)} } }, \label{eq:meanabsvrho} \\
 \mean{|v|}_{z} & = & \sqrt{\frac{\int_{\mathcal{U}_{y=0}} \text{d}^3 x q^{(D)} (v^{z})^2}
                              {\int_{\mathcal{U}_{y=0}} \text{d}^3 x q^{(D)} }}. \label{eq:meanabsvz}
\end{eqnarray}

Our velocity measurements can be interpreted in the following way: 
$\mean{\bar{v}}_{\rho,z}$ is the mean velocity of the ejected material inside the orbital, but not 
necessarily in the direction of the cylindrical radius, and perpendicular to it, 
$\mean{|v|}_{\rho,z}$ gives the average velocities, and  
$\mean{v}_{\rho}$ gives an estimate of the ``kick'' velocity produced by the 
ejecta.  

Furthermore we consider the entropy ``indicator'', 
\begin{equation}
 \hat{S} = \frac{p}{K_i \rho^{\Gamma_i}}, \label{eq:entropy}
\end{equation}
where $\Gamma_i$ and $K_i$ are locally determined by the density $\rho$
and the EOS. 
In cases where the additional thermal
contribution to the pressure $p_{th}$ is small $\hat{S}\sim 1$, 
while in presence of shock heating $\hat{S} \gg 1$.\\

\section{BNS Configurations}
\label{sec:config}

\begin{figure}[t]
  \includegraphics[width=0.5\textwidth]{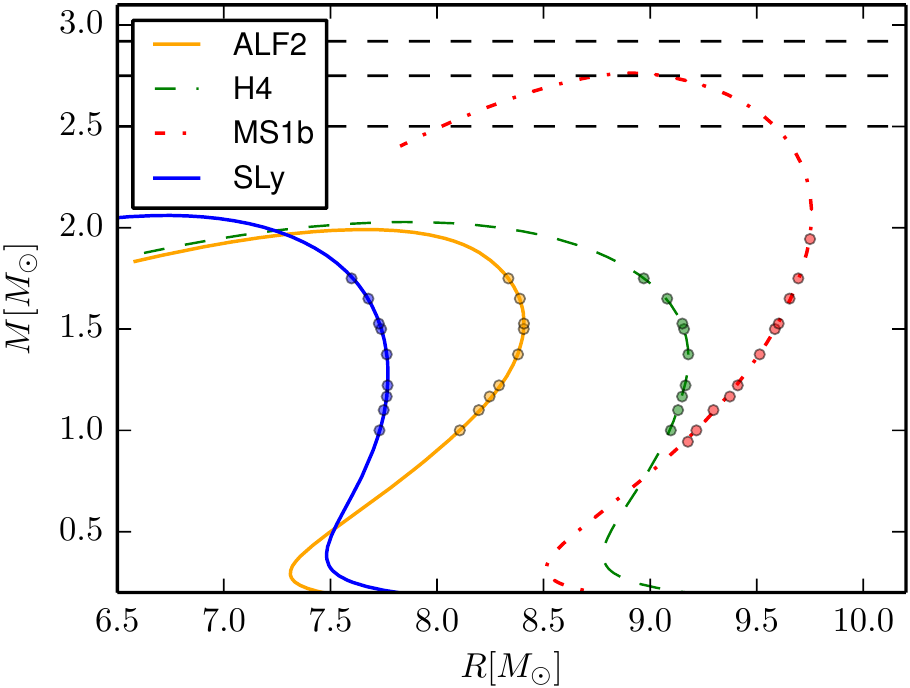}
  \caption{
    The four different EOSs employed in this work. 
    The EOSs are modeled by piecewise-polytropic fits to the tabulated EOSs. 
    The fits are taken from~\cite{Read:2008iy}, see also Tab.~\ref{tab:EOS}.
    Markers correspond to spherical individual stars in our
    configurations, and dashed black lines the total masses $M$ of our
    configurations for which we investigate all four EOSs. In the
    figure, we do not include the total mass of the setup MS1b-094194,
    since for those masses only the EOS MS1b was employed.}
   \label{fig:EOS}
\end{figure} 

Our configurations span the mass ratios $q=1.0,1.25,1.5,1.75$ and
total binary masses of $M=2.5,2.75,2.9167M_\odot$. For each pair
$(q,M)$ we simulate the four EOS ALF2, H4, MS1b, SLy EOS, cf.~Tab.~\ref{tab:EOS}.  
Additionally, we consider a $q=2.06$ configuration with the stiff EOS
MS1b, simulated at a single resolution in ~\cite{Dietrich:2015pxa}.
The mass-radius relations of each of the considered EOS are shown in
Fig.~\ref{fig:EOS}, where we also include as markers the isolated star
configurations, and as black dashed lines the total masses $M=M^A+M^B$ of the
systems.  

\begin{figure}[t]
  \includegraphics[width=0.5\textwidth]{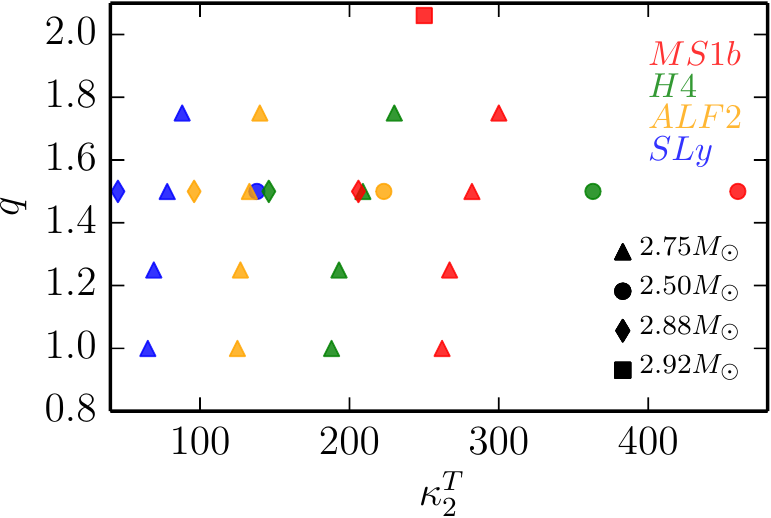}
  \caption{BNS parameter space in terms of $q$ and $\kappa^T_2$. 
   Different colors represent different EOSs, while different markers stand for 
   different total masses.}
   \label{fig:parspace}
\end{figure}

The properties of the initial configuration are listed in
Tab.~\ref{tab:ID_and_grid_irr}. Among various quantities, we report
the tidal polarizability coefficient
\begin{equation}
\kappa^T_2 = 2 \left( \frac{q^4}{(1+q)^5} \frac{k_2^A}{C_A^5}  +
\frac{q}{(1+q)^5} \frac{k_2^B}{C_B^5} 
\right)  \ ,
\label{eq:kappa}
\end{equation}
which describes at leading order the NSs tidal interactions, and depends on
the EOS via the quadrupolar dimensionless Love number $k_2$ of isolated
star configurations, e.g.~\cite{Damour:2009vw}. In
Eq.~\eqref{eq:kappa}, $\mathcal{C}^{A}$ is the compactness of star A
defined as the ratio of the gravitational mass in isolation $M^A$ with
the star's proper radius. It has been shown in
\cite{Bernuzzi:2014kca,Bernuzzi:2015rla} that $\kappa_2^T$ is
the relevant parameter encoding all the EOS information to
characterize the BNS dynamics during both the merger and the
postmerger phases.  
In addition, the GW of BNS is almost entirely  determined by the
mass ratio $q$ and the $\kappa^T_2$, because the binary total mass
scales trivially in absence of tides and its dependency in the tidal
waveform is hidden in the $\kappa^T_2$. Although other tidal
polarizability parameters, corresponding to higher-than-quadrupole
interactions, 
do play a role in the detailed modeling of the 
GW~\cite{Bernuzzi:2012ci,Bernuzzi:2014owa,Hinderer:2016eia}, the leading
order $\kappa^T_2$ encodes the main effect and it is the only tidal
parameter measurable in GW searches.
The coverage of the $q$-$\kappa^T_2$ parameter space by
our simulations is shown in Fig.~\ref{fig:parspace}.

All our configurations are simulated with at least two resolutions to
control numerical artifacts and to have (at least rough) estimate of
error bars for \textit{all} the quantities. The specific grid setups
can be found in Tab.~\ref{tab:ID_and_grid_irr}.   

\begin{table}[t]
  \centering
  \caption{\label{tab:EOS} Properties of the equations of state (EOSs)
    used in this work. Our EOSs use a crust with
    $\kappa_\text{crust}=\kappa_0=8.94989\times10^{-2}$ and
    $\Gamma_\text{crust}=1.35692$. The divisions for the
    individual parts are at $\rho_{1}={\rho}_\text{crust} \times 10^{-4}$,
    $\rho_{2}=8.11322\times 10^{-4}$ and $\rho_{3} =
    1.61880\times 10^{-3}$.  The columns 
    refer to: the name of the EOS, the maximum density in the crust,
    the three polytropic exponents $\Gamma$
    for the individual pieces, and
    the maximum supported gravitational mass $M^\text{max}$, maximum
    baryonic mass $M_\text{b}^\text{max}$, and maximum dimensionless compactness
    $\mathcal{C}^\text{max}$, respectively, of an isolated
    nonrotating star.  }
  \begin{tabular}{l|cccc|ccc}        
    \hline EOS & ${\rho}_\text{crust}$& $\Gamma_1$ & $\Gamma_2$ & $\Gamma_3$ & $M^\text{max}$ &
    $M_\text{b}^\text{max}$ & $\mathcal{C}^\text{max}$ \\ \hline 
    SLy & 2.36701 & 3.005 & 2.988 & 2.851 & 2.06 & 2.46 & 0.31 \\ 
    ALF2 & 3.15280 & 4.070 & 2.411 & 1.890 & 1.99 & 2.32 & 0.26 \\ 
    H4 & 1.43709 & 2.909 & 2.246 & 2.144 & 2.03 & 2.33 & 0.26 \\  
    MS1b & 1.83977 & 3.456 & 3.011 & 1.425 & 2.76 & 3.35 & 0.31 \\ \hline \hline
  \end{tabular}
\end{table}

\begin{table*}[t]
  \centering    
  \caption{Configurations and grid setups. 
    The first column defines the configuration name. 
    Next 11 columns describe the physical properties: EOS, 
    gravitational mass of the individual stars $M^{A,B}$, 
    baryonic mass of the individual stars $M_{b}^{A,B}$, 
    stars' compactnesses $\mathcal{C}^{A,B}$, 
    the tidal polarizability coefficient $\kappa^T_2$, 
    the initial dimensionless GW frequency $M \omega_{22}^0$, 
    the initial data ADM-Mass $M_{\rm ADM}$ and ADM-angular momentum $J_{\rm ADM}$.
    Next 8 columns describe the grid configuration:
    finest grid spacing $h_{L-1}$, radial resolution inside the
    shells $h_r$, number of points $n$ $(n^{mv})$ in the fix (moving) levels, 
    radial point number $n_r$ and azimuthal number of points
    $n_\theta$ in the shells, inradius $r_1$ up to which GRHD
    equations are solved , and the outer boundary $r_b$. 
    Notice that we divide most configurations in 3 different grid setups R1, R2, R3
    (compare the specific simulation name).
    }
    \begin{Tiny}
  \begin{tabular}{c|l|c|ccc|ccc|cccc|cccccccc}        
  $(q,M)$ & Name & EOS & $M^A$ & $M_b^A$ & $\mathcal{C}^A$ & $M^B$ & $M_b^B$ & $\mathcal{C}^B$ & 
    $\kappa^T_2$ & $M \omega_{22}^0$ &$ M_{ADM}$ &  $J_{ADM}$ &  
    $h_{L-1}$ & $h_{r}$ & $n$ & $n^{mv}$ & $n_r$ & $n_\theta$ & $r_1$ & $r_b$ \\
     \hline
     \hline
      \multirow{8}{*}{\rotatebox[origin=c]{90}{\textbf{(1.00,2.75)}}}
   &  ALF2-137137-R1s & ALF2 & 1.375 & 1.518 & 0.164 & 1.375 & 1.518 & 0.164 & 125 & 0.036 & 2.728 & 8.120 & 0.25 & 8.000 & 128 & 64  & 128 & 64 & 572 & 1564 \\
   &  ALF2-137137-R2s & ALF2 & 1.375 & 1.518 & 0.164 & 1.375 & 1.518 & 0.164 & 125 & 0.036 & 2.728 & 8.120 & 0.16 & 5.333 & 192 & 96  & 196 & 96 & 552 & 1555 \\     
   &  H4-137137-R1s   & H4   & 1.375 & 1.499 & 0.150 & 1.375 & 1.499 & 0.150 & 188 & 0.035 & 2.728 & 8.093 & 0.25 & 8.000 & 128 & 72  & 128 & 64 & 572 & 1564 \\
   &  H4-137137-R2s   & H4   & 1.375 & 1.499 & 0.150 & 1.375 & 1.499 & 0.150 & 188 & 0.035 & 2.728 & 8.093 & 0.16 & 5.333 & 192 & 108 & 196 & 96 & 552 & 1555 \\      
   &  MS1b-137137-R1s & MS1b & 1.375 & 1.497 & 0.145 & 1.375 & 1.497 & 0.145 & 262 & 0.035 & 2.729 & 8.158 & 0.25 & 8.000 & 128 & 72  & 128 & 64 & 572 & 1564 \\
   &  MS1b-137137-R2s & MS1b & 1.375 & 1.497 & 0.145 & 1.375 & 1.497 & 0.145 & 262 & 0.035 & 2.729 & 8.158 & 0.16 & 5.333 & 192 & 108 & 196 & 96 & 552 & 1555 \\ 
   &  SLy-137137-R1s  & SLy  & 1.375 & 1.526 & 0.177 & 1.375 & 1.526 & 0.177 & 65  & 0.036 & 2.728 & 8.006 & 0.23 & 7.392 & 128 & 64  & 128 & 64 & 529 & 1445 \\
   &  SLy-137137-R2s  & SLy  & 1.375 & 1.526 & 0.177 & 1.375 & 1.526 & 0.177 & 65  & 0.036 & 2.728 & 8.006 & 0.15 & 4.928 & 192 & 96  & 196 & 96 & 510 & 1437 \\   
     \hline
     \hline   
      \multirow{8}{*}{\rotatebox[origin=c]{90}{\textbf{(1.25,2.75)}}}
   &  ALF2-122153-R1s & ALF2 & 1.527 & 1.707 & 0.182 & 1.222 & 1.334 & 0.147 & 127 & 0.036 & 2.728 & 7.956 & 0.25 & 8.000 & 128 & 64  & 128 & 64 & 572 & 1564 \\
   &  ALF2-122153-R2s & ALF2 & 1.527 & 1.707 & 0.182 & 1.222 & 1.334 & 0.147 & 127 & 0.036 & 2.728 & 7.956 & 0.16 & 5.333 & 192 & 96  & 196 & 96 & 552 & 1555 \\          
   &  H4-122153-R1s   & H4   & 1.527 & 1.683 & 0.167 & 1.222 & 1.318 & 0.133 & 193 & 0.035 & 2.729 & 8.025 & 0.25 & 8.000 & 128 & 72  & 128 & 64 & 572 & 1564 \\
   &  H4-122153-R2s   & H4   & 1.527 & 1.683 & 0.167 & 1.222 & 1.318 & 0.133 & 193 & 0.035 & 2.729 & 8.025 & 0.16 & 5.333 & 192 & 108 & 196 & 96 & 552 & 1555 \\          
   &  MS1b-122153-R1s & MS1b & 1.527 & 1.680 & 0.159 & 1.222 & 1.318 & 0.130 & 267 & 0.035 & 2.729 & 8.032 & 0.25 & 8.000 & 128 & 72  & 128 & 64 & 572 & 1564 \\
   &  MS1b-122153-R2s & MS1b & 1.527 & 1.680 & 0.159 & 1.222 & 1.318 & 0.130 & 267 & 0.035 & 2.729 & 8.032 & 0.16 & 5.333 & 192 & 108 & 196 & 96 & 552 & 1555 \\          
   &  SLy-122153-R1s  & SLy  & 1.527 & 1.719 & 0.198 & 1.222 & 1.338 & 0.157 & 69  & 0.036 & 2.728 & 7.934 & 0.23 & 7.392 & 128 & 64  & 128 & 64 & 529 & 1445 \\
   &  SLy-122153-R2s  & SLy  & 1.527 & 1.719 & 0.198 & 1.222 & 1.338 & 0.157 & 69  & 0.036 & 2.728 & 7.934 & 0.15 & 4.928 & 192 & 96  & 192 & 96 & 510 & 1437 \\   
     \hline
     \hline   
      \multirow{8}{*}{\rotatebox[origin=c]{90}{\textbf{(1.50,2.5)}}}
   &  ALF2-100150-R1s & ALF2 & 1.500 & 1.672 & 0.178 & 1.000 & 1.073 & 0.123 & 223 & 0.031 & 2.482 & 6.637 & 0.25 & 8.000 & 128 & 64  & 128 & 64 & 572  & 1564 \\
   &  ALF2-100150-R2s & ALF2 & 1.500 & 1.672 & 0.178 & 1.000 & 1.073 & 0.123 & 223 & 0.031 & 2.482 & 6.637 & 0.16 & 5.333 & 192 & 96  & 196 & 96 & 552  & 1555 \\    
   &  H4-100150-R1s   & H4   & 1.500 & 1.649 & 0.164 & 1.000 & 1.063 & 0.110 & 363 & 0.030 & 2.283 & 6.664 & 0.25 & 8.000 & 128 & 72  & 128 & 64 & 572  & 1564 \\
   &  H4-100150-R2s   & H4   & 1.500 & 1.649 & 0.164 & 1.000 & 1.063 & 0.110 & 363 & 0.030 & 2.283 & 6.664 & 0.16 & 5.333 & 192 & 108 & 196 & 96 & 552  & 1555 \\      
   &  MS1b-100150-R1s & MS1b & 1.500 & 1.647 & 0.156 & 1.000 & 1.063 & 0.109 & 460 & 0.030 & 2.483 & 6.657 & 0.25 & 8.000 & 128 & 72  & 128 & 64 & 572  & 1564 \\
   &  MS1b-100150-R2s & MS1b & 1.500 & 1.647 & 0.156 & 1.000 & 1.063 & 0.109 & 460 & 0.030 & 2.483 & 6.657 & 0.16 & 5.333 & 192 & 108 & 196 & 96 & 552  & 1555 \\      
   &  SLy-100150-R1b  & SLy  & 1.500 & 1.672 & 0.193 & 1.000 & 1.074 & 0.129 & 138 & 0.031 & 2.482 & 6.587 & 0.23 & 7.392 & 160 & 64  &   -   &   -  & 1190 & 1190 \\
   &  SLy-100150-R2b  & SLy  & 1.500 & 1.672 & 0.193 & 1.000 & 1.074 & 0.129 & 138 & 0.031 & 2.482 & 6.587 & 0.15 & 4.928 & 240 & 96  &   -   &   -  & 1188 & 1188 \\
   \hline
      \multirow{8}{*}{\rotatebox[origin=c]{90}{\textbf{(1.50,2.75)}}}
   &  ALF2-110165-R1b & ALF2 & 1.650 & 1.862 & 0.197 & 1.100 & 1.190 & 0.134 & 133 & 0.036 & 2.729 & 7.686 & 0.25 & 8.000 & 160 & 64  &  -  & -  &1288 & 1288 \\
   &  ALF2-110165-R2b & ALF2 & 1.650 & 1.862 & 0.197 & 1.100 & 1.190 & 0.134 & 133 & 0.036 & 2.729 & 7.686 & 0.16 & 5.333 & 240 & 96  &  -  & -  &1285 & 1285 \\     
   &  H4-110165-R1s   & H4   & 1.650 & 1.835 & 0.182 & 1.100 & 1.177 & 0.121 & 209 & 0.035 & 2.729 & 7.820 & 0.25 & 8.000 & 128 & 72  & 128 & 64 & 572 & 1564 \\
   &  H4-110165-R2s   & H4   & 1.650 & 1.835 & 0.182 & 1.100 & 1.177 & 0.121 & 209 & 0.035 & 2.729 & 7.820 & 0.16 & 5.333 & 192 & 108 & 196 & 96 & 552 & 1555 \\
   &  MS1b-110165-R1s & MS1b & 1.650 & 1.830 & 0.171 & 1.100 & 1.177 & 0.118 & 282 & 0.035 & 2.729 & 7.799 & 0.25 & 8.000 & 128 & 72  & 128 & 64 & 572 & 1564 \\
   &  MS1b-110165-R2s & MS1b & 1.650 & 1.830 & 0.171 & 1.100 & 1.177 & 0.118 & 282 & 0.035 & 2.729 & 7.799 & 0.16 & 5.333 & 192 & 108 & 196 & 96 & 552 & 1555 \\
   &  SLy-110165-R1b  & SLy  & 1.650 & 1.878 & 0.215 & 1.098 & 1.190 & 0.142 & 78  & 0.036 & 2.727 & 7.700 & 0.23 & 7.392 & 160 & 64  &   -   &   -  & 1190 & 1190 \\
   &  SLy-110165-R2b  & SLy  & 1.650 & 1.878 & 0.215 & 1.098 & 1.190 & 0.142 & 78  & 0.036 & 2.727 & 7.700 & 0.15 & 4.928 & 240 & 96  &   -   &   -  & 1188 & 1188 \\
     \hline     
      \multirow{6}{*}{\rotatebox[origin=c]{90}{\textbf{(1.50,2.92)}}}
   &  ALF2-117175-R1b & ALF2 & 1.750 & 1.992 & 0.210 & 0.167 & 0.127 & 0.141 & 96  & 0.035 & 2.895 & 8.792 & 0.25 & 8.000 & 160 & 64  &  -  &  - & 1288 & 1288 \\
   &  ALF2-117175-R2b & ALF2 & 1.750 & 1.992 & 0.210 & 0.167 & 0.127 & 0.141 & 96  & 0.035 & 2.895 & 8.792 & 0.16 & 5.333 & 240 & 96  &  -  &  - & 1285 & 1285  \\     
   &  H4-117175-R1s   & H4   & 1.750 & 1.961 & 0.195 & 1.167 & 1.253 & 0.128 & 146 & 0.038 & 2.894 & 8.612 & 0.25 & 8.000 & 128 & 64  & 128 & 64 & 572 & 1564\\ 
   &  H4-117175-R2s   & H4   & 1.750 & 1.961 & 0.195 & 1.167 & 1.253 & 0.128 & 146 & 0.038 & 2.894 & 8.612 & 0.16 & 5.333 & 192 & 108 & 196 & 96 &  552 & 1555  \\    
   &  MS1b-117175-R1s & MS1b & 1.750 & 1.954 & 0.180 & 1.167 & 1.253 & 0.125 & 206 & 0.038 & 2.894 & 8.612 & 0.25 & 8.000 & 128 & 72  & 128 & 64 & 572 & 1564 \\
   &  MS1b-117175-R2s & MS1b & 1.750 & 1.954 & 0.180 & 1.167 & 1.253 & 0.125 & 206 & 0.038 & 2.894 & 8.612 & 0.16 & 5.333 & 192 & 108 & 196 & 96 &  552 & 1555  \\       
   &  SLy-117175-R1b  & SLy  & 1.750 & 2.012 & 0.230 & 1.167 & 1.272 & 0.150 & 53  & 0.033 & 2.893 & 8.850 & 0.23 & 7.392 & 160 & 64  &   -   &   -  & 1190 & 1190 \\
   &  SLy-117175-R2b  & SLy  & 1.750 & 2.012 & 0.230 & 1.167 & 1.272 & 0.150 & 53  & 0.033 & 2.893 & 8.850 & 0.15 & 4.928 & 240 & 96  &   -   &   -  & 1188 & 1188 \\
     \hline     
     \hline      
      \multirow{6}{*}{\rotatebox[origin=c]{90}{\textbf{(1.75,2.75)}}}
   &  ALF2-100175-R1 & ALF2 & 1.750 & 1.992 & 0.210 & 1.000 & 1.074 & 0.123 & 140 & 0.032 & 2.731 & 7.674 & 0.25 & 8.000 & 160 & 64  &  -  &  - & 1288 & 1288 \\
   &  ALF2-100175-R2 & ALF2 & 1.750 & 1.992 & 0.210 & 1.000 & 1.074 & 0.123 & 140 & 0.032 & 2.731 & 7.674 &  0.16 & 5.333 & 240 & 96  &  -  &  - & 1285 & 1285 \\
   &  H4-100175-R1   & H4   & 1.750 & 1.961 & 0.195 & 1.000 & 1.063 & 0.110 & 230 & 0.035 & 2.730 & 7.531 & 0.25 & 8.000 & 128 & 64 & 128 & 64 & 572 & 1564 \\ 
   &  H4-100175-R2   & H4   & 1.750 & 1.961 & 0.195 & 1.000 & 1.063 & 0.110 & 230 & 0.035 & 2.730 & 7.531 & 0.16 & 5.333 & 192 & 108 & 196 & 96 & 552 & 1555  \\       
   &  MS1b-100175-R1 & MS1b & 1.750 & 1.954 & 0.180 & 1.000 & 1.063 & 0.109 & 300 & 0.035 & 2.730 & 7.531 & 0.25 & 8.000 & 128 & 72 & 128 & 64 & 572 & 1564\\
   &  MS1b-100175-R2 & MS1b & 1.750 & 1.954 & 0.180 & 1.000 & 1.063 & 0.109 & 300 & 0.035 & 2.730 & 7.531 & 0.16 & 5.333 & 192 & 108 & 196 & 96 &  552 & 1555  \\         
   &  SLy-100175-R1b & SLy  & 1.750 & 2.012 & 0.230 & 1.000 & 1.075 & 0.129 & 88  & 0.031 & 2.735 & 7.756 & 0.23 & 7.392 & 160 & 64  &   -   &   -  & 1190 & 1190 \\
   &  SLy-100175-R2b & SLy  & 1.750 & 2.012 & 0.230 & 1.000 & 1.075 & 0.129 & 88  & 0.031 & 2.735 & 7.756 & 0.15 & 4.928 & 240 & 96  &   -   &   -  & 1188 & 1188 \\
     \hline
   \multirow{3}{*}{\rotatebox[origin=c]{90}{\textbf{(2.06,\ }}}
   \multirow{3}{*}{\rotatebox[origin=c]{90}{\textbf{\ 2.88)}}}   
   &  MS1b-094194-R1    & MS1b & 1.944 & 2.200 & 0.199 & 0.944 & 1.000 & 0.103 & 250 & 0.036 & 2.868 & 7.850 & 0.250  & 8.000 & 128 & 72 & 144  & 64 & 572 & 1692  \\  
   &  MS1b-094194-R2    & MS1b & 1.944 & 2.200 & 0.199 & 0.944 & 1.000 & 0.103 & 250 & 0.036 & 2.868 & 7.850 & 0.167  & 5.333 & 192 & 108 & 216 & 96 & 552 & 1683  \\  
   &  MS1b-094194-R3    & MS1b & 1.944 & 2.200 & 0.199 & 0.944 & 1.000
   & 0.103 & 253 & 0.036 & 2.868 & 7.850 & 0.125  & 4.000 & 256 & 144
   & 288 & 128 & 542 & 1678  \\  
\hline
\hline
  \end{tabular}
  \end{Tiny}
 \label{tab:ID_and_grid_irr}
\end{table*}

\section{Dynamics}
\label{sec:dynamics}

\subsection{Mass transfer} 

\begin{figure}[t]
  \includegraphics[width=0.45\textwidth]{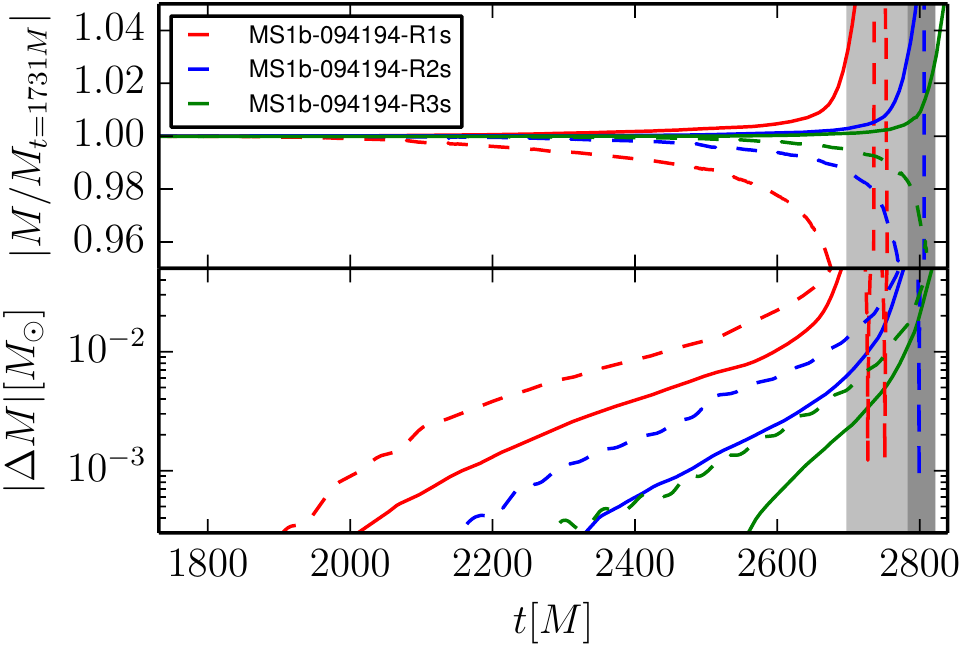}
   \includegraphics[width=0.45\textwidth]{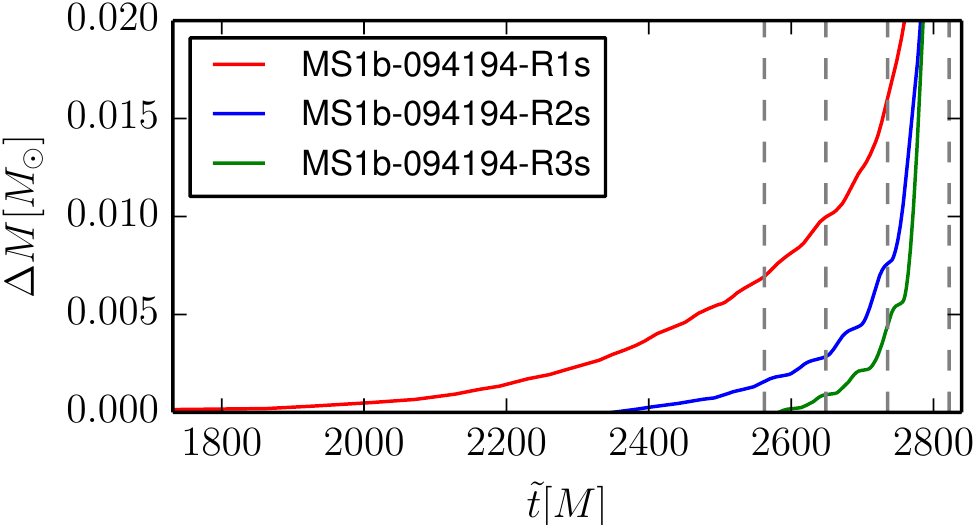}
  \caption{Mass-transfer between the two neutron stars.
    Top: The panels show he time evolution of the baryonic mass and the 
    ``mass defect'' $|\Delta M| = |M(t)-M(t=5000M_\odot)|$ inside coordinate 
    spheres of radius $10.5M_\odot$ around the two neutron stars as a function of the simulation time $t$. 
    While the mass is increasing for the primary star (solid lines) the secondary star ``looses'' mass (dashed lines). 
    For comparison we rescale everything to the mass at $t=5000M_\odot$, with
    this approach only influences at the late stage of the inspiral are taken
    into account. Note that the dashed line can be seen as an upper bound for the mass 
    transfer, since mass is also lost due to ejecta.
    Bottom: mass transfer as a function of the shifted time $\tilde{t}$, such that  
    the moment of merger happens at the same time $\tilde{t}^{\rm mrg}$ for all models. 
    The gray vertical dashed lines refer to the times shown in Fig.~\ref{fig:2d_rho_MS1b-094194}}
  \label{fig:mass_transfer_MS1b-094194}
  \label{fig:mass_transfer_MS1b-094194_rescaled}
\end{figure}

\begin{figure*}[t]
  \includegraphics[width=\textwidth]{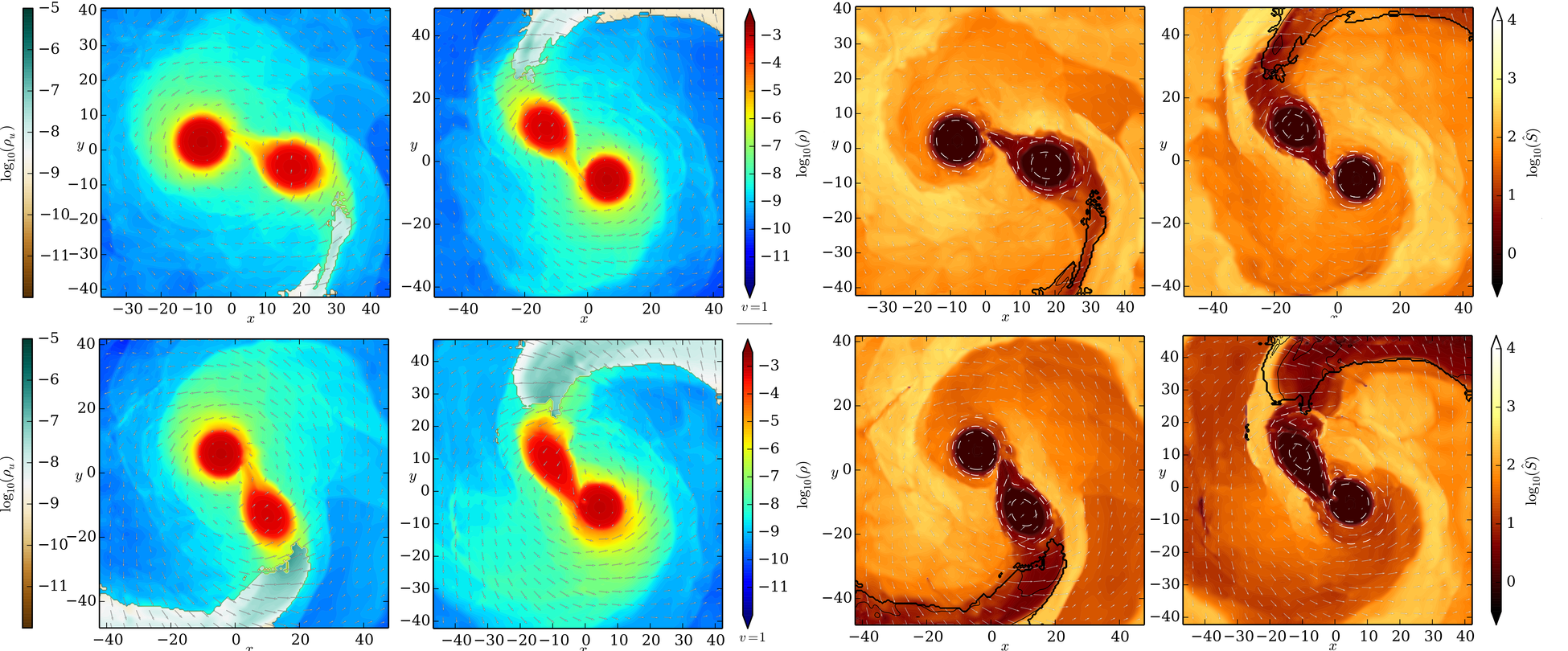}
  \caption{Density and entropy on the orbital plane for MS1b-094194 in
    the final merger phase. 
    Left plots: 
    Density profile inside the orbital plane. 
    The presented times are: (2562.3,2648.9,2735.5,2822.0)$M$ for the 
    upper left, upper right, lower left, lower right panel respectively. 
    Those times correspond to the times marked in Fig.~\ref{fig:mass_transfer_MS1b-094194_rescaled}
    as vertical dashed lines. 
    We color the density from blue to red and the unbound density from brown to green.
    During the inspiral mass is ejected from the tidal tail of the less massive star due to torque. 
    The upper right panel corresponds to the moment of merger.
    Right plots: Entropy indicator $\hat{S}$ inside the orbital plane for
    MS1b-094194. The
    times are the same as in the right panels. We also include white contour
    lines for densities $\rho=10^{-6},10^{-5},10^{-4}$ and black lines for the
    ejecta material with $\rho_u =
    10^{-12},10^{-11},10^{-10},10^{-9},10^{-8},10^{-7},10^{-6},10^{-5}.$
    During the simulation we observe that the entropy between the two neutron
    stars is small, i.e.~no shock heating produced ejecta can be observed
    for this setup at the chosen times.}
  \label{fig:2d_rho_MS1b-094194}
  \label{fig:2d_entropy_MS1b-094194}
\end{figure*}  

\paragraph*{MS1b-094194:}
We reported in~\cite{Dietrich:2015pxa} for MS1b-094194-R1s that a rest mass 
up to $\approx10^{-2}M_\odot$ was transfered between the stars during the 
last orbits before the actual merger. 
However, our previous investigations suffered from two facts, 
which we already mentioned in~\cite{Dietrich:2015pxa}:
(i) we used a relatively low resolution (quite similar to the R1 setup); 
(ii) the infrastructure to compute baryonic masses inside coordinate spheres around the 
NSs was not implemented and the mass transfer was estimated by investigating mass changes 
across Cartesian refinement boxes.
Here, we can investigate the mass transfer in more
detail using higher resolutions. We find that the mass transfer is not robust for
varying resolution, and the amount of mass decreases at higher resolutions. 

We compute the baryonic mass around the NSs
inside coordinate spheres of $r_{c1}=r_{c2}=10.5M_\odot$. While for the
more massive star a mass increase is directly related to the accretion of
material, a mass loss of the less massive star is also produced by
material ejected prior to merger.
Figure~\ref{fig:mass_transfer_MS1b-094194} summarizes our findings. The
upper panel shows the baryonic mass inside $r_{c1/2}$ rescaled to the
value at $t\approx5000M_\odot\approx1731M$. We decided not to rescale the
mass with respect to the initial data to study only mass changes within a
few revolutions prior to merger to allow an easier investigation. The
middle panel shows the mass difference in solar masses with respect
to the mass at $t\approx 1731M$. 
The solid lines indicate the mass gain of the
more massive star, while the dashes lines show the mass loss of the secondary NS.
For increasing resolution the transfered mass decreases, while
for R1s we observe a mass difference of $\sim 10^{-2}$, the
transfered mass decreases up to a factor of 10 for R3s.

Due to different numerical dissipation for different resolutions 
the moment of merger, i.e.~the peak in the GW amplitude, differs. 
For a more simplified and straight forward comparison of the different configurations, 
we compensate this effect by applying a time rescaling according to $\tilde{t}=\eta t$, 
where the factor $\eta$ is the quotient $t^{\rm mrg}_{R3}/t^{\rm mrg}_{RX}$ with $X={1,2}$.
$t^{\rm mrg}$ denotes the moment of merger, see also \cite{Hotokezaka:2015xka}.

Thus the merger happens for all simulations at the same time
$\tilde{t}^{\rm mrg}$
and we can easily compare the transfered mass for different resolutions.
To give an impression about the density profile during the simulation, 
we show the rest mass density inside the orbital plane
at the times 
$\tilde{t}=2562.3M (7400M_\odot),
2648.9M (7650M_\odot)$,
$2735.5M (7900M_\odot),
2822.0M (8150M_\odot)$ 
in Fig.~\ref{fig:2d_rho_MS1b-094194}. 
Those times are also marked inside Fig.~\ref{fig:mass_transfer_MS1b-094194_rescaled} (bottom panel)
as vertical dashed lines. The bottom panel of Fig.~\ref{fig:mass_transfer_MS1b-094194_rescaled}
shows the mass gain of the more massive star over the time $\tilde{t}$. 
In total, we estimate the transfered mass to be smaller than $5 \times 10^{-3}M_\odot$. 
Furthermore, while the lower resolution simulation suggested that the mass transfer sets in several
revolutions before the merger, mass is only transfered $\sim 2$ revolutions before merger for the R3 setup. 
In addition to the amount of transfered mass it is also interesting whether 
the mass transfer could heat up the more massive NS.
To investigate this, we are showing the entropy indicator $\hat{S}$, cf.~Eq.\eqref{eq:entropy}, 
together with contour density lines (white dashed) 
in Fig.~\ref{fig:2d_entropy_MS1b-094194}. 
According to the entropy indicator $\hat{S}$ no shocks are produced at 
the surface of the more massive star. 
Consequently this region of the NSs does not become significantly hotter 
than other parts. \\

\paragraph*{Other configurations:}
Our discussion about the mass transfer is based on MS1b-094194, 
but similar results can be obtained for other configurations. 
Most notably, we can verify that with increasing the resolution from R1 to R2 
the transfered mass is decreasing. Our simulations are also in agreement with the naive understanding 
that for stiff EOSs and higher mass ratios the transfered mass is increasing. 
Motivated by this observation, 
we conclude that the mass transfer estimated from the setup MS1b-094194
can be seen as an upper bound for astrophysical realistic systems. 
Because of this finding together with the fact that no shocks are produced due to mass 
accretion to the massive neutron star, we do not expect that for astrophysical realistic systems 
the energy released by the mass-transfer might lead to EM counterparts which 
could be observed before the actual merger of the BNS system.

\subsection{Ejecta}

The mass ejection in our simulations is caused by two effects. 
i) Part of the unbound mass is expelled either during the (partial) tidal
disruption or from the tidal tail
of the companion by a centrifugal effect,
see e.g.~Fig.~\ref{fig:2d_rho_MS1b-094194}. These ejecta are emitted
already during the last orbits in an essentially adiabatic way (small
entropy value). As shown in simulations that includes microphysics
e.g. \cite{Sekiguchi:2015dma,Lehner:2016lxy}, the composition is rich in neutrons and
has a small electron fraction.
ii) Part of the unbound mass is expelled when the two NS cores
collide. These
shock-triggered ejecta are characterized by large entropy values and
a higher electron fraction. 
The two components are clearly distinguishable by plotting the entropy
indicator $\hat{S}$, Eq.~\eqref{eq:entropy}. Mechanism i) is dominant
for configurations with large $q$.

In addition to the amount of ejecta, we also compute the kinetic energy, as well as the 
average velocity, cf.~Sec.~\ref{sec:diagno}. 
The results are summarized in Tab.~\ref{tab:ejecta}. In the following
we will present two exemplary cases before discussing the general influence of
the mass ratio and total mass. We often assign an uncertainty as given by the
difference between different resolutions. It is however important to
notice that this does not necessarily corresponds to the total
uncertainty since systematic errors are significant
\cite{Dietrich:2015iva}. Most notably, when the fluid expands it is
possible that the density falls below the artificial atmosphere
value producing mass losses in ejecta. Additionally material can also
leave the grid, see Appendix~\ref{sec:accuracy}. 
These are the reasons why the estimated ejecta quantities are evaluated at
a time, where most of the ejected material is rather close to the center
of the system, at a distance of $\approx 150$--$250M_\odot$. While the
ejected mass does not depend much on this fact, the momenta and velocities
of the ejecta are typically overestimated due to fact that the gravitational
potential of merger remnant is neglected in Eqs.~\eqref{eq:Tejecta}--\eqref{eq:meanabsvz}.
A simple Newtonian estimate shows that the ejecta velocities and the 
linear momenta are therefore overestimated by up to
$20$--$30\%$.
\\   

\paragraph*{MS1b-094194:}  
Figure~\ref{fig:2d_rho_MS1b-094194} (left panels) presents the 2D density profile of
MS1b-095194 ($\rho$ from blue to red) and the unbound density ($\rho_u$
from brown to green). The largest amount of ejecta comes from
the tidal tail of the less massive NS, i.e. mechanism i) described above.
We also present snapshots of
the entropy indicator $\hat{S}$ in Fig.~\ref{fig:2d_entropy_MS1b-094194} (right panels).
The density is marked as white thin, dashed contour lines and the
ejecta by black contours. From the plots it is clear that no shock
heating happens inside the tidal tail.
The low entropy value suggests that for such large mass ratios
and stiff EOS, the ejected material will be neutron rich as found in e.g.~\cite{Lehner:2016lxy}.\\

\begin{figure}[htpb]
  \includegraphics[width=0.42\textwidth]{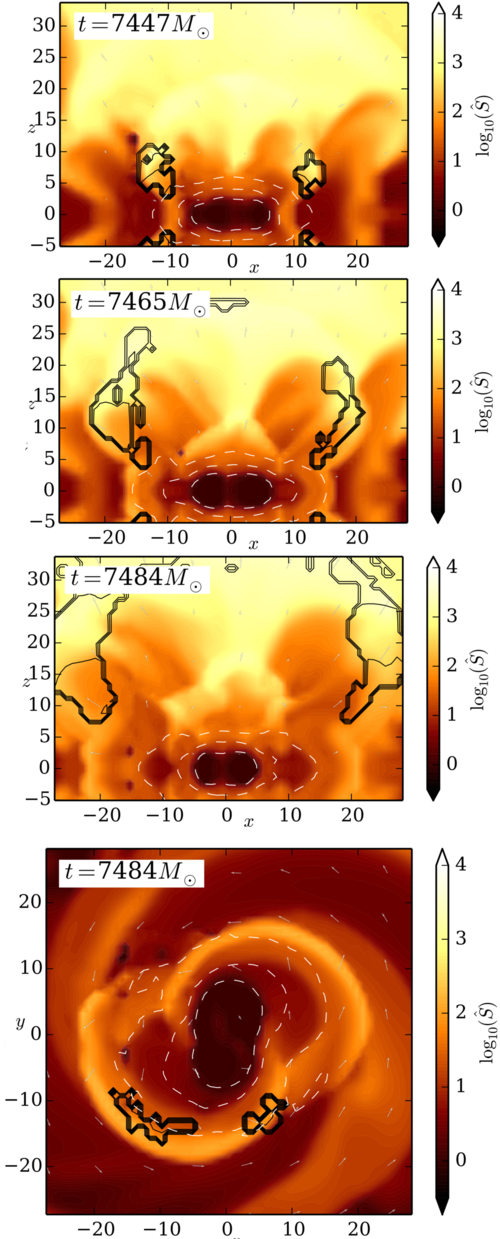}
  \caption{Entropy indicator $\hat{S}$ for SLy-137137 in the early
    postmerger phase.
    The panels represent
    snapshots at $t=2708M, t=2714, 2721M$ from top to bottom in the $x$-$z$-plane. 
    The last panel shows the entropy inside the orbital plane at $t=2721M$. 
    We also include white contour lines for densities $\rho=10^{-6},10^{-5},10^{-4}$ and black lines
    for the ejecta material with $\rho_u =
    10^{-12},10^{-11},10^{-10},10^{-9},10^{-8},10^{-7},10^{-6},10^{-5}$.}
  \label{fig:2d_entropy_SLy137137}
\end{figure}    

\paragraph*{SLy-137137:}    
As an opposite scenario, we present the results for SLy-137137, an
equal mass NSs described by a soft
EOS. Figure~\ref{fig:2d_entropy_SLy137137} shows $\hat{S}$ for
SLy-137137-R1 inside the $x$-$z$-plane for $t=2708M, 2714M, 2721 M$
and inside the orbital plane ($x$-$y$-plane) at $t=2721M$. 
The colors and contour lines are identical to the ones used in
Fig.~\ref{fig:2d_entropy_MS1b-094194}. The first three panels show an
ejection of material, colored by black contour lines in an angle $\gtrsim
45^\circ$ to the orbital plane. The ejection is triggered from high entropic
regions near the surface of the NS (white dashed lines).
Also, we find that after the collision of the cores the central region of the
HMNS in the orbital plane is low entropic; this suggests that no hot
core forms during the merger, but that instead the hot material is
confined in streams of matter expanding from the NS-NS interface and
at the HMNS surface  (see also \cite{Bernuzzi:2015opx}).
Type ii) ejecta are triggered from these regions. \\

\begin{figure}[t]
  \includegraphics[width=0.5\textwidth]{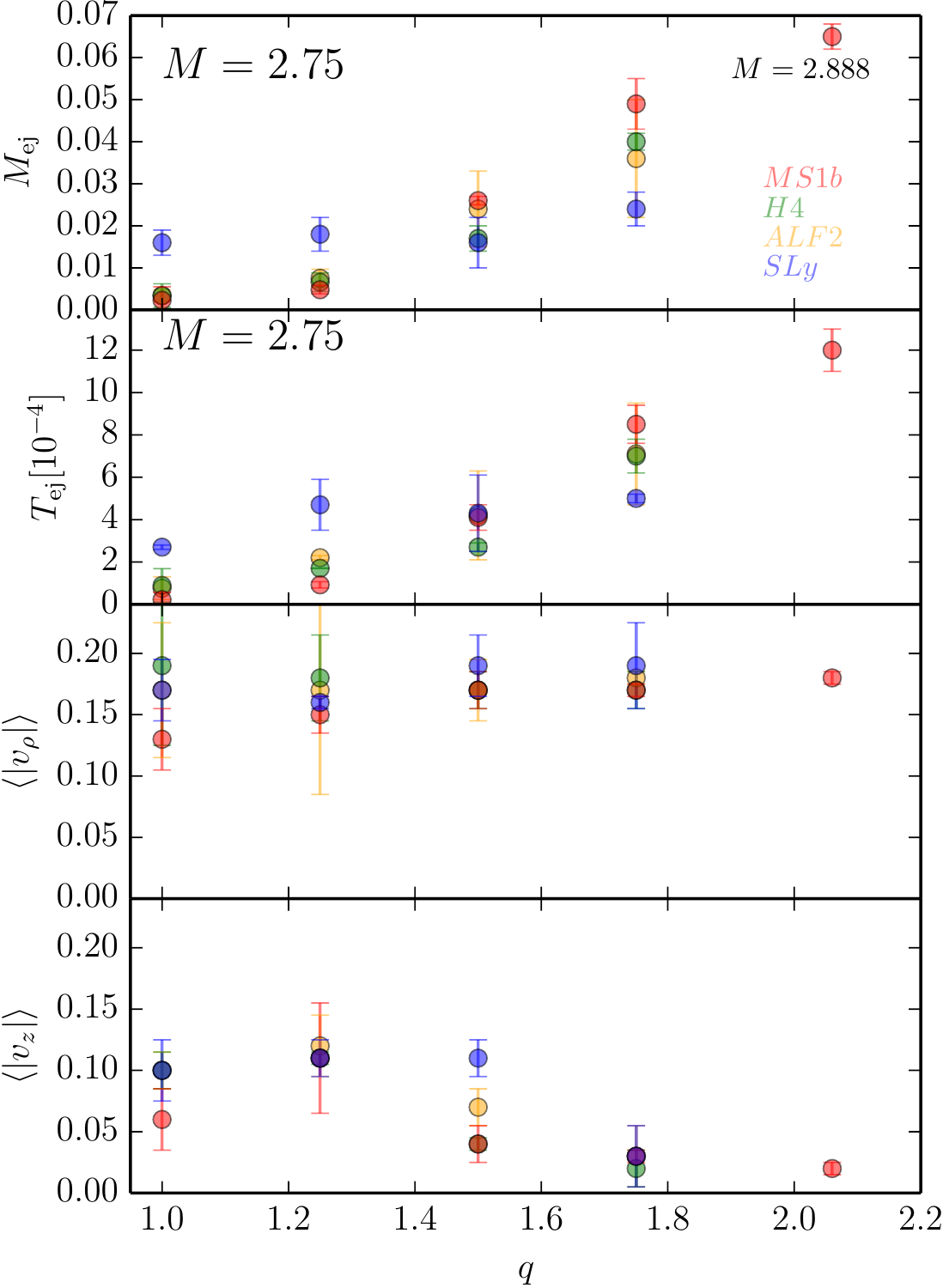}
  \caption{Ejected properties as a function of the mass ratio for masses $M=2.75$ and $M=2.888$. 
    We show the difference between two adjacent resolutions as an error bar in our plot. 
    Top: Ejecta mass, where we see that for an increasing mass ratio stiffer EOS produce significantly more ejecta, 
    while a similar effect can not be observed for softer EOS.
    Upper middle: Kinetic energy of the ejecta, where for an increasing mass ratio ejecta have significantly more kinetic energy, in particular for stiff EOSs.
    Lower middle: $\mean{|v_\rho|}$, Eq.~\eqref{eq:meanabsvrho}, as a function of q. 
    Bottom: $\mean{|v_z|}$, Eq.~\eqref{eq:meanabsvz}, as a function of q. 
    While the velocity inside the orbital plane seems to be almost constant independent of the mass ratio, 
    the velocity orthogonal to the orbital plane decreases for higher mass ratios.}
  \label{fig:M_ej(q)}
\end{figure}  

\paragraph*{Effect of the mass-ratio:}  
Figure~\ref{fig:M_ej(q)} (top panel) shows the dependence of the mass of the
ejected material for different mass ratios. We mainly focus on
simulations with a total mass of $M=2.75M_\odot$, but also include the 
MS1b-094194, which has a total mass of $M=2.888M_\odot$ to support our argumentation. 
Different EOSs are represented by different colors and 
the uncertainty is given as an error bar based on the  
difference between different resolutions solely (cf. error discussion above). 
We find that for stiff EOSs more mass is ejected for larger mass
ratios. This is in-line with previous studies \cite{Dietrich:2015iva,Hotokezaka:2012ze},
although the mass ratios considered  here are significantly larger. 
The trend in $q$ is almost linear regarding MS1b (red circles), H4
EOS (green circles), ALF2 (orange circles). 
A rather similar dependence on the mass ratio was already found for BHNS scenarios, 
e.g.~\cite{Foucart:2012nc,Kawaguchi:2016ana}, for 
the ejecta and the disk mass, but up to our knowledge, never presented for BNS configurations.  
For soft EOSs (SLy) no strong correlation between the mass-ratio and
the ejected mass is visible, all the configurations produce the same
amount of ejecta between
 $1$--$2\times 10^{-2} M_\odot$. 

Figure~\ref{fig:M_ej(q)} (middle panel) shows the dependence of the
kinetic energy as a function of the mass ratio for the  $M=2.75
M_\odot$ and $M=2.888M_\odot$ configurations. As for the amount of
ejected material it is obvious that for stiff EOSs a higher mass ratio
leads to ejecta with a higher kinetic energy.  
Again the relation between $T_{\rm ej}$ and $q$ is almost linear.

Figure~\ref{fig:M_ej(q)} (bottom panel) shows the estimated velocities of the ejecta inside the orbital plane and orthogonal 
to the orbital plane. We find that the velocity of the ejecta inside
the orbital plane are $\mean{|v_\rho|}\sim 0.17 \sim 50000{\rm km/s}$ 
for all configurations independent of the EOS and the mass ratio. 
However, we stress again that the ejecta velocities are typically overestimated with our current approach. 
The velocity orthogonal to the orbital plane is in general smaller than $\mean{|v_\rho|}$. Furthermore, 
we find that, independent of the EOS and above $q=1.25$, $\mean{|v_z|}$ decreases for an increasing mass ratio. 
This suggests that for large mass ratio configurations more ejection happens
inside the orbital plane than in other directions. The effect is
expected since ejecta due to shock heating play a subdominant role in
those cases. Furthermore this leads to a more oblate and less isotropic 
shape of the ejecta.\\

\begin{figure}[t]
  \includegraphics[width=0.5\textwidth]{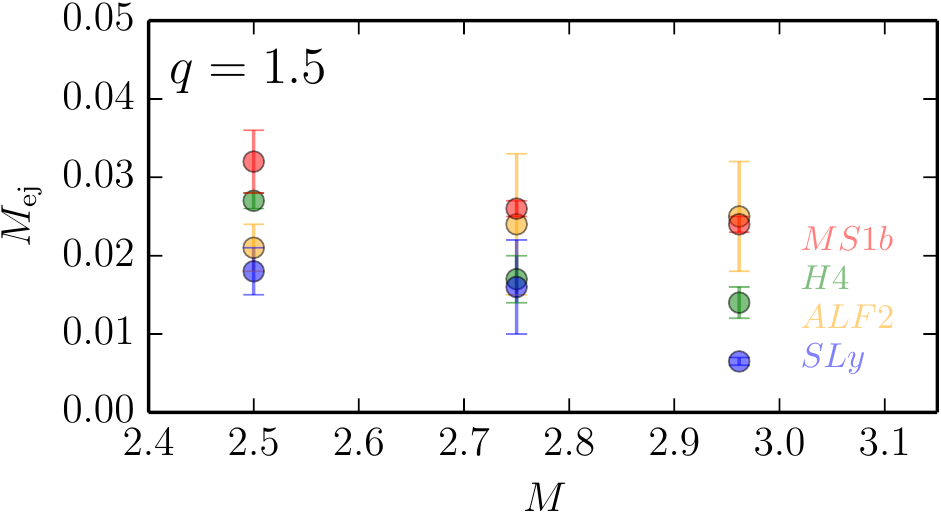}
  \caption{Ejected mass as a function of the total mass of the binary system for a mass ratio 
    of $q=1.5$. We show the difference between two adjacent resolutions as an error bar. 
    The influence of the total mass is smaller than the effect of the mass ratio, however, 
    less ejecta are produced for higher total masses in most cases.}
  \label{fig:M_ej(q=1.5)}
\end{figure}    

\paragraph*{Effect of the total mass:}
Finally, we study the influence of the total mass on the amount of ejected material. 
For this purpose we compare all our configurations with a mass ratio of $q=1.5$. 
We have evolved these configurations for total masses of $M=2.5M_\odot$, $M=2.75M_\odot$, and $M\approx2.92M_\odot$.
The data show that the influence of the total mass is smaller than the influence 
of the mass ratio, see Fig.~\ref{fig:M_ej(q=1.5)}. However, for
increasing total mass, less massive ejecta are produced in agreement
with previous studies, e.g.~\cite{Hotokezaka:2012ze}.  
The reason for this is that larger mass systems form more compact and more massive merger remnants 
and the material is more bound by the larger gravitational potential. 

The dependence of the kinetic energy on the total mass can be read off
from Tab.~\ref{tab:ejecta}. Similarly to the ejecta mass, the kinetic energy is decreasing for 
higher total masses, although the uncertainties are even larger than for the mass of the ejecta. 
For the velocities of the ejecta no strict statement about the influence of the total mass can be made 
according to our simulations because of the large uncertainties. 

\begin{table*}[t]
  \centering    
  \caption{
    Ejecta properties. The columns refer to: 
    the name of the configuration, the mass of the ejecta, the kinetic energy of the ejecta, 
    the $D$-weighted integral $v_\rho$, Eq.~\eqref{eq:barvrho}, 
    the average velocity of the ejecta inside the orbital plane $\mean{|v|}_\rho$ 
    and perpendicular to it $\mean{|v|}_z$, cf.\ Eqs.~\eqref{eq:meanabsvrho} and~\eqref{eq:meanabsvz}, 
    and the average of $v^2$ of fluid elements inside the orbital plane 
    $\mean{\bar{v}}^\rho$ and perpendicular to it $\mean{\bar{v}}^z$, Eqs.~\eqref{eq:meanbarvrho} 
    and~\eqref{eq:meanbarvz}.
    We present results for R2 resolved runs and give R1 results inside brackets for an error estimate.
    (With the exception of MS1b-094194, where we give results for R3 and R2.) }
  \begin{tabular}{l|cccccccc}        
    Name    & $M_{\rm ej} \ [10^{-2}M_\odot] $    & $T_{\rm ej}\ [10^{-4}]$ &  
              $v_\rho $       & $\mean{|v|}_\rho$ & $\mean{|v|}_z$  & $\mean{\bar{v}}^\rho$ & $\mean{\bar{v}}^z$  \\
     \hline
     \hline
ALF2-137137  & 0.34 (0.20) & 0.76 (0.22)   & 0.066 (0.058) & 0.17 (0.12) & 0.10 (0.11) & 0.17 (0.12) & 0.22 (0.15) \\
H4-137137    & 0.34 (0.06) & 0.89 (0.10)   & 0.039 (0.080) & 0.19 (0.13) & 0.10 (0.14) & 0.19 (0.13) & 0.23 (0.22) \\
MS1b-137137  & 0.23 (0.55) & 0.22 (0.56)   & 0.030 (0.032) & 0.13 (0.11) & 0.06 (0.08) & 0.13 (0.11) & 0.14 (0.14) \\
SLy-137137   & 1.6 (1.3)   & 2.7 (2.8)     & 0.060 (0.042) & 0.17 (0.19) & 0.10 (0.12) & 0.17 (0.19) & 0.17 (0.20) \\
     \hline
     \hline  
ALF2-122153  & 0.75 (0.97) & 2.2 (2.1)     & 0.077 (0.089) & 0.17 (0.09) & 0.12 (0.10) & 0.17 (0.09) & 0.23 (0.17) \\
H4-122153    & 0.66 (0.88) & 1.7 (1.7)     & 0.032 (0.002) & 0.18 (0.15) & 0.11 (0.11) & 0.18 (0.16) & 0.22 (0.28) \\
MS1b-122153  & 0.48 (0.58) & 0.92 (0.77)   & 0.038 (0.104) & 0.15 (0.14) & 0.11 (0.07) & 0.16 (0.14) & 0.20 (0.21) \\
SLy-122153   & 1.8 (1.4)   & 4.7 (3.5)     & 0.014 (0.076) & 0.16 (0.16) & 0.11 (0.10) & 0.16 (0.16) & 0.22 (0.21) \\
     \hline 
     \hline
ALF2-100150  & 2.1 (1.8)   & 2.7 (1.9)     & 0.095 (0.092) & 0.15 (0.14) & 0.07 (0.06) & 0.15 (0.14) & 0.15 (0.15) \\
H4-100150    & 2.7 (2.6)   & 4.5 (3.0)     & 0.130 (0.084) & 0.17 (0.14) & 0.03 (0.04) & 0.17 (0.15) & 0.16 (0.15) \\
MS1b-100150  & 3.2 (2.8)   & 4.4 (2.8)     & 0.124 (0.078) & 0.16 (0.14) & 0.03 (0.04) & 0.16 (0.14) & 0.17 (0.13) \\
SLy-100150   & 1.8 (1.5)   & 5.1 (1.8)     & 0.095 (0.023) & 0.19 (0.14) & 0.12 (0.06) & 0.19 (0.14) & 0.23 (0.14) \\
     \hline
     \hline
ALF2-110165  & 2.4 (1.5)   & 4.2 (2.1)     & 0.101 (0.088) & 0.17 (0.15) & 0.07 (0.08) & 0.17 (0.15) & 0.18 (0.16) \\
H4-110165    & 1.7 (2.0)   & 2.7 (2.9)     & 0.123 (0.105) & 0.17 (0.16) & 0.04 (0.04) & 0.17 (0.16) & 0.18 (0.17) \\
MS1b-110165  & 2.6 (2.5)   & 4.1 (3.5)     & 0.126 (0.101) & 0.17 (0.16) & 0.04 (0.05) & 0.17 (0.16) & 0.16 (0.17) \\
SLy-110165   & 1.6 (1.0)   & 4.3 (2.5)     & 0.064 (0.054) & 0.19 (0.17) & 0.11 (0.12) & 0.19 (0.18) & 0.21 (0.24) \\ 
     \hline
     \hline
ALF2-117175  & 2.5 (1.8)   & 6.0 (4.9)     & 0.100 (0.088) & 0.19 (0.16) & 0.06 (0.09) & 0.18 (0.18) & 0.20 (0.20) \\
H4-117175    & 1.4 (1.6)   & 2.6 (2.6)     & 0.125 (0.105) & 0.18 (0.17) & 0.05 (0.06) & 0.19 (0.17) & 0.19 (0.17) \\
MS1b-117175  & 2.4 (2.5)   & 4.3 (3.9)     & 0.108 (0.099) & 0.18 (0.16) & 0.05 (0.08) & 0.18 (0.16) & 0.17 (0.19) \\
SLy-117175   & 0.65 (0.60) & 3.1 (1.6)     & 0.118 (0.063) & 0.25 (0.14) & 0.11 (0.07) & 0.38 (0.15) & 0.24 (0.20) \\      
     \hline
     \hline     
ALF2-100175  & 3.6 (5.0)   & 7.1 (9.5)     & 0.118 (0.113) & 0.18 (0.18) & 0.03 (0.05) & 0.19 (0.19) & 0.21 (0.17) \\ 
H4-100175    & 4.0 (4.2)   & 7.0 (6.2)     & 0.131 (0.106) & 0.17 (0.16) & 0.02 (0.03) & 0.18 (0.17) & 0.29 (0.18) \\
MS1b-100175  & 4.9 (5.5)   & 8.5 (9.4)     & 0.134 (0.115) & 0.17 (0.17) & 0.03 (0.03) & 0.18 (0.17) & 0.19 (0.18) \\     
SLy-100175   & 2.4 (2.8)   & 5.0 (4.8)     & 0.096 (0.044) & 0.19 (0.16) & 0.03 (0.05) & 0.21 (0.16) & 0.21 (0.22)  \\      
     \hline 
     \hline
MS1b-094194  & 6.5 (6.8)   & 12 (13)       & 0.130 (0.126) & 0.18 (0.18) & 0.02 (0.02) & 0.18 (0.18) & 0.17 (0.21) \\
     \hline     
     \hline  
  \end{tabular}
  \label{tab:ejecta}
\end{table*}

\subsection{Merger Remnant}
\label{sec:merger_remnant}

\begin{table*}[t]
  \centering    
  \caption{Properties of the merger remnant. 
    The columns represent: 
  (i) the classification of the merger remnant; 
    (ii) the lifetime in case a HMNS has formed and collapsed during our 
    simulation (given in multiples of $100 M_\odot$) 
    or a description as `p.c.' for a prompt collapse 
  of the remnant; 
  (iii) the final mass of the BH $M_{BH}$;
  (iv) the dimensionless spin of the final BH $j_{BH}$; 
  (v) the mass of the disk surrounding the BH $M_{disk}$ measured 
  $250M_\odot$ after BH formation.
  We mark runs where the apparent horizon finder did not work reliably 
  with a * and runs where no BH 
  has formed for one resolution but for the other with **.}
  \begin{tabular}{l|ccccc}        
    Name & remnant &  $\tau \  [100 M_\odot] $& $M_{BH}[M_\odot]$ & $j_{BH}$ & $M_{disk} [M_\odot]$ \\
     \hline
     \hline
ALF2-137137  & HMNS$\rightarrow$BH & 26 (24) & 2.49 (2.50) & 0.65 (0.65) & 0.20 (0.17) \\
H4-137137    & HMNS$\rightarrow$BH & 42 (39) & 2.44 (*)    & 0.59 (*)    & 0.23 (0.11) \\
MS1b-137137  & MNS                 & -       &   -         & -           & - \\
SLy-137137   & HMNS$\rightarrow$BH & 77 (36) & 2.42 (2.43) & 0.60 (0.61) & 0.23 (0.21) \\
\hline
\hline  
ALF2-122153  & HMNS$\rightarrow$BH & 23 (**) & 2.48 (**)   & 0.64 (**)   & 0.23 (**)  \\
H4-122153    & HMNS                & -       &  -          & -           & - \\
MS1b-122153  & MNS                 & -       &  -          & -           & - \\
SLy-122153   & HMNS$\rightarrow$BH & 33 (29) & 2.44 (2.45) & 0.59 (0.60) & 0.20 (0.20)  \\
\hline 
\hline
ALF2-100150  & HMNS                & -       &  -          & -           & - \\
H4-100150    & HMNS                & -       &  -          & -           & - \\
MS1b-100150  & MNS                 & -       &  -          & -           & - \\
SLy-100150   & HMNS                & -       &  -          & -           & - \\
\hline
ALF2-110165  & HMNS$\rightarrow$BH & 21 (42) & 2.48 (2.40) & 0.61 (0.57) &  0.24 (0.30)  \\
H4-110165    & HMNS                & -       &  -          & -           & - \\
MS1b-110165  & MNS                 & -       &  -          & -           & - \\
SLy-110165   & HMNS$\rightarrow$BH & 58 (**) & 2.45 (**)   & 0.61 (**)   &  0.26 (**)  \\ 
\hline   
ALF2-117175  & BH                  & p.c.    &  2.70 (2.67)  & 0.71 (0.70)  & 0.20 (0.21) \\
H4-117175    & HMNS$\rightarrow$BH & 14 (17) &  2.65 (2.62)  & 0.68 (0.65)  & 0.26 (0.24) \\
MS1b-117175  & SMNS                & -       &  -            & -            & - \\
SLy-117175   & BH                  & p.c.    &  2.75 (2.73)  & 0.73 (0.72)  & 0.14 (0.15) \\     
\hline
\hline     
ALF2-100175  & HMNS$\rightarrow$BH & 25 (26) & 2.48 (2.42) & 0.61 (0.57)  & 0.26 (0.29) \\ 
H4-100175    & HMNS                & -       &  -          & -            & - \\
MS1b-100175  & MNS                 & -       &  -          & -            & - \\     
SLy-100175   & BH                  & p.c.    & * (*)       & * (*)        & 0.18 (0.21) \\      
\hline 
\hline
MS1b-094194  & SMNS                & -       &  -          & -            & - \\
\hline     
\hline  
\end{tabular}
\label{tab:remnant}
\end{table*}    

Tab.~\ref{tab:remnant} summarizes the merger outcome (second column)
and the properties of the merger remnant.  
In most cases ($15$ configurations) the merger remnant is a HMNS, which is temporary stabilized from collapse by
thermal pressure and centrifugal support. The HMNS will collapse
within a dynamical timescale to a BH; this happens within the
simulated time for $9$ configurations.  
$3$ configurations undergo a prompt collapse and for $7$ cases the centrifugal support is sufficient 
to support a MNS or SMNS. 
In the following we discuss the lifetime of the HMNSs and the 
properties of the final BH and disk structure.\\ 

\paragraph*{Lifetime of the merger remnant:}
The collapse time of the merger remnant is very sensitive to numerical
error and grid resolution, e.g.~\cite{Bernuzzi:2013rza}.
Precise numbers are difficult to obtain, and uncertainties can be of the
other of several milliseconds.
From Tab.~\ref{tab:remnant} one observes that a larger total mass of
the system leads to an earlier collapse, as expected by the fact that a larger
rest-mass star is closer to the collapse threshold.

The influence of the EOS is less clear than the effect of the total mass, 
however, as outlined in~\cite{Dietrich:2015iva} one can expect that systems with a softer EOSs
collapse earlier than systems with a stiff EOS. 
The EOSs SLy, ALF2, H4 EOS support approximately the same maximum mass regarding
single spherical stars, cf.~Fig.~\ref{fig:EOS} and Tab.~\ref{tab:EOS}. 
But, the stiffer the EOS, the longer is the lifetime of the HMNS.
Binaries with stiffer EOS are in general less bound at merger than soft EOSs
binaries. This fact has been quantified in \cite{Bernuzzi:2014kca},
that computed universal relations of the binary's binding energy and
angular momentum at formation of the merger remnant as functions of
the parameter $\kappa^T_2$, Eq. \eqref{eq:kappa}. A consequence of
those relations is that stiffer EOS lead to binaries with larger centrifugal support.  
In addition to the centrifugal support, the pressure support in the central regions 
is larger for stiffer EOS, and the merger remnant is even further stabilized. 

Focusing on the effect of the mass-ratio on the lifetime of the
remnant, we observe for several cases that larger $q$ give slightly larger lifetimes.
Regarding, e.g., setups employing the H4 EOS we find that the equal mass setup 
collapses to a BH, but the HMNSs formed during the merger of most of the
unequal mass systems 
survive until the end of the simulation. 
However, for a more conclusive statement we suggest that higher 
resolution are needed since for several setups the differences in the lifetime of the 
merger remnant caused by different mass ratios 
lie within the error bar obtained from different resolutions. 
For larger $q$ the merger remnant is typically more non-axisymmetric than 
for equal-mass configurations, and thus has a larger angular momentum support.
Also, looking again at Eq. \eqref{eq:kappa}, for fixed EOS ($k_2$) and fixed
masses, larger $q$ correspond to larger $\kappa^T_2$. 
Using the relations of \cite{Bernuzzi:2014kca}, this implies indeed
that remnants from binaries with larger $q$ have larger angular
momentum at formation.  
\\ 

\begin{figure}[t]
  \includegraphics[width=0.5\textwidth]{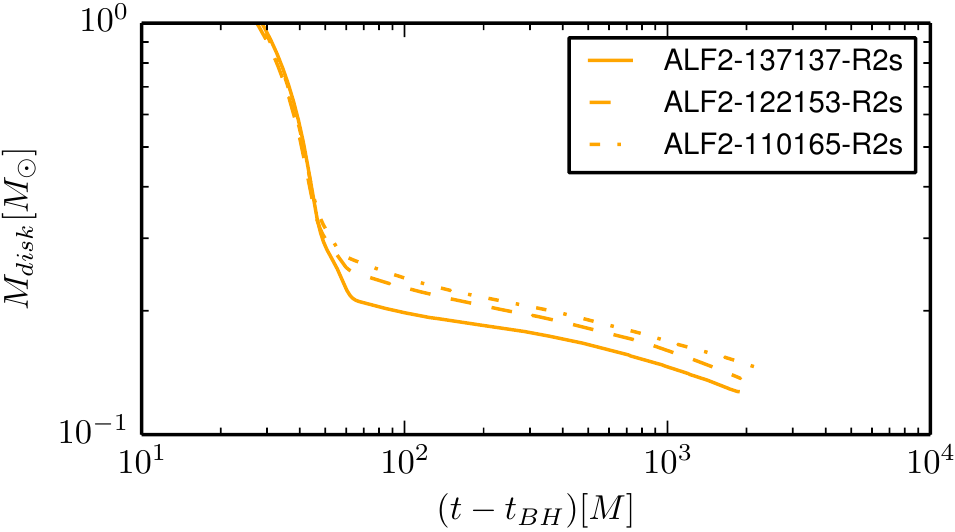}
  \caption{Rest mass of the disk surrounding the final BH for the ALF2 configurations with a mass of $M=2.75$ and mass ratios 
  $q=1.0;1.25,1.5$. Clearly visible is that for an increasing mass-ratio the disk is more massive.
  However, ALF2-100175 and ALF-110165 give almost the same disk masses. We have not included 
  ALF2-100175 in this figure since the simulation after the collapse to the BH is much shorter than 
  for the other configurations.}
  \label{fig:Mdisk}
\end{figure}     

\paragraph*{Final black hole and disk:}  

For $12$ configurations a black hole has formed during our simulations, and we expect that if 
we evolved our configurations for longer the remaining $6$ HMNS would have formed a BH as well. 
In cases where a BH forms we also report the BH mass $M_{BH}$, the dimensionless spin $j_{BH}$ 
of the BH and the mass of the accretion disk $M_{disk}$. 

The mass of the BH depends trivially on the mass of the binary systems, 
i.e.~more massive systems produce more massive BHs. 
Regarding $M_{BH}/M$, we find that for systems with $M=2.75 M_\odot$ 
at a time $t = t^{\rm mrg} + 250 M_\odot$ 
$M_{BH}$ is $\sim 88\%$- up to $91\%$ of the total mass.  
For systems with a total mass of $M\approx2.92$ at $t=t^{\rm mrg} + 250 M_\odot$ 
the BH mass is $89\%$--$94\%$ of the total mass. 
Interestingly for the irreducible mass $M_{\rm irr}=\sqrt{A_{\rm BH}/(16 \pi)}$ (with $A_{\rm BH}$ being the horizon area)
we find that $M_{\rm irr}/M$ is almost independent of the EOS, mass, and mass ratio and takes a value around 
$\sim0.85$ for all setups. 
As the final BH mass, also the final dimensionless spin of the BH $j_{BH}$ 
is larger for more massive systems, see Tab.~\ref{tab:remnant} and lies around $\sim 0.7$
for our systems with a mass of $2.92M_\odot$.
On the other hand, the imprint of the mass-ratio on the 
mass and spin of the BH is less clear. We find that the 
black hole mass is almost independent of the mass-ratio and that 
the dimensionless spin is slightly decreasing for an increasing mass ratio, 
cf.~in particular the systems employing the ALF2 EOS.
The decrease in the dimensionless spin is caused by the fact that 
outer region and the formed disk has larger angular momentum. 
Among the different EOS considered and for fixed mass ratio and
total mass, the ALF2 EOS produces in most cases a faster rotating
black hole than the other configurations.  

We find in our simulations that, for configurations undergoing prompt
collapse, the system has no sufficient time to redistribute angular
momentum to outer regions. Thus, the formed disk is less massive. 
In other configurations that form a HMNS, larger $q$ give more
massive disks \cite{Shibata:2006nm,Rezzolla:2010fd}.
In Fig.~\ref{fig:Mdisk} we present the disk mass as a function of time
after BH formation for the ALF2 configurations up to $q=1.50$ with
$M=2.75$. 
The disk mass increases up $30\%$ comparing $q=1$ and $q=1.75$.
We find that in general disks produced by larger mass ratios 
have smaller maximum densities, but a larger radii. 
For SLy no monotonic trend is present in our simulations.

\section{Gravitational Waves}  
\label{sec:GW}

\begin{figure*}[t]
  \includegraphics[width=1\textwidth]{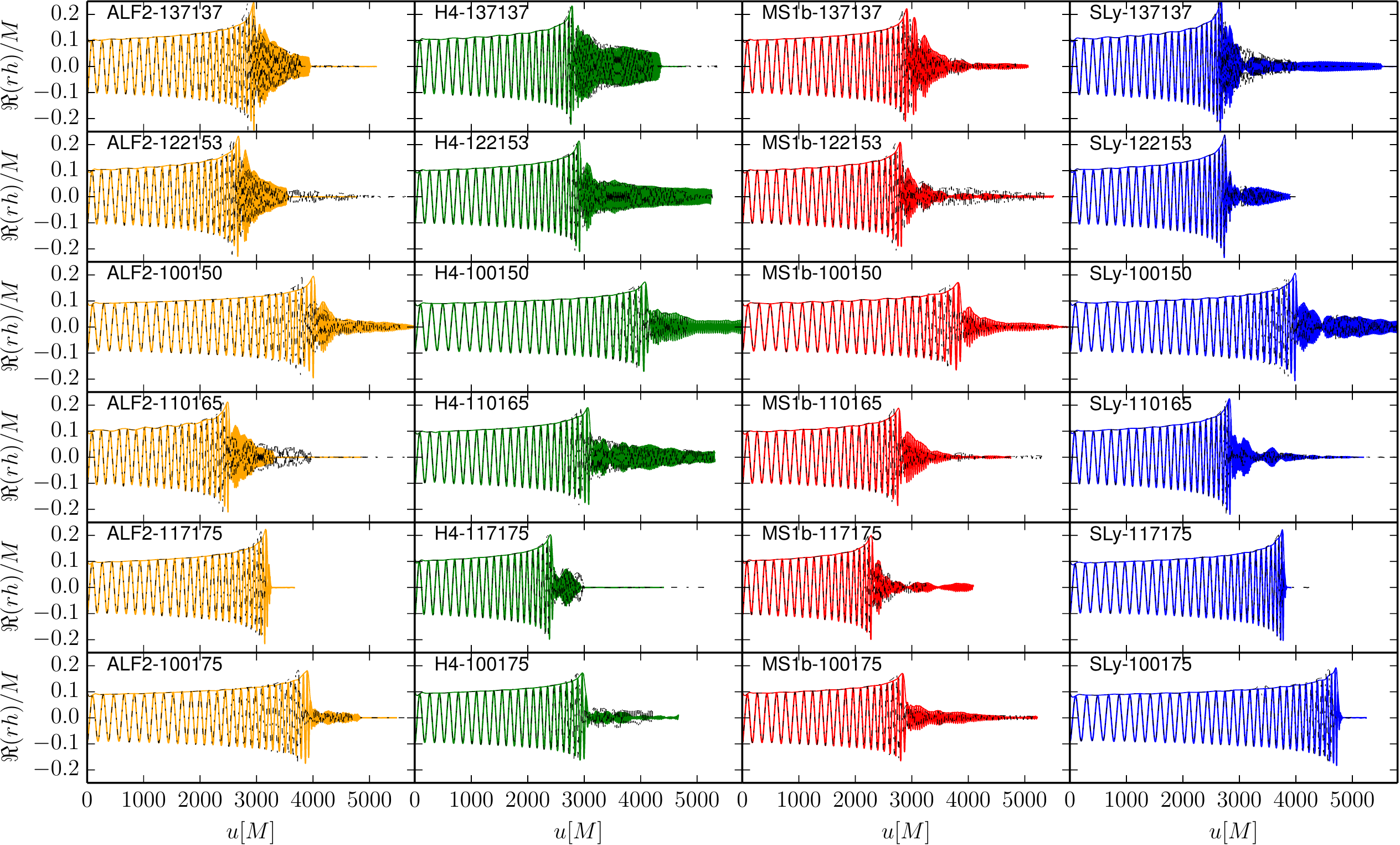}
  \caption{Real part and amplitude of the (2,2)-multipole $h_{lm}(u)$
    of the GWs for different 
    configurations and the highest resolution.
    Different colors correspond to different EOS:
    ALF2 is orange, H4 green, MS1b red, and SLy blue. The lower
    resolution runs are shown as a black dotted line to understand  
    the influence of the resolution in our configurations.   }
  \label{fig:GW_h22} 
\end{figure*}  

GWs are calculated extracting the curvature invariant
$\Psi_4$ on coordinate spheres and computing projections on the
spin-weighted spherical harmonics for 
spin $-2$, ${}^{-2}Y_{lm}$, see e.g.~\cite{Brugmann:2008zz}. The metric multipoles $r 
h_{\ell m}$ are reconstructed from the curvature multipoles using the frequency domain
integration of~\cite{Reisswig:2010di}. The initial circular GW frequency 
is used as a cutting frequency. The GW strain is then 
given by 
\begin{equation}
h(t,\theta,\phi) = \sum_{l=2}^{l_\text{max}} \sum_{m=-l}^{l} r\,
h_{lm}(t) {}^{-2}Y_{lm}(\theta,\phi) \ ,
\end{equation}  
where we include individual modes up to $l_\text{max}=4$.
All waveforms are plotted against the retarded time, 
\begin{equation}
 u=t-r_*=t-r_{\rm extr}-2M\ln\left(r_{\rm extr}/2M-1\right),
\end{equation}
where the extraction radius is set to $r_{\rm extr}\sim 1000\,M_\odot$. 
As explained in~\cite{Bernuzzi:2016pie} for such an extraction radius the error compared to 
extrapolated waveforms is $\lesssim0.5\%$ for the amplitude and 
below $0.1$~rad for the phase of the GW.
The GW energy radiated during the simulations is calculated as
\begin{align}
\mathcal{E}_{\rm rad} &= \sum_{l,m}^{l_{\rm max}} E_{l,m} = \label{eq:GW_Erad}
\dfrac{1}{16\pi}\sum_{l,m}^{l_{\rm max}} \int_{0}^t
dt'\left|r\,\dot{h}_{lm}(t')\right|^2 \ ,  
\end{align}
with $l_{\rm max}=8$. We also define the $m$-mode contributions as 
\be
E_m := \sum_{l=m}^{l_{\rm max}} E_{lm} \ .
\ee
The \textit{total} GW energy emitted during the binary history is given by 
\be
E^{\rm tot}_{\rm rad} = M - M_{\rm ADM}(t=0) + \mathcal{E}_{\rm rad}
\ .
\ee

In addition to the waveforms, we compute spectra and
spectrograms by applying the Fourier transform $\mathcal{F}$ (in fact,
the discrete FFT) and compute the power spectral density (PSD)  
of the GWs as in  \cite{Bernuzzi:2013rza}, 
\begin{eqnarray}
 \tilde{h}_{lm}(f) & = & |\mathcal{F}[h_{lm}(t)]|, \label{eq:htilde} \\
 \tilde{h}(f; \theta,\phi) & =  & |\mathcal{F}[h(t; \theta,\phi)]| \ . \label{eq:htilde}
\end{eqnarray}
The spectrogram allows to identify to which part
of the wave [and so of the dynamics] the spectrum peaks correspond to.
They are computed with chunks of width $\Delta t = 400M_\odot$. 
Note also that $\tilde{h}$ depends on the inclination $\theta$ and
azimuthal angle $\phi$,  
i.e.~the location of the detector with respect to our source. 
For most of our analysis we pick $\theta=\pi/4$ and $\phi=0$, i.e., 
the line of sight to the detector is $45^\circ$ above the orbital plane. 
But, we will also consider one BNS configuration for which we investigate the influence of 
$\theta$ on $\tilde{h}$. 

Figure~\ref{fig:GW_h22} shows the real part of the dominant (2,2)
multipole for our configurations, where 
the solid colored lines represent the highest resolved runs (R2) and the dashed black
lines the lower resolution (R1). The signal is composed by the
well-known chirp corresponding the inspiral-to-merger transition,
which formally ends at the wave amplitude's peak, followed by the
post-merger emission which corresponds to the HMNS/MNS-phase. 
In some simulations the signal also contains the collapse to the BH and
the quasi-normal ringing.  

Table~\ref{tab:GWs} reports the number of orbits from the start of the
simulation to the end of the chirp, $u:=u_\text{mrg}$, defined as the
time of the 
$h_{22}$ amplitude peak. We also report there the GW frequency at
$u_\text{mrg}$ as $f_\text{mrg}$. 
We find that the dimensionless frequency at merger
$M \omega_\text{mrg}$ depends on the EOS and the mass ratio. While stiffer EOSs merge
with a lower frequency, softer EOSs merge at higher frequencies.  
Furthermore, higher mass ratios lead to smaller merger frequencies. 
This behavior is understood in terms of the leading order tidal
coupling constant $\kappa^T_2$, and can be encoded with high precision
in the quasi-universal relations proposed in \cite{Bernuzzi:2014kca}.

In the following, we discuss spectra, spectrograms, and GW energy focusing
on the post-merger phase. Similar analysis have been reported in
e.g.~\cite{Bernuzzi:2013rza,Clark:2015zxa,Foucart:2015gaa,Rezzolla:2016nxn}. 
GW from unequal-masses BNS, in particular, have been computed recently
in e.g. \cite{Hotokezaka:2012ze,Takami:2014tva,Bernuzzi:2015rla,Lehner:2016lxy}.

\begin{table*}[t]
  \centering    
  \begin{small}
    \caption{GW quantities. The columns refer to: 
      the name of the configurations, the number of orbits until merger from the beginning of the simulation estimated as 
      $N_\text{orb}=\Phi^\text{mrg}/4\pi$ with $\Phi^\text{mrg}$ being the accumulated phase,  
      the dimensionless merger frequency $M \omega_{\rm mrg}$, 
      the merger frequency in kHz $f_{\rm mrg}$, 
      the dominant postmerger frequencies for 
      extracted from the (2,1),(2,2),(3,3)-mode, and a possible secondary peak in the (2,2)-mode. 
      Also these frequencies are stated in kHz. In cases where no secondary peak is found, we mark this 
      simulations with $-$. We abbreviate the prompt collapse of some configurations with p.c..
      Results in brackets refer to R1 resolved runs expect for
      MS1b-094194, where we show results for R2 in brackets.}
    \begin{tabular}{l|cc|cccccc}        
      Name & $N_\text{orb}$ & $M\omega_{\rm mrg}$ & $f_{\rm mrg}$  & $f_1$ &  $f_{2}$ & $f_3$ & $f_{s}$   \\
      \hline
      \hline
      ALF2-137137  & 11.7 (11.2) & 0.144 (0.142) &  1.72 (1.70) & 
      1.55 (1.46) & 2.80 (2.77) &  4.30 (4.06) & 1.70 (1.70)\\
      H4-137137    & 10.9 (10.7) & 0.133 (0.127) & 1.59 (1.52) &   
      1.27 (1.38) & 2.50 (2.58) & 3.74 (3.84)     & 1.61 (1.74) \\
      MS1b-137137  & 10.9 (10.7) & 0.121 (0.118) & 1.45 (1.41)& 
      1.06 (1.06) & 2.14 (2.06) & 3.10 (3.07) & 1.58 (1.57) \\
      SLy-137137   & 11.4 (11.2) &  0.167 (0.163) & 2.00 (1.95) & 
      1.83 (1.76)& 3.66 (3.47) & 5.39 (5.16) & 2.80 (2.56) \\               
      \hline
      \hline  	
      ALF2-122153  & 10.9 (10.4) & 0.133 (0.131) & 1.59 (1.57) &
      1.42 (1.44) & 2.72 (2.68) & 4.11 (4.13) & 2.41 (2.43) \\
      H4-122153    & 11.1 (10.9) & 0.114 (0.115) & 1.36 (1.38) & 
      1.28 (1.24) & 2.42 (2.38) & 3.78 (3.70) & 1.75 ( - )   \\
      MS1b-122153  & 10.6 (10.2) & 0.110 (1.08) & 1.32 (1.29) &  
      1.07 (1.08) & 2.02 (2.16) & 3.12 (3.24) & 1.82 (1.80) \\ 
      SLy-122153   & 11.3 (11.3) & 0.143 (0.144) & 1.71 (1.72) & 
      1.65 (1.66) & 3.30 (3.36) & 5.09 (5.04) & 2.91 (3.00) \\
      \hline 
      \hline
      ALF2-100150  & 13.8 (12.9)   & 0.106 (0.103) & 1.37 (1.33) & 
      1.19 (1.20) & 2.44 (2.38) & 3.69 (3.60) & - \\
      H4-100150    & 13.4 (12.8)   & 0.086 (0.084) & 1.11 (1.09) & 
      1.15 (1.05) & 2.20 (2.10) & 3.25 (3.05) & - \\ 
      MS1b-100150  & 13.4 (12.8)   & 0.084 (0.084) & 1.11 (1.11) & 
      1.00 (0.95) & 1.95 (1.95) & 2.90 (2.95) & - \\
      SLy-100150   & 14.3 (13.6)  &  0.114 (0.111) & 1.51 (1.47) & 
      1.46 (1.43)  & 2.96 (2.87) & 4.40 (4.23) & 2.59 (2.54) \\
      \hline
      ALF2-110165  & 10.2 (9.8) & 0.119 (0.118) & 1.40 (1.39) & 
      1.45 (1.32) & 2.74 (2.74) & 4.17 (4.06) & 1.89 (1.95) \\
      H4-110165    & 11.6 (10.9) & 0.101 (0.098) & 1.19 (1.15) & 
      1.28 (1.24) & 2.57 (2.43) & 3.83 (3.54) &  - \\
      MS1b-110165  & 10.4 (10.0) & 0.098 (0.096) & 1.15 (1.13) & 
      0.99 (1.06) & 1.94 (1.98) & 3.00 (3.02) & - \\
      SLy-110165   & 11.6 (11.3) & 0.133 (0.130) & 1.56 (1.53) & 
      1.76 (1.62) & 3.51 (3.51) & 5.25 (5.04) & 2.31 (-)  \\ 
      \hline   
      ALF2-117175  & 12.4 (12.0) & 0.131 (0.127) & 1.45 (1.41)& 
      p.c. & p.c. & p.c. & p.c.   \\ 
      H4-117175    & 9.9 (9.6) & 0.110 (0.109) & 1.22 (1.21) & 
      1.38 (1.36) &  2.82 (2.80) & 4.21 (4.05) & 1.95 (1.90) \\
      MS1b-117175  & 9.3 (9.0) & 0.109 (0.110) & 1.21 (1.22) &
      1.11 (1.04)  & 2.09 (1.99) & 3.20 (3.10) & - \\
      SLy-117175   & 14.6 (14.2) & 0.148 (0.137) & 1.64 (1.52)& 
      p.c. & p.c. & p.c. & p.c.   \\     
      \hline
      \hline          
      ALF2-100175  & 14.0 (13.4)  & 0.106 (0.109) & 1.27 (1.31)& 
      p.c. & p.c. & p.c. & p.c.   \\ 
      H4-100175    & 11.0 (10.6) &  0.089 (0.090) & 1.07 (1.07) & 
      1.25 (1.21) & 2.44 (2.50) & 3.77 (3.68) & - \\
      MS1b-100175  & 10.4 (10.0) & 0.088 (0.087) & 1.05 (1.04) & 
      0.96 (0.93) & 1.98 (2.05) & 3.03 (2.99) & - \\     
      SLy-100175   & 16.6 (15.9) & 0.122 (0.105) & 1.35 (1.25) & 
      p.c. & p.c. & p.c. & p.c.   \\                
      \hline 
      \hline
      MS1b-094194  & 10.7 (10.5) & 0.087 (0.088) & 0.96 (0.97) & 
      0.92 (1.01) & 2.14 (2.00) & 3.20 (3.23)  & - \\
      \hline     
      \hline  
    \end{tabular}
    \label{tab:GWs}
  \end{small}
\end{table*}

 \subsection{Spectrograms}
 \label{sec:GW_spectra}
 
 \begin{figure*}[t]
   \includegraphics[width=1\textwidth]{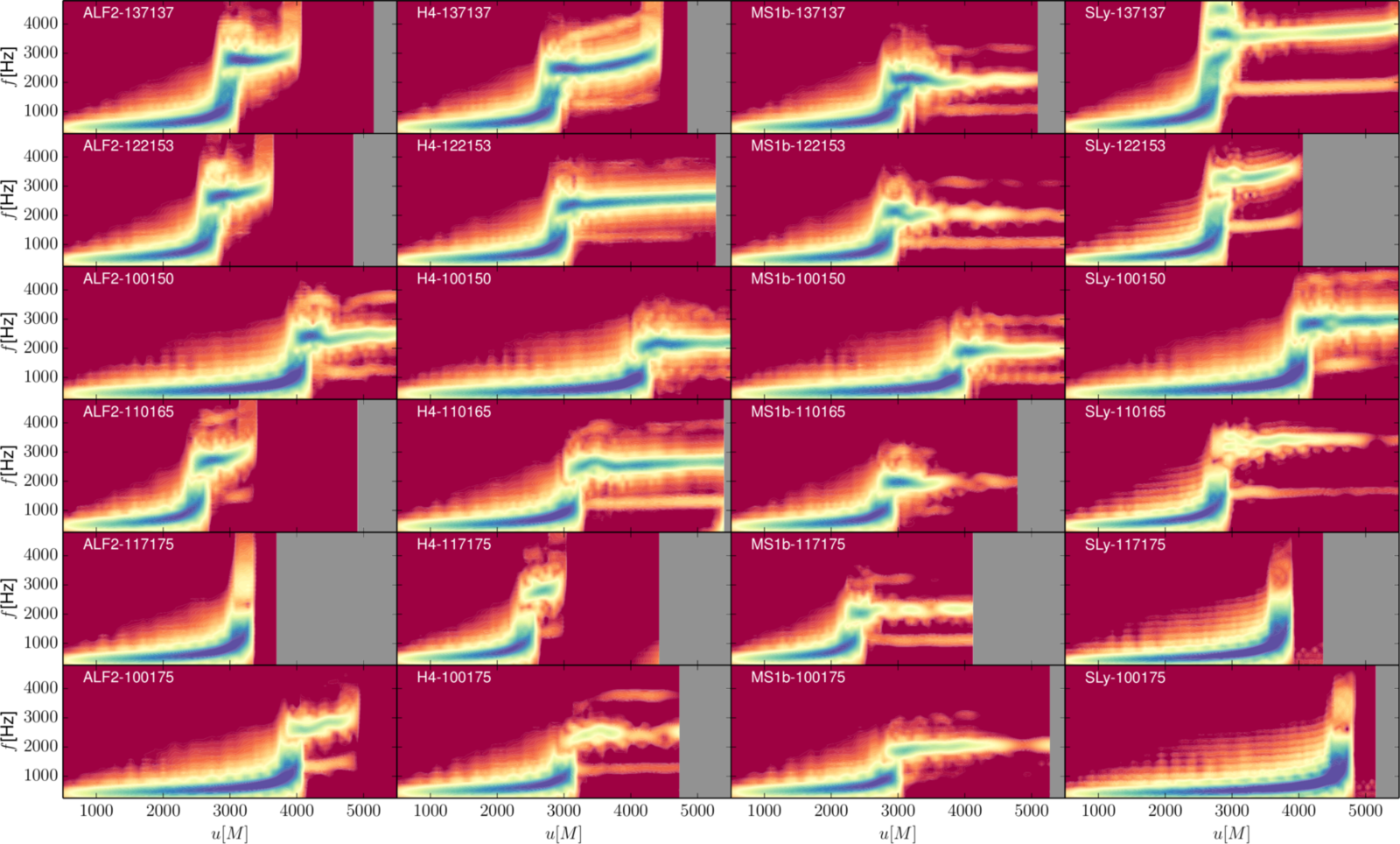}
   \caption{Spectra $\tilde{h}(\theta=\pi/4,\phi=0)$ for all configurations.  
     Clearly visible is the chirp signal of the inspiral-merger phase
     followed by the post-merger signal (if no prompt collapse). 
     In most of the cases, the spectrogram highlights that several
     frequencies, other than the $f_2$, are excited in the postmerger.
     Note also that configurations like ALF2-137137, H4-137137, and
     SLy-122153, in which a HMNS forms and collapse within dynamical
     times, clearly shows that the postmerger spectrum is not
     discrete, but rather continuous, and similar to the
     inspiral-merger chirp.}
   \label{fig:GW_h22_spectra} 
 \end{figure*}    
  
We present the spectrograms for our configurations in
Fig.~\ref{fig:GW_h22_spectra}.
The color bar goes from red to blue and is given in arbitrary units, since  
we are only interested in the frequencies and the relative strength.
Additional information are given in Table~\ref{tab:GWs}.

During the inspiral-merger one observes the typical chirp signal, with
frequency and amplitude increasing monotonically over time. 
The end of the chirp, $u=u_\text{mrg}$, is marked by an amplitude
peak followed by a sharp amplitude's cut-off. 
The (2,2)-mode is by far the most dominant mode emitting more than 99\%
of the total energy. Thus, higher modes can not be seen when
$\tilde{h}$ is considered, but can be studied looking at
$\tilde{h}_{lm}$ (see below).

The postmerger spectra are mainly characterized by a dominant emission
frequency $f_2$, related to the (2,2)-mode. A prominent secondary peak
in the (2,2) channel is also visible at frequency $f_s<f_2$ for several configurations (see also
Fig.~\ref{fig:hspectra} below). Interpretation of the latter have
been proposed in e.g.~\cite{Takami:2014tva,Rezzolla:2016nxn,Bauswein:2015yca,Clark:2015zxa} 
and references therein. Our spectrograms indicate the peak originates
during the very early postmerger phase, right after the GW amplitude peak
that marks the end of the chirping signal (inspiral-merger), see 
e.g.~the panel of MS1b-137137 for $u\sim3000M$. During this short
period the waveforms are characterized, in addition to the oscillation
at frequency $f_2$, by an amplitude modulation corresponding to the
fluid's mass axisymmetric mode.

Notably, for larger mass ratios the spectra becomes more complicated
as shown in Fig.~\ref{fig:hspectra}. At fixed EOS and $M$, the peak of
the spectra for larger $q$ has less power and more peaks appear at frequencies
$f_\text{mrg}<f<f_2$. The secondary peak, in particular, at $f_s$ has
maximum power for $q=1$ and progressively disappears for larger values
$q>1$. Correspondingly, we checked that, for large $q$, the fluid's
mass axisymmetric oscillations diminish (see also the wave's amplitude
in Fig.~\ref{fig:GW_h22}). 

\begin{figure}[t]
  \includegraphics[width=0.5\textwidth]{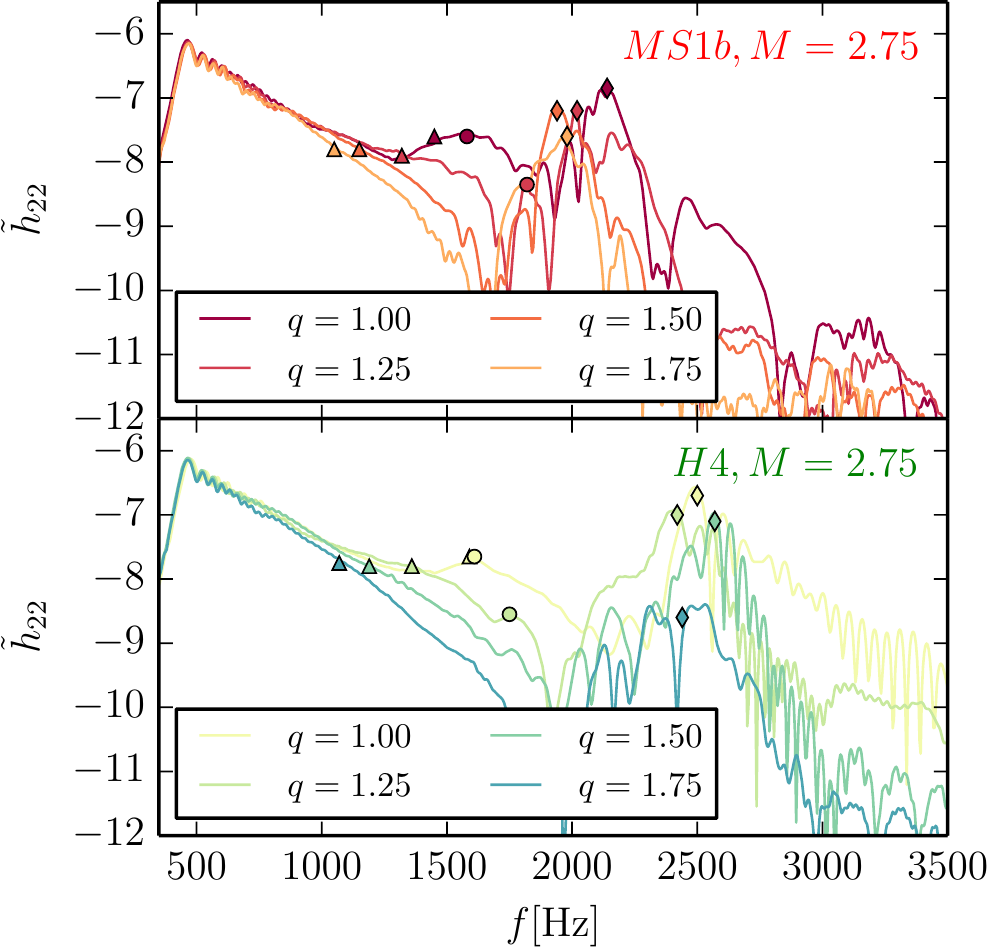}
  \caption{Effect of mass-ratio on the GW spectra. 
  Top: $\tilde{h}_{22}$ for MS1b setups for mass ratios $q=1.00;1.25;1.50;1.75$. 
  Bottom: $\tilde{h}_{22}$ for H4 setups for mass ratios $q=1.00;1.25;1.50;1.75$.
  We mark the $f_2$-frequencies as diamonds, the $f_{\rm mrg}$ as triangles and 
  the secondary peak frequencies $f_s$ as circles.}
  \label{fig:hspectra}
\end{figure} 

By considering the multipolar waveforms $\tilde{h}_{lm}$, and not only the $(2,2)$ mode,
we observe other modes excited during the postmerger. 
These mode frequencies are named $f_1,f_3,...,f_m$ and roughly
correspond to the azimuthal $m=1,2,3,...$ modes of
oscillations of the fluid\footnote{To avoid confusion, we explicitly mention the difference between 
  our $f_1$ and $f_3$ frequency and the one defined in~\cite{Takami:2014tva}. 
  However, we decided to stick to the notation used in~\cite{Dietrich:2015iva,Radice:2016gym}, since this 
  notation is better suited once more than just the dominant $(2,2)$-mode is considered.}.
These frequencies are more robustly extracted in the ``late''
postmerger phase, i.e.~few milliseconds after the waveform amplitude
peak. For a clear interpretation we extract the $f_1$ frequency from the $(2,1)$ and the
$f_3$-frequency from the $(3,3)$ mode, but they are present in all the
GW multipoles unless killed for symmetry reason in the projection
integrals.  

\begin{figure}[t]
  \includegraphics[width=0.5\textwidth]{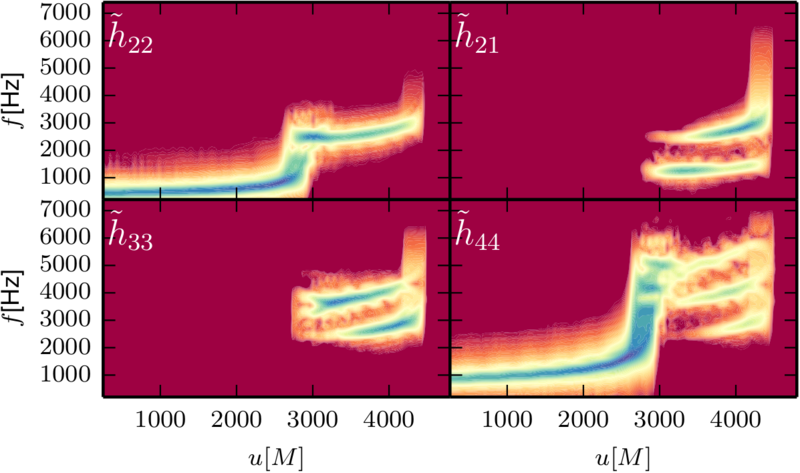}
  \caption{Spectra of individual GW modes for H4-137137-R2 setup. We
    present $\tilde{h}_{lm}$ for  
    $(lm)=(22),(21),(33),(44)$. The amplitudes are rescaled for better
    visibility, i.e.\  
    $\tilde{h}_{21}$ and $\tilde{h}_{33}$ are rescaled by a factor of $10^{2.5}$ and $\tilde{h}_{44}$ by $10^{3.5}$.}
  \label{fig:GW_h22_spectra_H137137} 
\end{figure}

As a further illustration of the multipoles effect, we present
$\tilde{h}_{lm}$ for H4-137137-R2 in Fig.~\ref{fig:GW_h22_spectra_H137137}. We have rescaled the 
individual contributions for better visibility (we multiplied $\tilde{h}_{21}$ 
and $\tilde{h}_{33}$ by $10^{2.5}$ and $\tilde{h}_{44}$ by $10^{3.5}$).
In this case it is clear that the (2,2) mode is dominated by the 
$f_2$ frequency, the (2,1)-mode by $f_2$ and $f_1$, the (3,3)-mode by $f_2$ and $f_3$, and 
for the (4,4)-mode we find peaks at the $f_2$,$f_3$,$f_4$ frequency. 
As expected the frequency of the (4,4)-mode during the inspiral-merger is
approximately twice the frequency of the (2,2)-mode.  
This harmonicity of the frequencies is also present in the post-merger phase.

Let us specify the spectrogram features for each of the different
postmerger scenarios. 

\paragraph*{MNS/SMNS waveforms:}
For the configurations employing the MS1b EOS, the merger remnant is a
stable MNS or a SMNS, see lines for the total mass of the systems in Fig.~\ref{fig:EOS}. 
The spectrogram is dominated by the frequency $f_2$, but one can clearly
see contributions from other multipoles in the $m=1,3$ and even the
$m=4$ channels.  The emission of
energy and angular momentum decreases over time, as
also visible in Fig.~\ref{fig:GW_h22} by a decreasing amplitude.
Characteristic timescales for the GW emission have been identified in
\cite{Bernuzzi:2015opx}.

\paragraph*{HMNS waveforms:}
In cases of a HMNS and, in particular for configurations undergoing
gravitational collapse within dynamical times, the postmerger signal
is shorter and peaks at specific frequencies $f_1,f_2,f_3$ are more
difficult to extract than for MNS/SMNS.
Considering the $f_2$-frequency one clearly observes a ``postmerger-chirp'', i.e.~that the 
frequency increases over time up to the formation of the BH, cf.~Fig.~\ref{fig:hspectra} 
for $q=1$ H4\footnote{Note that the 
wiggles for $f>2500$~Hz are just due to the Fourier transform of the finite-length signal.}.
But contrary to the inspiral the postmerger-chirp is characterized by a decreasing amplitude. 
The feature is physically expected from the increase of rotational velocity and compactness of the star over time
\cite{Bernuzzi:2015rla}. We observe it in all our configurations.
We stress that this indicates that spectra are actually \textit{continuous}, and
they can be modeled with discrete frequencies only for cases in which
the remnant lifetime is sufficiently long so that most of the GW
energy is radiated at frequencies close to $f_2$. This timescale is 
$\gtrsim20$~ms~\cite{Bernuzzi:2015opx}.

\paragraph*{Prompt Collapse waveforms:}
In cases of a prompt collapse to a BH, the spectra/spectrograms have
cut-off after the chirp with no other signal. 
Since no additional refinement levels are added once the BH forms, 
the resolution around the puncture is lower than in our BH simulations~\cite{Brugmann:2008zz}. 
Thus although the quasi-normal ringing is visible, we can only resolve between five to eight 
local maxima of $r\Psi_{4,22}$ after the merger. This is not sufficient to extract 
accurately the quasi-normal modes of the newly formed BH.

\subsection{Source sky location}

\begin{figure}[t]
  \includegraphics[width=0.48\textwidth]{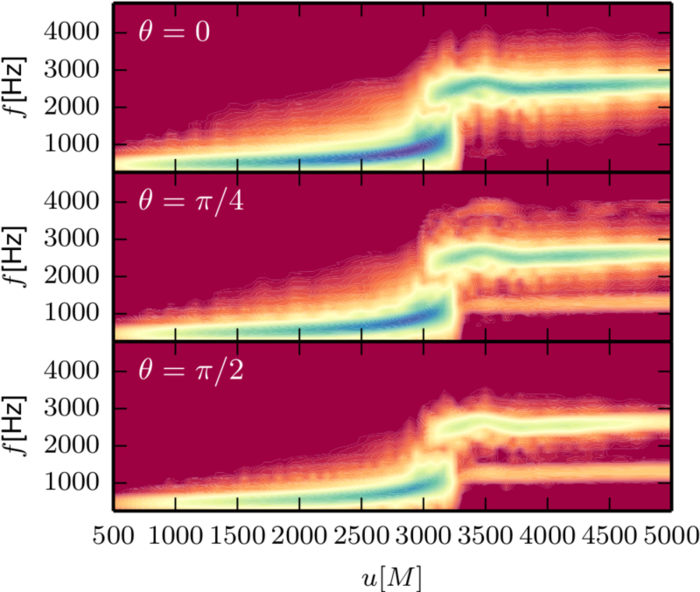}
  \caption{Angular dependence of the GW spectra for H4-110165-R2. We
    present $\tilde{h}$ for:  
    $(\theta=0,\phi=0)$ (top panel), 
    $(\theta=\pi/4,\phi=0)$ (middle panel), and $(\theta=\pi/2,\phi=0)$ (bottom panel).}
  \label{fig:spectra_detector} 
\end{figure}      

We discuss the influence of the source's sky location on the measured GW spectrum
for the model H4-110165. We pick as fiducial angles 
$(\theta=0,\phi=0)$, $(\theta=\pi/4,\phi=0)$, $(\theta=\pi/2,\phi=0)$ and 
present the results in Fig.~\ref{fig:spectra_detector}.
 
The inspiral-merger signal is strongest for $\theta=0$, which can be explained by the 
fact that the (2,2)-mode contribution to $h$ is largest for this angle
since ${}^{-2}Y_{22}(\theta)$ has its maximum  
for $\theta=0$. The post-merger signal is dominated by the $f_2$-mode,
and no $f_1$ frequency could be detected because
${}^{-2}Y_{21}(0)={}^{-2}Y_{2-1}^{-2}(0)=0$. 
Increasing the inclination $\theta$, the contribution from the 
$f_2$-frequency decreases, and the detectability of the
$f_1$-frequency increases. Note however that is unlikely the $m=1$
modes will be detected in GW observations \cite{Radice:2016gym}.

\subsection{GW Energy} 
\label{sec:GW_energy}

\begin{figure}[t]
  \includegraphics[width=0.5\textwidth]{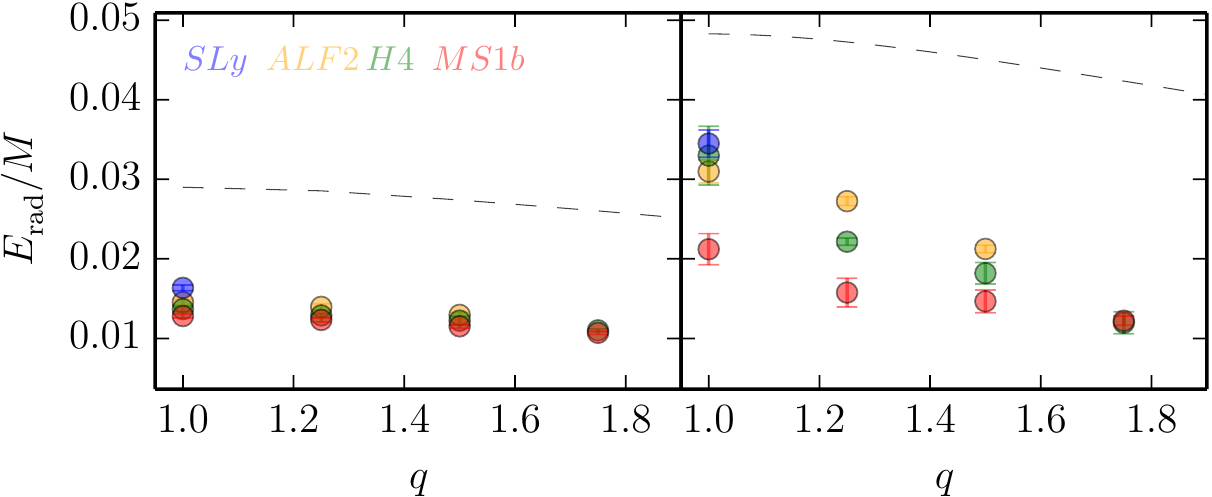}
  \caption{Total GW energy emitted for different EOSs and mass ratios. 
     For the total energy we also take into account contributions before the beginning of the
     simulation ($M-M_{\rm ADM}(t=0)$). Shown as dashed lines are 
     estimates for the released energy during BBH mergers (see text for details). 
     The left panel shows the energies at the moment of merger, i.e.~peak of the GW amplitude, 
     right panels represent data $20$ ms after merger.
     We have removed those setups for which the initial linear momentum, Tab.~\ref{tab:initPandecc}, 
     was large and an artificial drift of the center of mass was present.
     }
  \label{fig:Erad_modes_q_total} 
\end{figure} 

\begin{figure}[t]
  \includegraphics[width=0.5\textwidth]{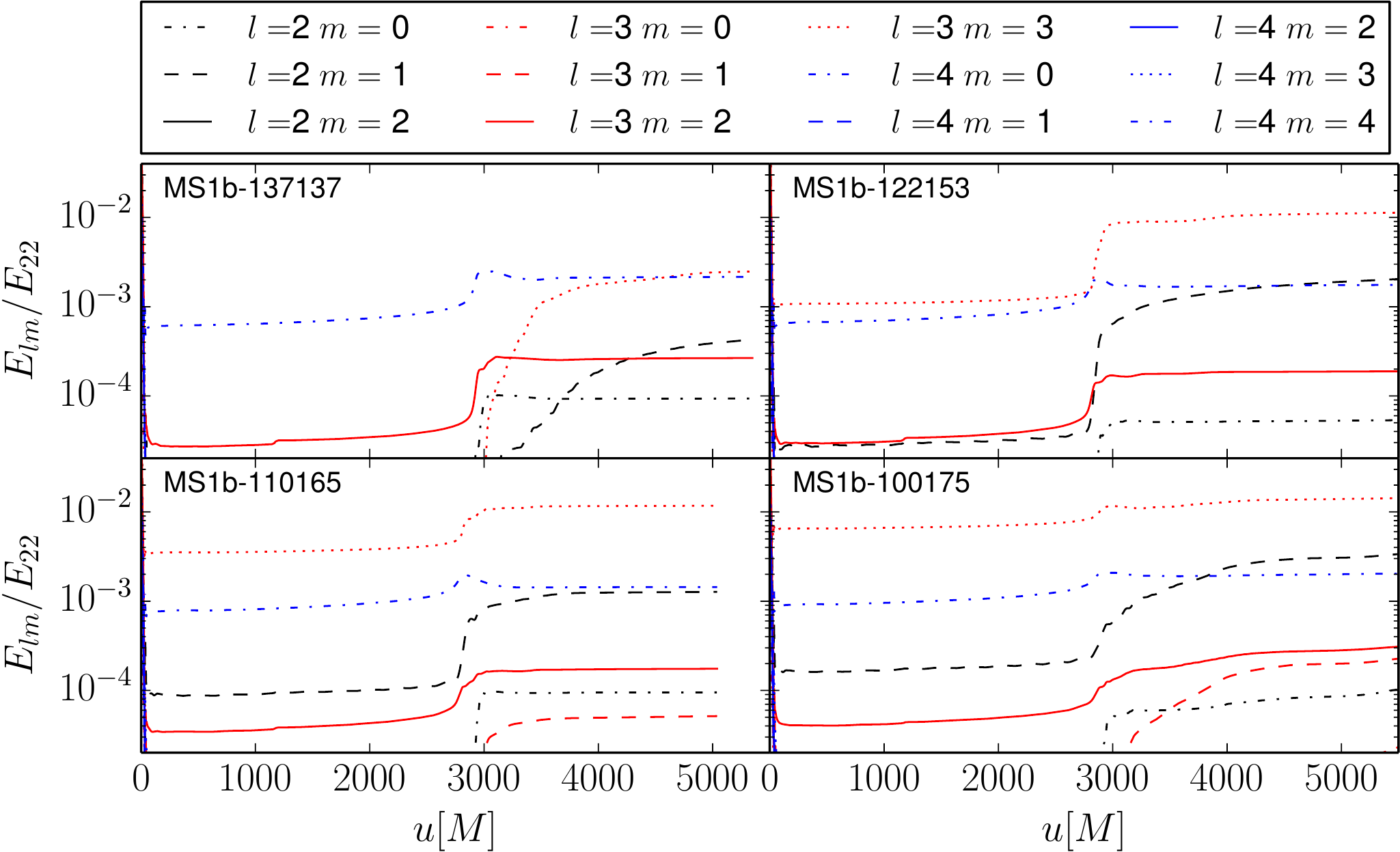}
  \caption{GW energy emitted by different modes (up to
    $l=m=4$) and rescaled by the amount of energy emitted by the (2,2)-mode. 
    Black lines correspond to $l=2$ contributions, red lines the
    $l=3$ contributions, and $l=4$ is represented by blue lines. Different
    dashing corresponds to different $m$. We present the configurations 
    MS1b-137137 (upper left panel), MS1b-122153 (upper right panel), 
    MS1b-110165 (lower left panel), MS1b-100175 (lower right panel).}
  \label{fig:Erad_modes_rescaled} 
\end{figure}   

\begin{figure}[t]
  \includegraphics[width=0.5\textwidth]{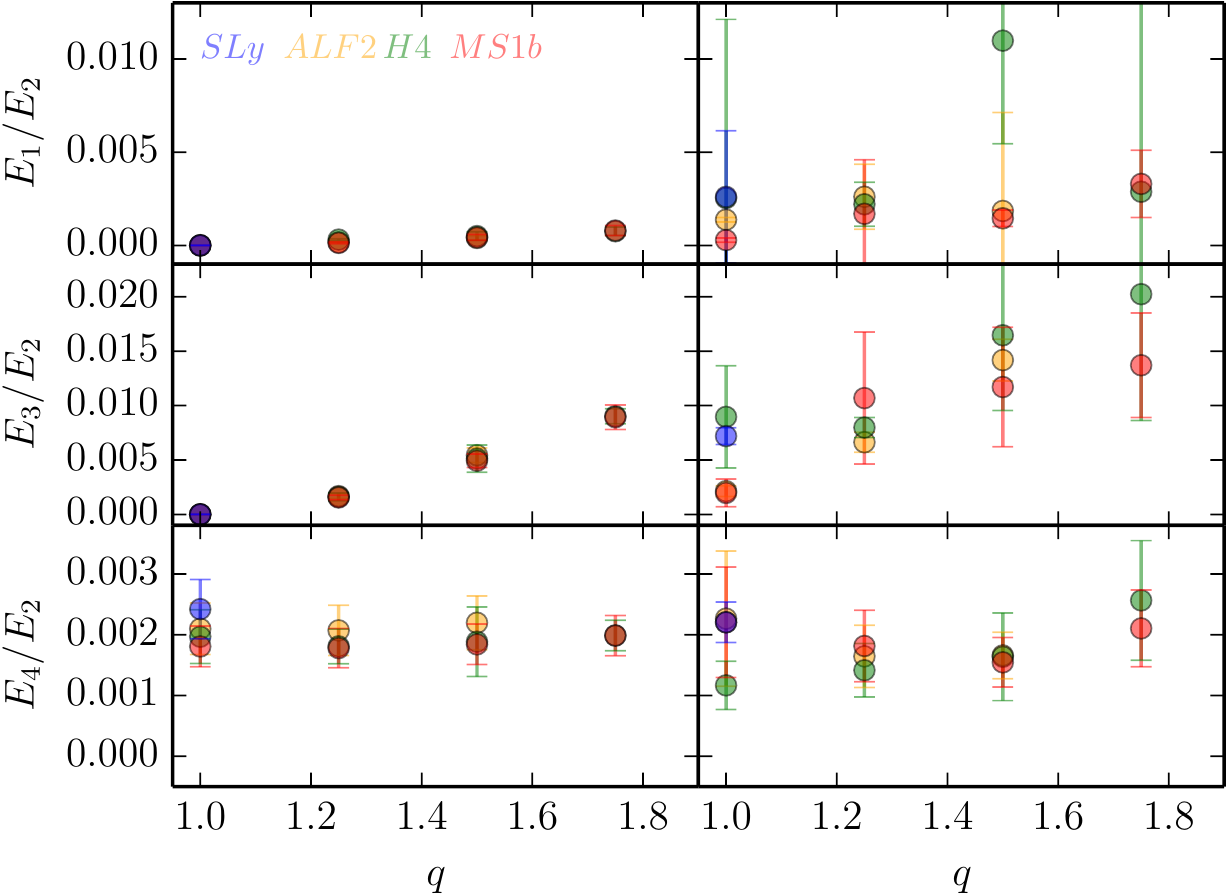}
  \caption{GW energy emitted by different modes $E_m= \sum_{l=m}^{8} E_{lm}$. 
     We show $E_m$ for $m=1;3;4$ rescaled by $E_2$. 
     Left panels show the energies at the moment of merger, i.e.~peak of the GW amplitude, 
     right panels represent data $20$ ms after merger.
     We have removed those setups for which the initial linear momentum, Tab.~\ref{tab:initPandecc}, 
     was large and an artificial drift of the center of mass was present.}
  \label{fig:Erad_modes_q} 
\end{figure}  
  
Finally, let us discuss the influence of the mass ratio on the emitted GW energy. 
The emitted energy is calculated according to Eq.~\eqref{eq:GW_Erad} and 
we compute the contributions for all modes individually. 
To the end of the simulation $\sim1$--$5\times 10^{-2} M_\odot$ is
radiated. The largest amount of energy is emitted by the (2,2)-mode, 
which emits 2-3 orders of magnitude more energy than other modes. 
The total radiated energy $E^{\rm tot}_{\rm rad}$ is
shown in Fig.~\ref{fig:Erad_modes_q_total} at the moment of merger,
i.e.~$u_{\rm mrg}$, (left panel), and $\sim 4000M_\odot$,
i.e.~$20$ms  after the merger (right panel).  
The thin dashed lines represent the emitted energy for a non-spinning BBH systems.
To estimate the emitted energy at merger, we use the effective-one-body model 
of~\cite{Damour:2012ky}. The final total energy is instead computed by
the fitting formula of~\cite{Barausse:2012qz}. 
As for BBHs, the BNS emitted energy decreases for an increasing mass ratio.  
Since NSs merge at larger separations and are less relativistic than corresponding BBH systems, 
the emitted energy at merger is smaller than the corresponding emitted energy for 
BBHs~\cite{Bernuzzi:2015opx}.
In the postmerger phase the influence of the mass ratio becomes even more prominent than
during merger. More energy is released for equal mass systems than for
unequal mass setups. We also find an imprint of the EOS, 
where in general NSs employing a soft EOS emit 
more energy than for a stiff EOS. 
Both observations can be explained in terms of the quasi-universal relations:
systems with larger $\kappa^T_2$ produce a more bound merger remant and release 
more energy~\cite{Bernuzzi:2014kca}. 

We present in Fig.~\ref{fig:Erad_modes_rescaled} the 
rescaled energies $E_{lm}/E_{22}$ for simulations with mass $M=2.75 M_\odot$ and the EOS MS1b. 
For equal mass configurations the second dominant mode during the inspiral-merger is the (4,4) mode followed by the 
(3,2)-mode, see the upper left panel of Fig.~\ref{fig:Erad_modes_rescaled}. 
All other modes do not produce a significant contribution to the total
emitted energy during this phase and contribute  only $\sim0.1\%$ to the total energy up to the merger.  
For unequal mass configurations the important subdominant modes are (3,3), (4,4), (2,1), (3,2) in 
descending order\footnote{With the exception of ALF2-100175, SLy-117175,Sly-100175, for which the 
  (2,1) is particularly large because of center of mass drift due to
  the initial residual linear momentum, see
  Tab.~\ref{tab:initPandecc}.}.  
This is in qualitative accord to the post-Newtonian theory:
the amplitudes of (3,3) and (2,1) modes are nonzero for unequal mass
configurations and they are proportional to the mass ratio at leading order,
e.g.~\cite{Blanchet:2013haa}.
In our simulations we find that for a mass ratio of $q=1.75$ around
$1\%$ of the emitted energy at merger comes from the (3,3) mode. 

In the post-merger phase several different modes are excited.
The (2,2)-mode is still dominant. At the end of our simulations the 
(3,3) mode is the second strongest followed by (2,1), (4,4), (3,2), (2,0), where 
the exact ordering depends on the mass ratio and EOS. 
We find that also for $q=1$ the (2,1) and (3,3) mode are non-zero, 
cf.~\cite{Paschalidis:2015mla,Radice:2016gym}.
Computing the luminosity $dE_{lm}(t)/dt$ (not shown in the plot) we
find that while $ {\rm d} E_{22} / {\rm d}t$ decreases
over time due to the very efficient emission in this channel
\cite{Bernuzzi:2015opx}, other modes actually increase up to the
collapse time, in particular $ {\rm d} E_{33} / {\rm d}t$ and $ {\rm d}
E_{21} / {\rm d}t$, e.g.~\cite{Radice:2016gym,Lehner:2016wjg}.  

In Fig.~\ref{fig:Erad_modes_q} we present the energy released for $E_m$ with $m=1,3,4$ divided by $E_2$. 
We see that at the merger (left panels) for an increasing mass ratio more energy is emitted for $m=1,3$
with respect to the total energy. For $m=4$ the energy mode is approximately constant and contributed $\sim0.2\%$ 
to the total emitted energy.
The clear imprint of the mass-ratio for $m=1,3$ is lost after the merger (right panels).
However, a small trend towards more energy release for unequal mass
ratios for larger mass ratios 
is still present for $m=3$. 
In addition to the imprint of the mass ratio we find 
that during the postmerger phase the amount
of energy emitted in the subdominant modes is much higher compared to the inspiral 
independent of the mass ratio. In general up to $\sim 3\%$ 
of the total released energy can be emitted in the subdominant modes.

\section{Electromagnetic counterparts}
\label{sec:EM}

\begin{table*}[t]
  \centering    
  \caption{Electromagnetic Counterparts.
  The columns refer to: the name of the configuration, 
                        the time in which the peak in the near infrared occurs $t_{\rm peak}$, 
                        the corresponding peak luminosity $L_{\rm peak}$, 
                        the temperature at this time $T_{\rm peak}$, 
                        the time of peak in the radio band $t_{\rm peak}^{\rm rad}$, 
                        and the corresponding radio fluence.
                        As in other tables, we present results for R2 and R3 resolved simulations and 
                        results for the second highest resolution for all configurations 
                        are given in brackets.}
  \begin{tabular}{l|ccccc}        
    Name    & $t_{\rm peak}$ & $L_{\rm peak}$ & $T_{\rm peak}$ & $t^{\rm rad}_{\rm peak}$ & $F^{\nu {\rm rad}}_{\rm peak}$ \\
            & [days] & [$10^{40}{\rm erg/ s}$] & [$10^3$ K] & [years] & [mJy] \\
     \hline
     \hline
ALF2-137137  & 2.0 (1.8) & 2.6 (1.9) & 2.5 (2.7) & 6.4 (6.1)   & 0.041 (0.007) \\
H4-137137    & 1.9 (0.9) & 2.8 (1.4) & 2.5 (3.3) & 5.9 (3.5)   & 0.058 (0.005) \\
MS1b-137137  & 2.0 (3.1) & 1.9 (2.5) & 2.7 (2.4) & 7.3 (10.6)  & 0.006 (0.013) \\
SLy-137137   & 4.5 (3.7) & 4.5 (4.6) & 1.9 (2.0) & 10.0 (7.9)  & 0.143 (0.203) \\
     \hline
     \hline
ALF2-122153  & 2.9 (4.2) & 3.7 (2.9) & 2.2 (2.2) & 8.0 (17.6)  & 0.139 (0.046) \\
H4-122153    & 2.7 (3.4) & 3.5 (3.5) & 2.2 (2.1) & 7.3 (9.6)   & 0.105 (0.074) \\
MS1b-122153  & 2.5 (3.0) & 2.9 (2.8) & 2.3 (2.3) & 7.2 (9.1)   & 0.046 (0.026) \\
SLy-122153   & 4.7 (4.2) & 4.7 (4.3) & 1.9 (2.0) & 12.3 (11.3) & 0.237 (0.173) \\
     \hline
     \hline
ALF2-100150  & 5.4 (5.4) & 4.5 (4.0) & 1.9 (1.9) & 12.6 (13.6) & 0.100 (0.056) \\
H4-100150    & 6.1 (6.5) & 5.1 (4.6) & 1.8 (1.8) & 14.0 (15.7) & 0.187 (0.090) \\
MS1b-100150  & 6.9 (6.9) & 5.1 (4.5) & 1.8 (1.8) & 15.9 (17.2) & 0.152 (0.072) \\
SLy-100150   & 4.5 (4.9) & 5.2 (3.8) & 1.9 (2.0) & 9.9 (13.3)  & 0.359 (0.055) \\
     \hline
     \hline
ALF2-110165  & 5.6 (4.6) & 5.0 (4.1) & 1.8 (2.0) & 12.8 (11.4) & 0.190 (0.083) \\
H4-110165    & 4.8 (5.4) & 4.3 (4.5) & 1.9 (1.9) & 11.8 (12.9) & 0.111 (0.109) \\
MS1b-110165  & 6.1 (6.0) & 5.0 (4.8) & 1.8 (1.8) & 14.1 (14.1) & 0.161 (0.131) \\
SLy-110165   & 4.3 (3.3) & 4.9 (4.0) & 1.9 (2.1) & 9.8 (8.3)   & 0.288 (0.159) \\
     \hline
     \hline
ALF2-117175  & 5.5 (4.9) & 5.3 (4.6) & 1.8 (1.9) & 12.8 (13.0) & 0.317 (0.233) \\
H4-117175    & 4.2 (4.7) & 4.2 (4.3) & 2.0 (1.9) & 10.3 (11.5) & 0.129 (0.108) \\
MS1b-117175  & 5.6 (5.8) & 5.0 (5.0) & 1.8 (1.8) & 12.8 (13.0) & 0.195 (0.170) \\
SLy-117175   & 2.4 (3.0) & 4.1 (2.8) & 2.2 (2.3) & 6.0 (11.5)  & 0.340 (0.055) \\
     \hline
     \hline
ALF2-100175  & 6.9 (8.0) & 5.9 (6.6) & 1.7 (1.6) & 14.7 (16.0) & 0.340 (0.460) \\
H4-100175    & 7.4 (7.7) & 5.9 (5.7) & 1.7 (1.7) & 16.1 (16.8) & 0.293 (0.234) \\
MS1b-100175  & 8.2 (8.7) & 6.3 (6.5) & 1.6 (1.6) & 17.1 (18.0) & 0.361 (0.386) \\
SLy-100175   & 5.5 (6.3) & 5.2 (5.0) & 1.8 (1.8) & 12.5 (15.6) & 0.254 (0.177) \\
     \hline
     \hline
MS1b-094194  & 9.4 (9.5) & 7.0 (7.2) & 1.5 (1.5) & 18.8 (18.5) & 0.489 (0.581) \\
     \hline
     \hline
  \end{tabular}
 \label{tab:EM}
 \end{table*}

Since we do not simulate the evolution of the electron fraction and of
the internal composition of the fluid, we rely on simplified models
for estimating the EM luminosity, fluxes and light curves. 
We use in particular the analytical model of \cite{Grossman:2013lqa} which 
assumes that the diffusion time is less than the dynamical times 
for computing the peak luminosity and
temperature. Light-curves are computed with the model described in \cite{Kawaguchi:2016ana},
originally introduced for the merger of BHNS systems.
Radio flares peak fluxes are instead computed following \cite{Nakar:2011cw}. 

Simulations including microphysics have been presented 
e.g.~in~\cite{Sekiguchi:2016bjd,Lehner:2016lxy,Palenzuela:2015dqa,Radice:2016dwd,Just:2014fka,Wanajo:2014wha,Radice:2016dwd}. Although more detailed
in term of simulated physics, none of these work have explored the effect of large $q$ as we do here.

\subsection{Macronovae} 
 
\begin{figure}[t]
  \includegraphics[width=0.5\textwidth]{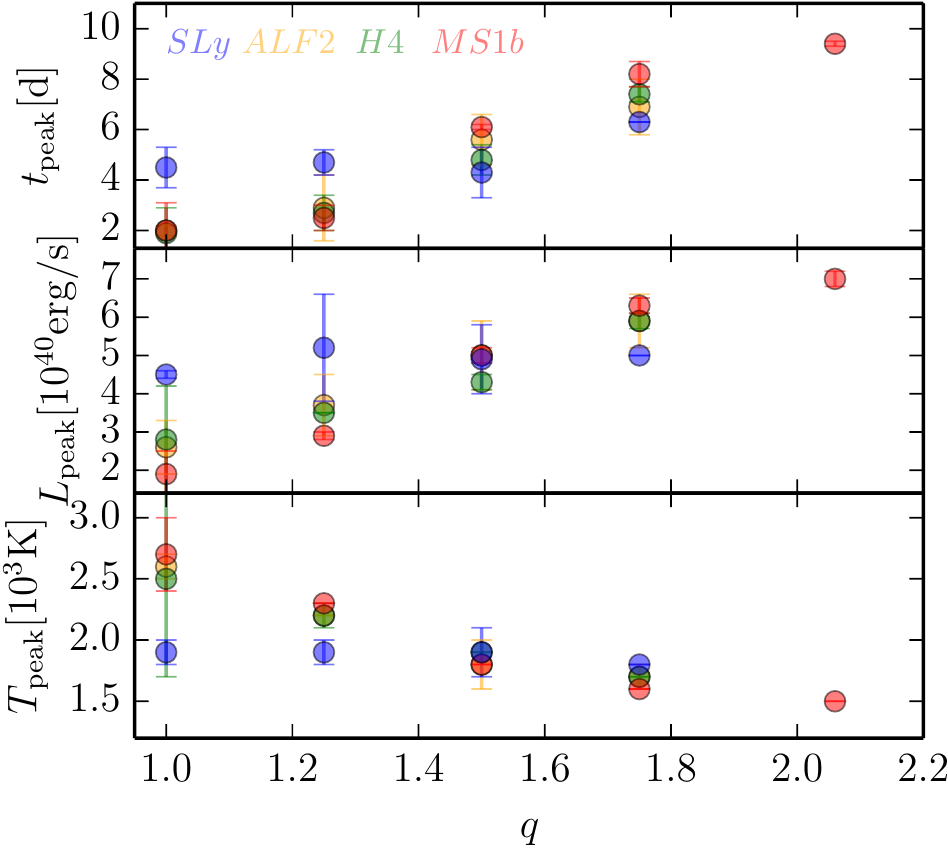}
  \caption{Peak time $t_{\rm peak}$ (top panel), peak luminosity $L_{\rm peak}$ (middle panel), 
           and peak temperature $T_{\rm peak}$ (bottom panel) of marcronovae produced 
           by the BNS mergers considered in this article as a function of the mass ratio. 
           As before we consider configurations with $M=2.75M_\odot$ and $M=2.888M_\odot$.}
  \label{fig:Lpeak(q)}
  \end{figure}  
 
   \begin{figure}[t]
  \includegraphics[width=0.5\textwidth]{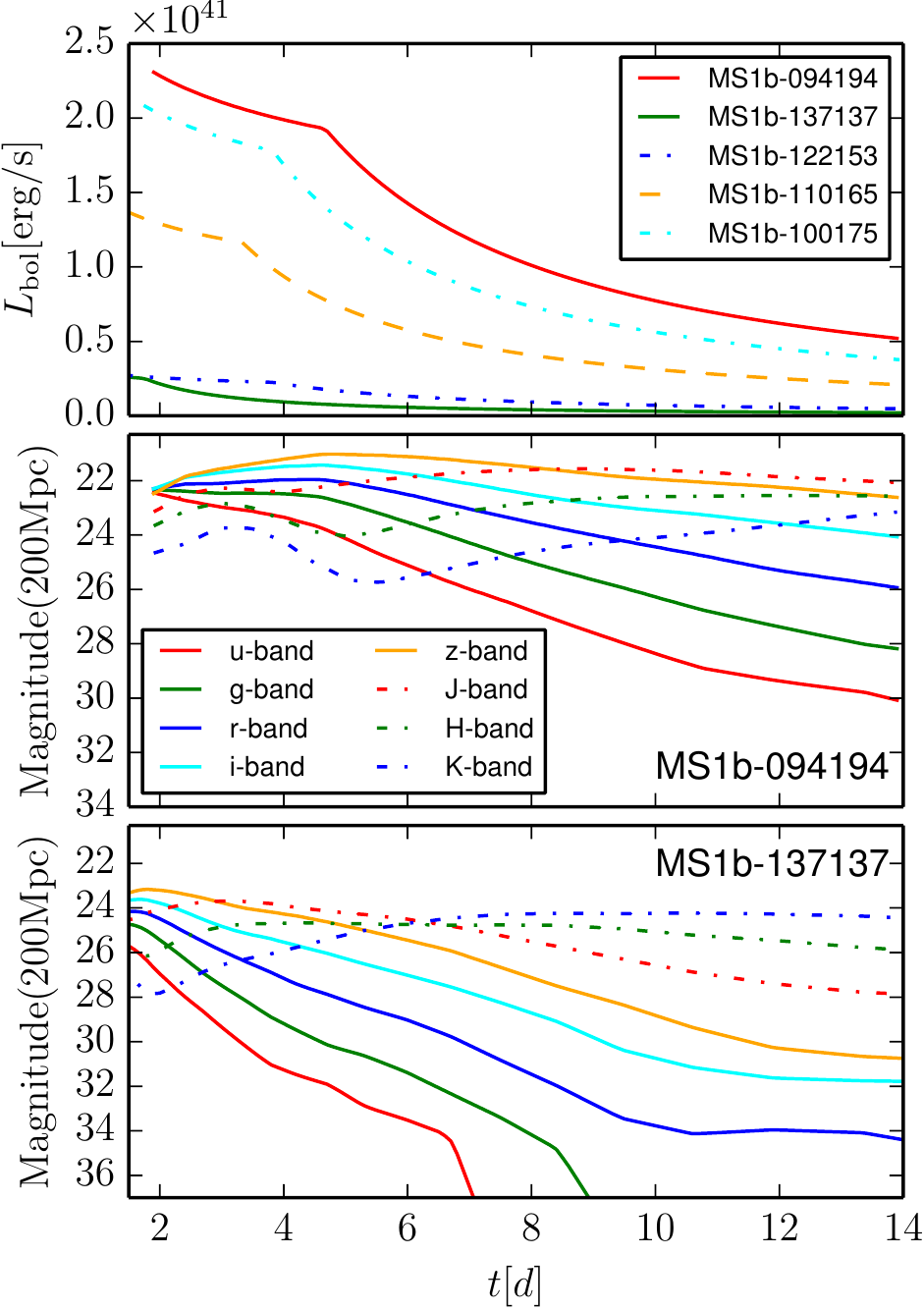}
  \caption{Top panel: Predicted lightcurve (absolute bolometric luminosity) for the 
           predicted macronova produced by the merger of the configurations employing the MS1b EOS. 
           The lightcurve is produced with the publicly available code of~\cite{Kawaguchi:2016ana}.
           Middle and bottom panel: absolute magnitudes for the ugrizJHK-band assuming that the macronova 
           is produced 200\Mpc\ away for MS1b-094194 (middle panel) and MS1b-137137 (bottom panel).}
  \label{fig:lightcurve} 
\end{figure}   

Following~\cite{Grossman:2013lqa}, we can estimate the 
time $t_{\rm peak}$ at which the peak in the near-infrared occurs,
the bolometric luminosity at this time $L_{\rm peak}$, and the corresponding temperature
$T_{\rm peak}$:
\begin{align}
   t_{\rm peak} & = 4.9 \ {\rm d} \nonumber \\ 
                & \times \left( \frac{M_{ej}}{10^{-2} M_\odot} \right)^{\frac{1}{2}} 
                      \left( \frac{\kappa}{10 {\rm cm^2 g^{-1}} } \right)^{\frac{1}{2}}
                      \left( \frac{v_{\rm ej}}{0.1} \right)^{-\frac{1}{2}} , \label{eq:tpeak} \\
   L_{\rm peak}&  = 2.5 \  10^{40} {\rm erg \, s^{-1}} \nonumber  \\
               & \times       \left( \frac{M_{ej}}{10^{-2} M_\odot} \right)^{1-\frac{\alpha}{2}} 
                      \left( \frac{\kappa}{10 {\rm cm^2 g^{-1}} } \right)^{-\frac{\alpha}{2}}
                      \left( \frac{v_{\rm ej}}{0.1} \right)^{\frac{\alpha}{2}} , \label{eq:Lpeak} \\
   T_{\rm peak} & = 2200 {\rm K}   \ \nonumber  \\
                & \times \left( \frac{M_{ej}}{10^{-2} M_\odot} \right)^{-\frac{\alpha}{8}} 
                      \left( \frac{\kappa}{10 {\rm cm^2 g^{-1}} } \right)^{-\frac{\alpha+2}{8}}
                      \left( \frac{v_{\rm ej}}{0.1} \right)^{\frac{\alpha-2}{8}} . \label{eq:Kpeak} 
\end{align}
Ref.~\cite{Grossman:2013lqa} assumes that the energy release due to
the radioactive decay is proportional to $\sim t^{-\alpha}$ with
$\alpha=1.3$. Furthermore, we set the average opacity to $\kappa =
10~{\rm cm^2 g^{-1}}$ as in~\cite{Grossman:2013lqa}.

We summarize the time of the peak, the corresponding luminosity and
temperature for our configurations in Tab.~\ref{tab:EM}. 
Results are also shown in Fig.~\ref{fig:Lpeak(q)}.
We find that for an increasing mass ratio the peak time increases from
$\sim2$ days up to $\sim 10$ days. This effect is larger for stiffer EOSs.
Also the peak luminosity increases with an increasing mass
ratio. Except for the SLy setup the peak luminosity scales almost
linear to the mass ratio. Finally (lower panel of
Fig.~\ref{fig:Lpeak(q)}) we see that the peak temperature decreases
from $\sim2500$K  to $\sim1500$K when the mass ratio increases, except
for SLy for which the temperature for all mass ratios is around $\sim1700$K.

Light curves are computed following \cite{Kawaguchi:2016ana}.
The model assumes homologous expansion of the ejecta, a gray opacity,
diffusion approximation for the radiation transfer and that the
photons diffuse only from the latitudinal edge.  
It was originally developed for the EM radiation
produced during the merger of a BHNS system. The main difference
between the ejecta of BNS and BHNS mergers is that during BHNS mergers
it is unlikely that shocks form. However, we have shown in
Fig.~\ref{fig:2d_entropy_MS1b-094194} that the entropy of the ejected
material for MS1b-094194 is relatively small. Similar statements hold
for other configurations employing stiff EOSs. For this reason we
expect that the model of~\cite{Kawaguchi:2016ana} is
appropriate for the MS1b as well as other stiff configurations and 
systems without shocks. 

We use the publicly available program of~\cite{Kawaguchi_web}.
Input parameters are taken from Tab.~\ref{tab:ejecta}. The latitude opening angles are estimated 
by evaluation of Eq.~(19) and Eq.~(20) of~\cite{Dietrich:2015iva}. 
Furthermore, we use as longitudinal opening angle $\pi$(rad), 
$\kappa = 10 {\rm cm^2 g^{-1}}$ for the opacity, 
$\dot{\epsilon} = 1.58 \times 10^{10}$erg/g/s for the heating rate coefficient, 
$\alpha = 1.2$ for the heating rate power and  
$\epsilon_{th}=0.5$ for the thermalization 
efficiency as in~\cite{Kawaguchi:2016ana,Kawaguchi_web}. 

Figure~\ref{fig:lightcurve} presents our results. 
We find that the considered configurations will have 
a luminosity between $10^{39}-10^{42}$erg/s (upper panel).
Since the luminosity strongly correlates to the mass of the ejecta,
we observe that for an increasing mass ratio the luminosity 
increases for more than one order of magnitude for our configurations.
Because of the increasing luminosity also the observed magnitude of the macronova 
at a hypothetical distance of $200\Mpc$ is larger for MS1b-094194 (middle and bottom panel). 
But also for MS1b-137137 counterparts~\footnote{Notice that the model of~\cite{Kawaguchi:2016ana}
was tested for ejecta masses above $0.01M_\odot$ and that higher uncertainties and errors 
might be present for low mass ejecta.} 
are observable for the first days after merger with 8-m
class telescopes assuming that magnitudes between 26 and 27 are
detectable. Contrary, for high mass ratio configurations like
MS1b-094194 macronovae can be observed for several days up to 
weeks after the merger. 
Thus, the mass ratio has a clear observational imprint on the EM
counterparts.  

\subsection{Radio Flares}

\begin{figure}[t]
  \includegraphics[width=0.5\textwidth]{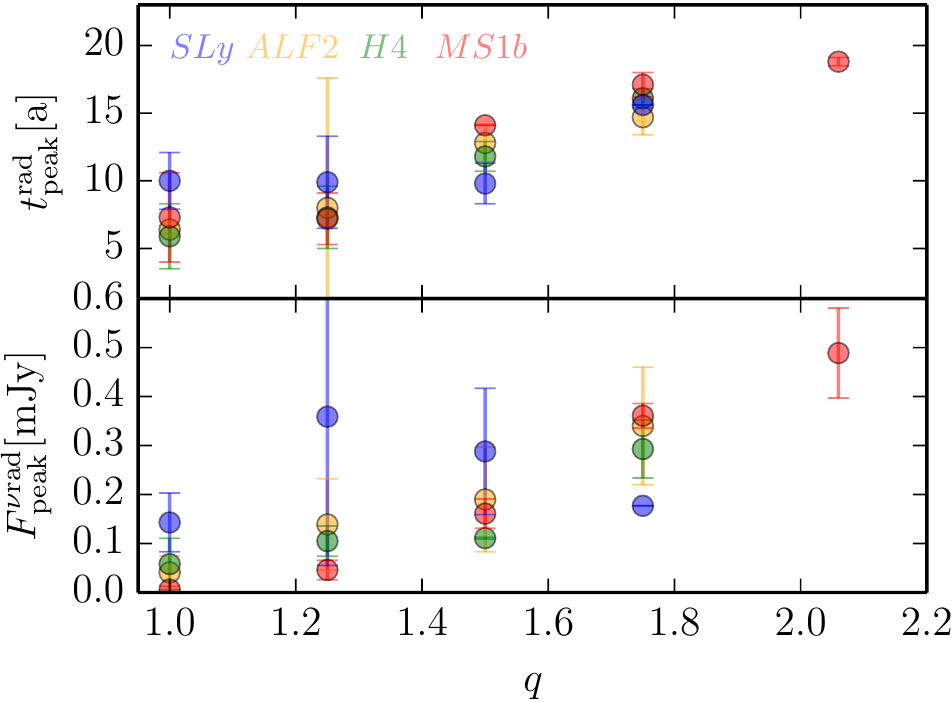}
  \caption{Peak time in the radio band $t_{\rm peak}^{\rm rad}$ (top panel) and corresponding 
           radio fluence $F^{\nu {\rm rad}}_{\rm peak}$ (bottom panel), 
           as a function of the mass ratio. We consider BNS configurations 
           with $M=2.75M_\odot$ and $M=2.888M_\odot$.}
  \label{fig:Rad2(q)}
  \end{figure} 

In order to estimate the radio emission, we use the model of~\cite{Nakar:2011cw} 
in which the strongest signal is expected at a time 
\begin{align}
   t_{\rm peak}^{\rm rad} & = 1392 \ {\rm d} \nonumber \\ 
                  & \times \left( \frac{T_{\rm ej} }{10^{49}{\rm erg}} \right)^{\frac{1}{3}} 
                  \left( \frac{n_0}{\rm cm^{-3}}\right)^{-\frac{1}{3}} 
                  \left( \frac{v_{\rm ej}}{0.1} \right)^{-\frac{5}{3}} 
\end{align}
with a radio fluence of 
\begin{align}
   {F^\nu}^{\rm rad}_{\rm peak} & = 0.3 \ {\rm mJy} \times
                  \left( \frac{T_{\rm ej} }{10^{49}{\rm erg}} \right)
                  \left( \frac{n_0}{\rm cm^{-3}}\right)^{\frac{p+1}{4}} 
                  \left( \frac{\epsilon_B}{\rm 0.1}\right)^{\frac{p+1}{4}} \nonumber \\
                  & \times \left( \frac{\epsilon_e}{\rm 0.1}\right)^{p-1}                  
                  \left( \frac{v_{\rm ej}}{1} \right)^{\frac{5p-7}{2}}
                  \left( \frac{D}{10^{27} {\rm cm}} \right)^{-2} \nonumber \\
                  & \times \left( \frac{\nu_{\rm obs}}{1.4 {\rm GHz}} \right)^{-\frac{p-1}{2}}
\end{align}   
for an observation frequency $\nu_{\rm obs}$, which is expected to be higher than the 
self-absorption and synchrotron peak frequency at a distance $D$.
The parameters $\epsilon_B$ and $\epsilon_e$ determine how efficiently the
energy of the blast wave is transfered to the magnetic field and to
electrons. We set both parameters to $\epsilon_B=\epsilon_e=0.1$ in our analysis as in~\cite{Nakar:2011cw}.
The variable $n_0$ denotes the surrounding particle density and is set to $0.1 {\rm
cm^{-3}}$. Following~\cite{Nakar:2011cw} we additionally set $p=2.3$ and
$\nu_{\rm obs}=1.4 {\rm GHz}$. 

We report our results for $t_{\rm peak}^{\rm rad}$ and ${F^\nu}^{\rm
  rad}_{\rm peak}$ in Tab.~\ref{tab:EM}  and in Fig.~\ref{fig:Rad2(q)}.
As for $t_{\rm peak}$, $t_{\rm peak}^{\rm rad}$ increases with an
increasing mass ratio. The peak time $t_{\rm peak}^{\rm rad}$ is in
the range of a few years up to two decades for very high mass
ratios. The corresponding radio fluence is characterized by large
uncertainties, however, for stiff EOS an almost linear growth for an
increasing mass ratio is observed.

\section{Summary}
\label{sec:conclusion}

In this article we studied the effect of the mass-ratio by a large set
of new numerical relativity simulations spanning, for the first time,
up to values $q\sim2$. Our findings are summarized in what follows.  \\

\paragraph*{Mass transfer:}
A resolution study of simulations with $q=2$ showed that mass transfer
during the last orbits is very dependent on the grid
resolution. In particular, by increasing resolution the amount of
transfered mass decreases. We conclude that no significant mass  
transfer happens during the merger of BNS, even in cases with large
mass-ratio. \\

\paragraph*{Mass ejection:}
Mass ejection in large $q$ systems is primary due to a centrifugal effect
and originates from the companion's tidal tail (or partial disruption)
during the late inspiral and merger. Ejecta components due to
shock-driven ejecta are only dominant for configurations with a soft
EOS and rather independent of the mass ratio. 
We showed, for the first time in the context of BNS, that the
dependence of the ejecta mass and kinetic energy is essentially linear
on the mass-ratio $q$ for stiff EOSs, see Fig.~\ref{fig:M_ej(q)}.
Also the velocity of the ejecta depend significantly on the mass
ratio. In particular, the component perpendicular to the orbital plane
decreases for increasing $q$, since torque becomes the dominant ejecta mechanism. 
For large $q$ the ejecta are almost entirely about the orbital plane.
Overall, these ejecta properties for large $q$ resamble those of black
hole - neutron star binaries and lead to characteristic features of
electromagnetic counterparts (see below).
Finally, the total mass of the configuration plays a minor role and is 
less important than the EOS or the mass-ratio. \\

\paragraph*{Merger remnant:}
The lifetime of the merger remnant depends strongly on the EOS, 
in most cases softer EOSs lead to an earlier collapse. The mass-ratio
is a secondary effect, but larger $q$ lead to delay collapse. 
We also showed that in most cases for which a black hole forms the rest
mass of the accretion disk increases. In cases of a prompt collapse 
no massive accretion disk forms. \\

\paragraph*{Gravitational Waves:}
Varying the mass ratio leads to quantitative changes to the GW
frequency, and to qualitative changes of the postmerger spectra.
The GW merger frequency is in general largest for equal masses and 
decreases for increasing $q$. For MS1b $f_{\rm mrg}$ decreases from
1.45~kHz to 0.09~kHz when $q$ goes from $1$ to $1.75$ respectively. 
No significant effect are instead on the postmerger frequency $f_2$.
We believe the latter is due to the limited accuracy the peak
frequency can be extracted, ultimately related to the broad character
of the spectra peak.
However, we find that for unequal masses the characteristic 
secondary peaks $f_s$ in the spectrum tend to disappear for large $q$,
and they are actually absent for high mass ratio systems, 
see Fig.~\ref{fig:hspectra}. 

Our spectrograms highlight the rich structure of the multipolar GW
waveform. Modes with azimuthal number $m=1,3$, in particular, become
progressively more relevant for larger $q$, although it is unlikely
their effect will be observed in next LIGO/Virgo observations,
e.g. \cite{Radice:2016gym}. 
Furthermore, for configurations producing a MNS merger remnant, the
spectrograms show that the postmerger frequencies increase over time
in a chirp-like fashion as the merger remnant 
becomes more compact. This implies that the postmerger spectrum is 
in fact continuous and not discrete as anticipated in
\cite{Bernuzzi:2015rla}.

The total emitted GW energy is a decreasing function of $q$ during
both the inspiral and the post-merger phase.  
This qualitative behavior is already known from binary black hole systems, but 
here we extend the result to BNS. This is nontrivial since for BNS
tidal interactions play an important role during inspiral-merger and
the postmerger phase has different physics from black hole binaries.
We find that neutron stars with softer EOS emit more energy. 
In addition to the total energy, also the mode hierarchy changes by 
varying $q$. We find that the energy in the $m=3$ (and $m=1$) emission channel
increase for larger $q$ and contribute up to $1\%$ to the total emitted 
energy up to merger. \\

\paragraph*{Electromagnetic counterparts:}
A GW detection of BNSs will trigger observations to capture
follow-up electromagnetic emissions. We used simplified models to
estimate the luminosity,  
peak time, and light curves of macronovae counterparts and the peak
time and fluence of radio flares. 
We showed that the peak luminosity, peak time, and persistency of these
counterparts are strongly dependent on the mass ratio $q$. Unequal mass
BNS systems are more luminous in the EM, than equal mass systems 
because of more massive ejecta. The larger the mass ratio, the more delayed 
is the luminosity peak. Also our estimated macronova lightcurves are 
more persistent for larger mass ratios; they could be detected up
to a few weeks, see Fig. \ref{fig:lightcurve}. 
Similarly to black hole neutron star mergers and differently from
$q\sim1$ BNS, the dynamical ejecta of large $q$ BNS are confined to
the equatorial plane and will not obscure optical emissions from the
disk wind~\cite{Kasen:2014toa}. Thus, the latter might be detectable
for face-on binaries \footnote{We thank D.Radice for pointing this out.}.
\\ 

Numerical uncertainties have been carefully evaluated on multiple quantities,
see Appendix~\ref{sec:accuracy}, and we are confident on the presented
results. However, our simulations do not account for microphysics as
done in other works 
e.g.~\cite{Just:2014fka,Bernuzzi:2015opx,Palenzuela:2015dqa,Lehner:2016wjg,Lehner:2016lxy,Sekiguchi:2016bjd,Radice:2016dwd}. 
The simplified assumptions in our work have been necessary to simulated a large number of
BNS configurations, and in order to better control the numerical
errors. We believe our results hold at least at a qualitative level;
dedicated simulations of selected configurations that include a more
sophisticated treatment of microphysics should be performed in the future  
to validate our predictions.

\appendix

\section{Accuracy}  
\label{sec:accuracy}

Our simulations are pushed in a region of 
the BNS parameter space, which received little or no attention in the past. 
Thus, we present some diagnostics for the accuracy of our results considering the 
initial configurations and the dynamical simulations. 
However, we refer the reader to~\cite{Tichy:2009yr,Dietrich:2015pxa} for a more detailed discussion 
about the convergence and accuracy of SGRID 
and~\cite{Bernuzzi:2011aq,Dietrich:2015iva,Bernuzzi:2016pie} for the accuracy of BAM. 

 \subsubsection{Initial configurations}

\begin{table*}[t]
  \centering    
  \caption{Eccentricity and initial linear momentum of the configurations. 
           Columns: Eccentricity computed from the proper density according 
           to~\cite{Dietrich:2015pxa} (result stated for the highest resolution and 
           in brackets for the second highest resolution),
           initial x,y,z-component of the linear momentum measured 
           by SGRID.}
  \begin{tabular}{l|cccc}        
    Name & $|\hat{e}_d [10^{-3}]|$ & $P^x_{\rm ADM}$ & $P^y_{\rm ADM}$ & $P^z_{\rm ADM}$\\
     \hline
     ALF2-137137  & 6.8 (6.3) & \PE{3.5}{6}   & \PE{-4.4}{10} & \PE{1.8}{9}  \\
     H4-137137    & 13 (13)   & \PE{-9.5}{7}  & \PE{-7.4}{11} & \PE{3.8}{9}  \\
     MS1b-137137  & 3.9 (3.9) & \PE{-1.7}{7}  & \PE{4.0}{10}  & \PE{2.0}{9}  \\
     SLy-137137   & 15 (15)   & \PE{5.8}{8}   & \PE{-2.3}{10} & \PE{1.3}{9}  \\
     \hline
     \hline
     ALF2-122153  & 10 (10)   & \PE{1.2}{6}   & \PE{-1.9}{4}  & \PE{5.5}{8}  \\
     H4-122153    & 6.6 (6.7) & \PE{-1.9}{7}  & \PE{6.6}{4}   & \PE{2.4}{7}  \\
     MS1b-122153  & 8.9 (9.0) & \PE{-2.2}{7}  & \PE{-9.0}{6}  & \PE{2.5}{6}  \\
     SLy-122153   & 8.2 (8.3) & \PE{3.9}{7}   & \PE{-2.5}{3}  & \PE{3.3}{8}  \\
     \hline
     \hline
     ALF2-100150  & 4.4 (4.8) & \PE{-4.3}{6}  & \PE{2.5}{4}   & \PE{1.9}{7}  \\
     H4-100150    & 12 (8.9)  & \PE{-2.5}{8}  & \PE{1.3}{4}   & \PE{5.7}{7}  \\
     MS1b-100150  & 16 (16)   & \PE{9.3}{7}   & \PE{-4.0}{5}  & \PE{3.5}{8}  \\
     SLy-100150   & 12 (12)   & \PE{1.9}{7}   & \PE{1.7}{4}   & \PE{-8.6}{8} \\
     \hline
     ALF2-110165  & 23 (22)   & \PE{2.2}{7}   & \PE{1.6}{4}   & \PE{-5.9}{8} \\
     H4-110165    & 3.3 (3.5) & \PE{4.2}{7}   & \PE{-4.9}{4}  & \PE{1.1}{6}  \\
     MS1b-110165  & 11 (14)   & \PE{1.2}{6}   & \PE{1.3}{4}   & \PE{1.4}{6}  \\
     SLy-110165   & 8.1 (8.3) & \PE{-2.1}{6}  & \PE{1.2}{3}   & \PE{9.8}{8}  \\
     \hline
     ALF2-117175  & 3.9 (4.1) & \PE{-7.5}{6}  & \PE{-1.4}{3}  & \PE{2.5}{8}  \\
     H4-117175    & 3.7 (3.8) & \PE{-3.5}{7}  & \PE{6.7}{7}   & \PE{9.3}{5}  \\
     MS1b-117175  & 6.2 (6.4) & \PE{-1.4}{7}  & \PE{1.7}{4}   & \PE{2.6}{8}  \\
     SLy-117175   & 4.1 (4.5) & \PE{-1.8}{7}  & \PE{-4.6}{5}  & \PE{1.0}{8}  \\
     \hline     
     \hline
     ALF2-100175  & 1.4 (1.6) & \PE{-9.5}{6}  & \PE{-3.7}{3}  & \PE{3.5}{6}  \\
     H4-100175    & 4.4 (4.6) & \PE{6.6}{7}   & \PE{1.3}{4}   & \PE{1.9}{6}  \\
     MS1b-100175  & 8.2 (8.4) & \PE{1.6}{7}   & \PE{3.5}{4}   & \PE{1.1}{7}  \\
     SLy-100175   & 6.7 (8.2) & \PE{-2.5}{7}  & \PE{3.0}{5}   & \PE{-6.1}{10}\\   
     \hline
     \hline
     MS1b-094194  & 3.4 (3.4) & \PE{4.2}{7}   & \PE{-1.1}{5}  & \PE{-2.0}{6} \\
     \hline
     \hline
  \end{tabular}
 \label{tab:initPandecc}
\end{table*} 
 
We start with giving the initial linear ADM-momentum and 
the eccentricity of our configurations in Tab.~\ref{tab:initPandecc}. 
We have not performed any eccentricity reduction. Thus the final 
eccentricities lie in a range of $10^{-3}$ to $10^{-2}$. 
The initial linear momentum of the system (Tab.~\ref{tab:initPandecc}) is in
most cases of the order of $10^{-5}$ to $10^{-4}$. For a few cases, those which
employ a large mass ratio and use a soft EOS, the linear momentum is rather
large.
The large linear momentum leads to
a movement of the center of mass and therefore results obtained from those
systems possess a larger error/uncertainty.

As already explained in~\cite{Dietrich:2015pxa}, it is difficult to construct
initial data for very compact stars. In this case the elliptic solver 
cannot
always solve the equation for the conformal factor $\psi$ within the main
iteration, because the values from the previous iterations are not good
enough as an initial guess for the Newton-Raphson scheme we use.
In~\cite{Dietrich:2015pxa} we could circumvent this issue by skipping the
elliptic solve for $\psi$ when this problem occurs. The overall iteration
still succeeds if we do not skip this elliptic solve too often. However, for
e.g. SLy-100175 this skip occurs so often that we cannot find a good
solution for $\psi$ at all. In this case we observe spurious oscillations in
$\psi$ near spatial infinity. We were able to cure this problem by changing
the boundary condition for $\psi$. Normally the boundary condition for
$\psi$ is simply
\begin{equation}
\lim_{r\to\infty}\psi = 1 .
\end{equation}
When we encounter the problems mentioned above we change this condition to
\begin{equation}
\psi = 1 \ \mbox{if} \ r>10000M ,
\end{equation}
so that $\psi=1$ for all grid points that are further away than $10000M$.
We do the same for all other elliptic fields.
While this leads to a less accurate boundary conditions, it avoids spurious
oscillations in $\psi$ near spatial infinity, and it allows the iteration to
succeed. We apply these modified boundary conditions only for
configurations where the above mentioned problems occur.

  \subsubsection{Mass conservation}

  \begin{figure}[t]
  \includegraphics[width=0.5\textwidth]{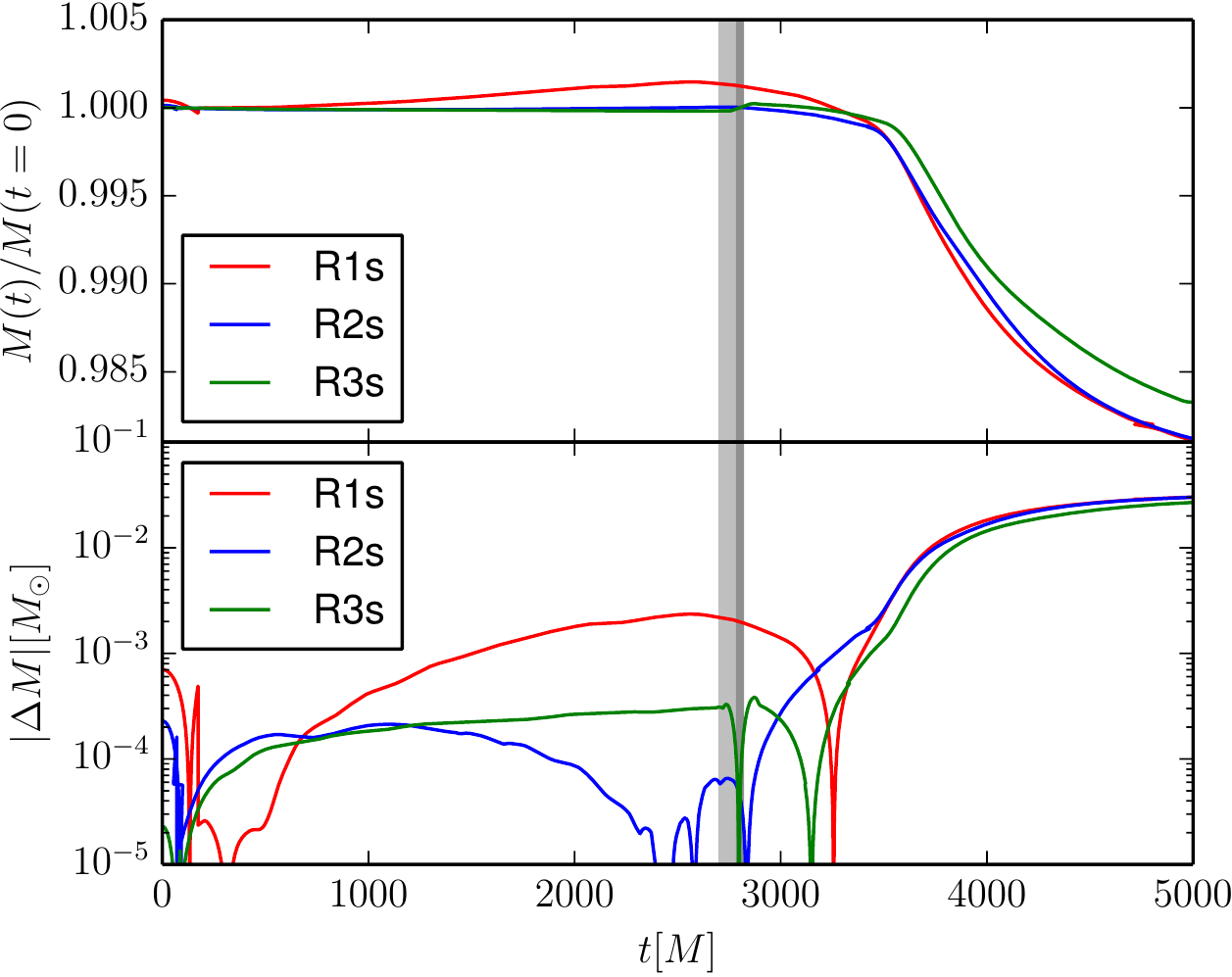}
  \caption{Mass conservation of MS1b-094194. Top panel shows the mass evolution of the baryonic mass 
  rescaled to its initial value. Bottom panel shows $\Delta M= M(t)-M(t=0)$. We mark the moment of merger 
  as gray shaded regions, where the light gray region marks the interval $[t_{R1s}^{mrg},t_{R2s}^{mrg}]$, 
  and where the dark shaded region is $[t_{R2s}^{mrg},t_{R3s}^{mrg}]$.
  The error in the total mass stays below $0.2\%$ even for the lowest resolution until the time where 
  material leaves the computational domain as ejecta.}
  \label{fig:D_convergence_MS1b-094194}
  \end{figure}  
  
  We are not only interested in the correctness and accuracy of our initial data,  
  of big importance is the performance during the dynamical evolutions. 
  We present here some diagnostics variables for the most 
  extreme case for which we also have three different resolutions: MS1b-094194. 
  We will start with discussing the mass conservation. 
  
  In Ref.~\cite{Dietrich:2015iva} it was shown in detail that an additional correction step during the 
  Berger-Oliger~\cite{Berger:1984zza} or Berger-Collela~\cite{Berger:1989} time stepping 
  improves the mass conservation significantly (up to several orders in some test cases). 
  We do not want to repeat the analysis here and refer 
  the reader to~\cite{Dietrich:2015iva,Bernuzzi:2016pie} for a detailed discussion. 
  Nevertheless, we want to present mass conservation for at least one of our simulations. 
  Figure~\ref{fig:D_convergence_MS1b-094194} shows the mass conservation for MS1b-094194. 
  Clearly visible is that up to the moment of merger
  mass conservation improves for increasing resolution~\footnote{
  We present the merger as a time interval. Due to 
  different numerical dissipation for different resolutions the merger happens at different times. 
  The light gray area corresponds to the interval $[t_{R1s}^{mrg},t_{R2s}^{mrg}]$, where 
  the dark region corresponds to $[t_{R2s}^{mrg},t_{R3s}^{mrg}]$.}.
  Also after the merger no considerable mass loss is present 
  and the difference to the initial mass stays even for the lowest resolution below $0.2\%$.
  At later times $\sim3500M$ material is leaving the 
  computational domain where the hydrodynamical variables are evolved
  and therefore the total baryonic mass is decreasing for all resolutions. 
  This feature is present for most of our simulations using spherical shells.
  The situation changes for simulations without shells, since the 
  GRHD equations are solved for those simulations also on level $l=0$.  
  
  \subsubsection{ADM Constraints}
  
  \begin{figure}[t]
  \includegraphics[width=0.5\textwidth]{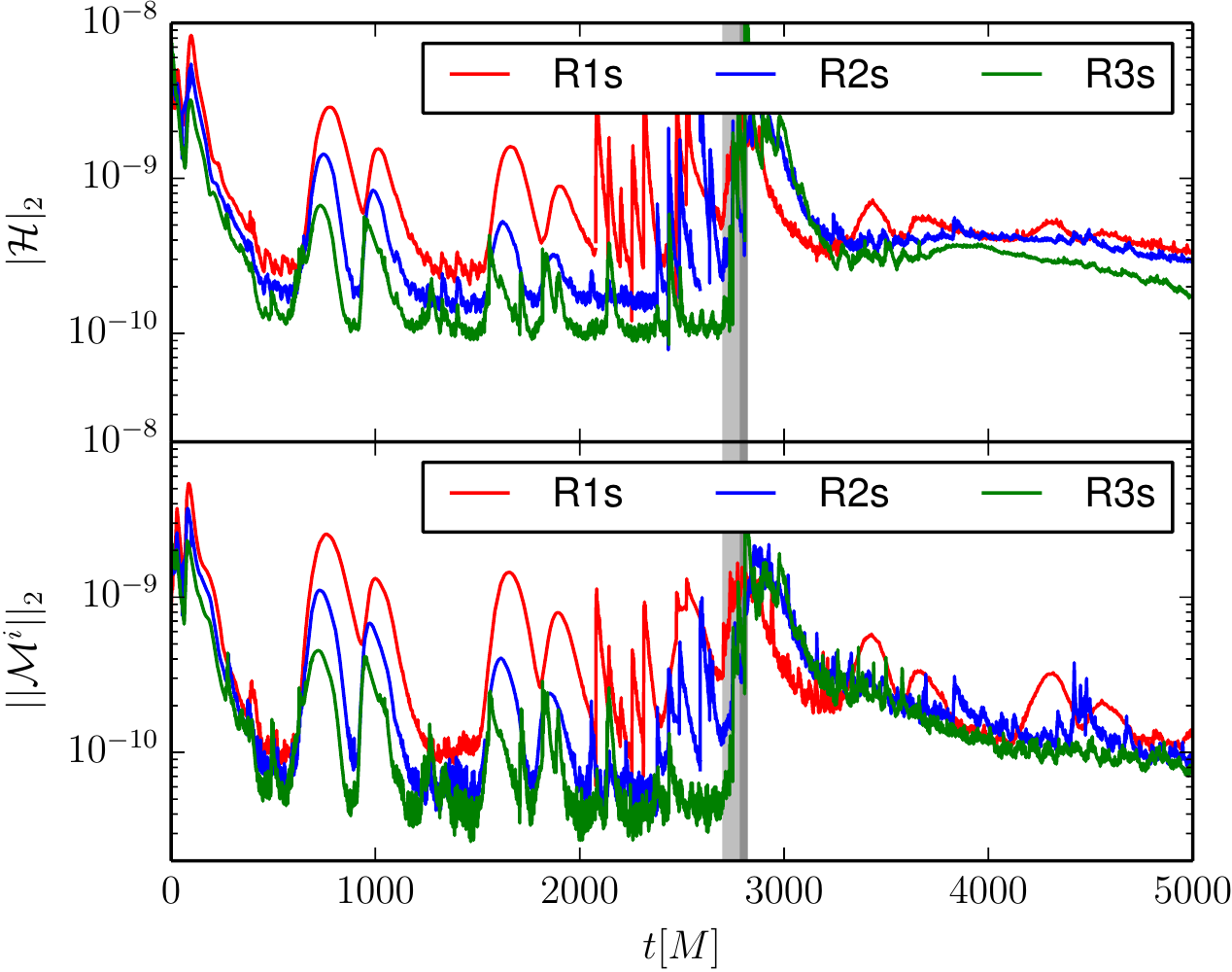}
  \caption{ADM Constraints for MS1b-094194. 
  Top panel shows the $L_2$-volume norm of the Hamiltonian constraint $|\mathcal{H}|_2$. 
  The lower panel shows the square magnitude of the momentum constraint 
  $||\mathcal{M}^i||_2 = \sqrt{|\mathcal{M}^x|_2^2+ |\mathcal{M}^y|_2^2+ |\mathcal{M}^z|_2^2}$.
  The constraints are evaluated on level $l=1$. 
  The constraints are decreasing for increasing resolution during the inspiral of the neutron stars. 
  For the whole simulation the constraints stay below or 
  at the constraint violation of the initial slice due to 
  the constraint damping of the Z4c scheme. 
  We have filtered our data with an average filter to allow a better 
  visualization and reduce high frequency noise.
  }
  \label{fig:Constraints_MS1b-094194}
  \end{figure} 
  
  Since we use the 3+1 decomposition, 
  see~\cite{Alcubierre:2008,baumgarteShapiroBook} for textbook explanations, 
  we have to ensure that our numerical evolution is a solution
  to Einstein's field equations, i.e., the Hamiltonian and the momentum
  constraints have to be zero over the entire simulation. While we explicitly
  solve the constraints to obtain our initial data, 
  the constraints are not solved during the simulation. 
  Figure~\ref{fig:Constraints_MS1b-094194} shows the $L_2$ volume norm of the Hamiltonian constraint 
  (top panel) and the $L_2$ volume norm of the square magnitude of the momentum constraint (bottom panel)
  during the simulation. 
  We see that due to the constraint damping properties of the Z4c evolution system the 
  constraints stay at or below the value of the initial data.
  Oscillations during the inspiral are mostly caused by the interface between the shells and the boxes
  and the movement of inner refinement levels which follow the motion of the neutron stars. 
  After merger those oscillations are absent since the stars stay near
  the center or only move with a small velocity compared to the inspiral.
  During the inspiral an increasing resolution shows the expected convergence, 
  see~\cite{Bernuzzi:2016pie} for a detailed discussion. After the merger the situation is less clear due to the 
  highly nonlinear evolution and the possible formation of shocks. 
 
  \subsubsection{ADM Energy and angular momentum}

  \begin{figure}[t]
  \includegraphics[width=0.5\textwidth]{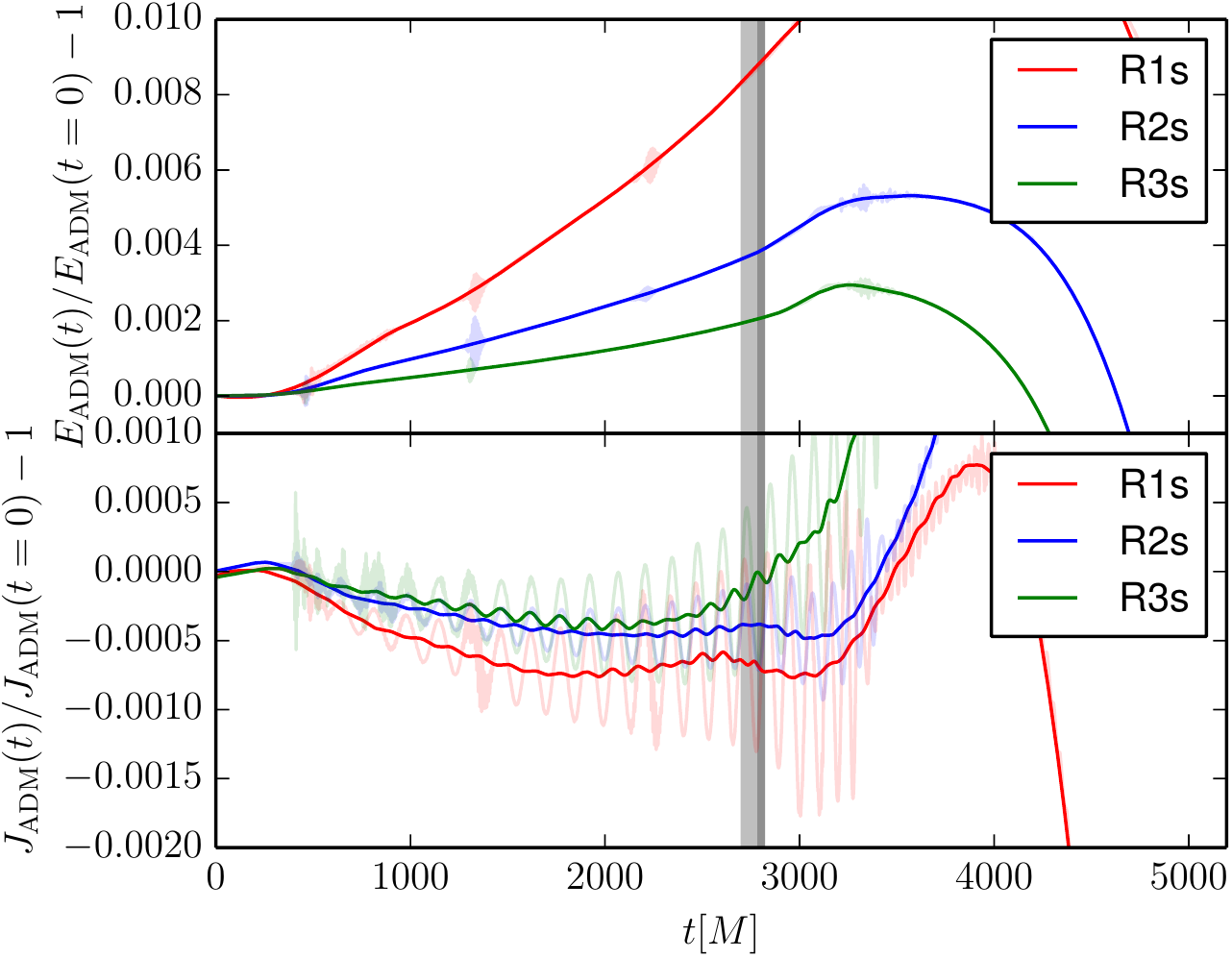}
  \caption{ADM quantities for MS1b-094194. Top panel shows the ADM energy, 
  which we correct with the amount of GW energy emitted by the system. 
  The quantity is conserved at or below the $0.5\%$ level for R2s and R3s 
  until material is leaving the computational domain. 
  Bottom panel shows the ADM-angular momentum corrected by the amount of angular momentum 
  extracted via GW emission. For R2s and R3s the ADM-angular momentum is conserved better 
  than $0.05\%$. 
  In both panels we show the numerical data as semi-transparent lines and the results after
  applying an average filter as solid lines.
  This filtering reduces oscillations caused 
  by the orbital motion and allows an easier interpretation of the data.}
  \label{fig:ADM_MS1b-094194}
  \end{figure}
  
  In principle, the ADM energy and angular momentum have to be conserved for isolated systems. 
  Because we are computing the ADM quantities at a finite distance to the binary system, 
  and energy is emitted via GWs, the ADM quantities do not stay constant. 
  To compensate this fact we compute the amount of radiated energy and 
  angular momentum and include this in our calculation. 
  Another aspect, not included in our energy/angular momentum budget is material 
  which leaves the computational domain, cf.~the discussion about the total baryonic mass. 
  Figure~\ref{fig:ADM_MS1b-094194} (upper panel) shows the deviation of the ADM-energy from the initial value. 
  Up to the moment of merger the error is $\approx 0.8\%$ for 
  the lowest and $\approx 0.2\%$ for the highest resolution. 
  Figure~\ref{fig:ADM_MS1b-094194} (bottom panel) visualizes the conservation of the ADM-angular momentum 
  $J_{\rm ADM}$. In both panels we show as a shaded line the actual numerical data and 
  as a solid line the result after applying an averaging filter to oscillations caused by the motion of the 
  two NSs. However, we find overall errors below $1\%$ even for the lowest resolution. 
  For higher resolutions the error goes down to $\approx 0.2\%$.

  \subsubsection{Waveforms}

\begin{figure}[t]
  \includegraphics[width=0.5\textwidth]{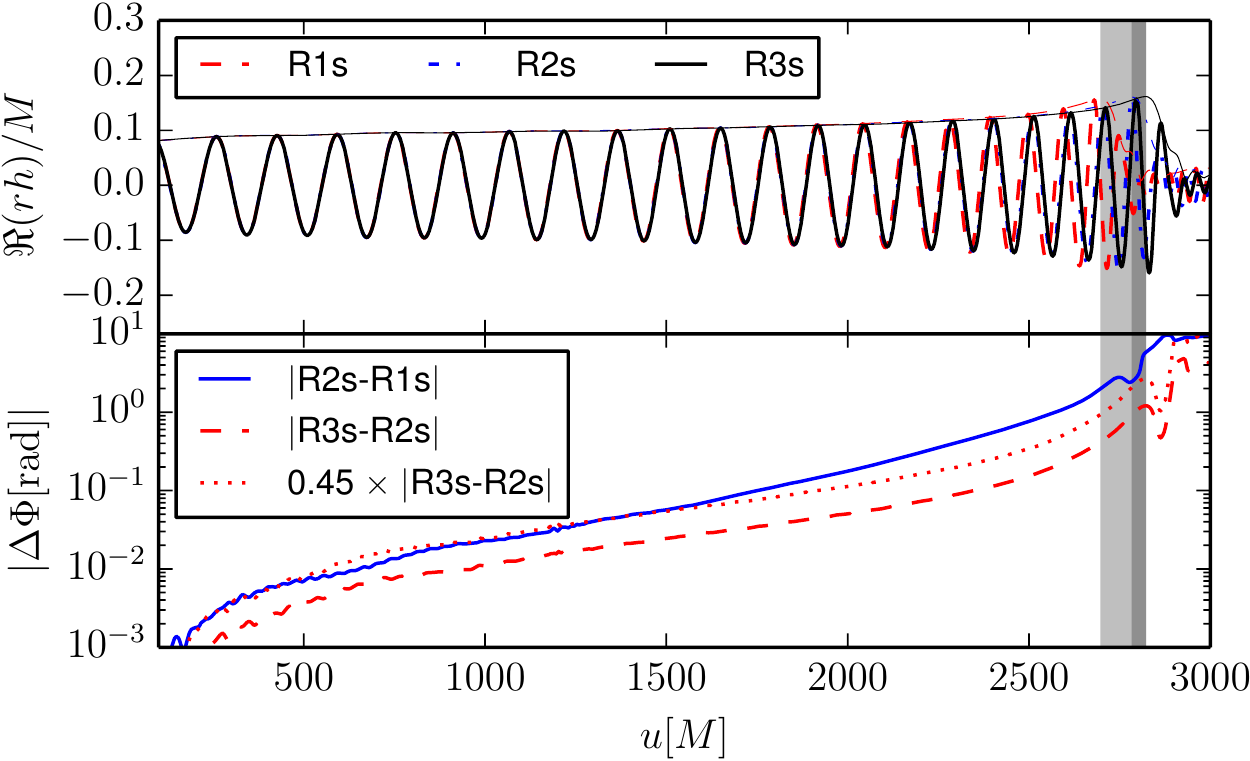}
  \caption{Top panel: Real part of the dominant (2,2) GW mode of MS1b-094194.
    Bottom panel: Phase difference between different resolutions. 
    The dotted red line corresponds to a convergence order of 1.3. 
    As for the previous discussions the different moments of merger 
    are represented as a shaded region.
}
  \label{fig:GW_q206} 
\end{figure}   

Although we have investigated the convergence behavior of BAM
waveforms many times in the past
\cite{Bernuzzi:2011aq,Hilditch:2012fp,Bernuzzi:2016pie}, we present here
a convergence test for the waveform of MS1b-094194. 
The reason is that this configuration has new features with respect
previous equal-masses runs, including new initial data, the large
mass-ratio, the significant mass transfer at 
low resolutions. Overall we find that the numerical computation is more challenging
and higher resolutions would be required to achieve the same accuracy
of $q\sim1$ runs.  

Figure~\ref{fig:GW_q206} (upper panel) shows the real part and the amplitude of the 
dominant (2,2) mode for MS1b-094194 for three different resolutions. 
Up to $\sim 1600M$ we find a convergence order $\approx 1.3$, cf.~red
dotted line in bottom panel of Fig.~\ref{fig:GW_q206}. The bottom
panel of Fig.~\ref{fig:GW_q206} also present the difference between
R2s and R1s as a dashed red line and the difference between R3s and
R2s as a solid blue line. The convergence order seems to increase
during the end of the simulation ($t>1800M$) and overconvergence is
present,
cf.~\cite{Bernuzzi:2016pie,Radice:2013xpa,Radice:2015nva}. Beside the
resolutions employed, this might be also caused by the mass
transfer which is much stronger for the R1s simulation than for the
higher resolved simulations.

\begin{acknowledgments}
  It is a pleasure to thank Roland Haas, Ian Hinder, Tanja Hinderer, Nathan~K.~Johnson-McDaniel,
  Charalampos~M.~Markakis, Serguei Ossokine, Justin Vines for helpful discussions and 
  Casey Handmer and David Radice for comments on the manuscript. 
  We are very grateful to Kyohei Kawaguchi for discussions about the 
  ejecta during BHNS mergers and for making his model and 
  code publicly available~\cite{Kawaguchi:2016ana}. 
  Computations where performed on SuperMUC at the LRZ (Munich) under 
  the project numbers pr87nu and pr48pu, 
  Juropa/Jureca (J\"ulich) under the project number HJN26 and HPO21,
  Stampede (Texas, XSEDE allocation - TG-PHY140019), and 
  the Jena group local cluster \texttt{quadler}.
\end{acknowledgments}


\bibliographystyle{apsrev}   
\bibliography{paper20160722.bbl}

\end{document}